\documentclass[a4paper,11pt]{article}
\usepackage{jheppub}

\usepackage[T1]{fontenc}
\usepackage[english]{babel}

\usepackage{lineno}
\usepackage{comment}

\usepackage{amsmath,amsfonts,amssymb}
\usepackage{mathtools}
\usepackage{braket,slashed}
\usepackage{mathrsfs}
\usepackage{nicematrix}

\usepackage{graphicx}
\usepackage{adjustbox}
\usepackage{placeins}
\usepackage{tikz}
\usepackage{tikz-feynman}
\usepackage{dsfont}
\usepackage{float}
\usepackage{subcaption}

\usepackage{multirow}
\usepackage{booktabs}

\usepackage[normalem]{ulem}

\usepackage[dvipsnames,table]{xcolor}

\hypersetup{colorlinks, urlcolor=BlueViolet, citecolor=Plum, linkcolor=PineGreen}

\newcommand{\mB}{\mathcal{B}}
\newcommand{\mH}{\mathcal{H}}
\newcommand{\mM}{\mathcal{M}}
\newcommand{\mN}{\mathcal{N}}
\newcommand{\mO}{\mathcal{O}}
\newcommand{\mU}{\mathcal{U}}

\newcommand{\mX}{\mathcal{X}}
\newcommand{\mY}{\mathcal{Y}}

\newcommand{\mC}{\mathscr{C}}

\newcommand{\tC}{\tilde{\mathcal{C}}}
\newcommand{\tE}{\tilde{\mathcal{E}}}
\newcommand{\tA}{\tilde{\mathcal{A}}}
\newcommand{\ve}{\varepsilon}

\allowdisplaybreaks

\title{\boldmath Minimal Flavoured DFSZ Axion Models: Predictive Yukawa Textures and Flavour Constraints}

\author{Peter Cox,}
\author{Maaz Hayat\footnote{Corresponding author},}
\author{Raymond R. Volkas}

\affiliation{ARC Centre of Excellence for Dark Matter Particle Physics,\\
School of Physics, The University of Melbourne,\\
Victoria 3010 Australia}

\emailAdd{peter.cox@unimelb.edu.au}
\emailAdd{maazh@student.unimelb.edu.au}
\emailAdd{raymondv@unimelb.edu.au}

\abstract{We study a predictive class of flavoured DFSZ axion models in which the Peccei--Quinn symmetry is fully horizontal and the domain wall number is unity. These \textsc{Minimal F-DFSZ} models simultaneously resolve the strong CP problem and avoid the post-inflationary domain wall problem, while enforcing a predictive set of texture-zero quark mass matrices. Remarkably, the entire quark Yukawa sector can be reconstructed in terms of the measured quark masses and Cabibbo--Kobayashi--Maskawa parameters. We derive analytic expressions for the resulting flavour-violating Higgs and axion couplings, thereby making explicit how the new-physics effects are determined by Standard Model observables. We then analyse the full flavour phenomenology of the framework and obtain bounds on the scale of new physics from precision flavour observables.}

\begin{document}

\maketitle
\flushbottom

%============================================================
\section{Introduction}
\label{sec:Introduction}
%============================================================

The absence of a natural explanation for why the CP-violating QCD parameter, $\bar{\theta}$, is constrained to extremely small values, $\bar{\theta} \lesssim 10^{-10}$~\cite{Abel:2020pzs}, is one of the most enduring puzzles of the Standard Model (SM), known as the strong CP problem.

One elegant and widely studied resolution involves promoting the parameter $\bar{\theta}$ to a dynamical field, known as the QCD axion~\cite{Peccei:1977ur,Peccei:1977hh}. The axion arises naturally as a pseudo-Nambu--Goldstone boson (pNGB) from the spontaneous breaking of a global, anomalous $U(1)$ symmetry, known as the Peccei--Quinn (PQ) symmetry.

For the PQ symmetry to be anomalous under QCD, it must act on chiral coloured fermions. This can be realised in two ways: by introducing heavy, vector-like PQ-charged quarks as in the Kim-Shifman-Vainshtein-Zakharov (KSVZ)~\cite{Kim:1979if,Shifman:1979if} model, or by embedding the PQ symmetry directly into the SM quarks by extending the scalar sector to a two-Higgs doublet model (2HDM) as in the Dine-Fischler-Srednicki-Zhitnitsky (DFSZ) model~\cite{Dine:1981rt,Srednicki:1985xd,Zhitnitsky:1980tq}.

While the KSVZ model is minimal, in principle only requiring one additional heavy quark with the SM fields remaining PQ-neutral, the DFSZ model offers a richer phenomenology through its intrinsic 2HDM structure, as well as tree-level couplings of the axion to the SM fermions. This extended scalar sector allows intriguing possibilities for flavour phenomenology, especially when the PQ symmetry is considered as a potential \textit{flavour symmetry}~\cite{Davidson:1981zd,Davidson:1983fy,Davidson:1984ik,Geng:1988nc}, distinguishing quarks of different generations.

Another strong reason to examine \textit{flavoured} DFSZ models is the well-known cosmological domain wall problem inherent in the original (flavour-blind) DFSZ scenario~\cite{Sikivie:1982qv}. This problem exists because the flavour-universal assignment of PQ charges of the original DFSZ model results in a discrete subgroup of the global $U(1)_{\text{PQ}}$ symmetry remaining unbroken by the QCD anomaly. The spontaneous breaking of this discrete symmetry in the early universe produces regions with topologically distinct vacua, separated by domain walls~\cite{Vilenkin:1982ks}. In the standard DFSZ formulation, the non-trivial domain wall number $(\mathcal{N}_{\text{DW}}>1)$ means that these domain walls are stable and come to dominate the energy density of the universe, if the PQ symmetry is broken after inflation.

An appealing resolution to the domain wall problem involves introducing flavour-dependent PQ charges that ensure the PQ symmetry is \textit{completely} broken by the QCD anomaly, setting the domain wall number to unity $(\mathcal{N}_{\text{DW}}=1)$. In previous work, we have systematically catalogued all such viable \textit{flavoured} DFSZ models, highlighting their potential to connect the origin of flavour with the axion solution to the strong CP problem~\cite{Cox:2023squ}.

In this paper, we explore a particularly well-motivated benchmark \textit{class} of models -- hereafter referred to as \textsc{Minimal Flavoured-DFSZ Models (F-DFSZ)} -- which sit at the intersection of axion and flavour physics. This class, originally identified in our previous classification of domain wall-free flavoured DFSZ models~\cite{Cox:2023squ}, as well as earlier phenomenological studies of flavourful PQ symmetries~\cite{Bjorkeroth:2018ipq, Rocha:2025ade}, features $\mathcal{N}_{\text{DW}} = 1$ implementations of the DFSZ framework with completely horizontal PQ symmetries. These symmetries enforce a highly predictive set of texture-zero mass matrices\footnotemark, for which we analytically reconstruct all Yukawa couplings in terms of the physical quark masses and Cabibbo--Kobayashi--Maskawa (CKM) parameters.\footnotetext{The same textures have previously been studied in the context of 2HDMs~\cite{Rocha:2024twm} and as SM textures~\cite{Ludl:2015lta,Ramond:1993kv}.}

The \textsc{Minimal F-DFSZ} framework encompasses six phenomenologically distinct models, each with a unique and highly predictive flavour structure. These models exhibit flavour-changing neutral currents (FCNCs) mediated both by the axion and the scalar fields of the underlying two-Higgs-doublet model. Crucially, the horizontal PQ symmetry of the \textsc{Minimal F-DFSZ} framework ensures that FCNCs are controlled by the hierarchical structure of CKM elements and fermion masses. Nevertheless, we find that there are important, potentially observable contributions to precision flavour observables, particularly in neutral meson mixing. Furthermore, meson decays such as $K \to \pi a$ are already probing the post-inflationary axion dark matter window.

This mechanism of suppressing FCNCs has clear analogues in the flavour physics literature and aligns with the broader principle of Minimal Flavour Violation (MFV)~\cite{DAmbrosio:2002vsn,Botella:2009pq}, where all flavour and CP violation arises from the SM Yukawa couplings. A particularly relevant example in the context of the general type-III 2HDM is BGL models \cite{Branco:1996bq}, where a symmetry enforces that FCNCs are controlled by CKM matrix elements.

The \textsc{Minimal F-DFSZ} models derive the same predictive control from a single, flavoured $U(1)_{\text{PQ}}$ symmetry -- solving the strong CP problem, furnishing a viable axion dark matter candidate, and shaping the flavour sector in one unified construction.

%============================================================
\section{Minimal F-DFSZ Models}
\label{sec:mfd}
%============================================================

The \textsc{Minimal F-DFSZ} models we study in this paper form a distinctive subclass of flavoured DFSZ models free of a cosmological domain wall problem that were comprehensively classified in our previous work~\cite{Cox:2023squ}. This subclass of models was first identified in~\cite{Bjorkeroth:2018ipq} and subsequently studied in~\cite{Rocha:2025ade}. Here, we briefly summarise the defining features of these models.

The PQ symmetry of a flavoured DFSZ model is defined to act on the matter fields as follows:
\begin{align}
q_{Li} &\rightarrow e^{i\mX(q_i)\alpha} q_{Li}, & u_{Ri} &\rightarrow e^{i\mX(u_i)\alpha} u_{Ri}, & d_{Ri} &\rightarrow e^{i\mX(d_i)\alpha} d_{Ri}, \nonumber \\
\Phi_r &\rightarrow e^{i\mX_r\alpha} \Phi_r, & S &\rightarrow e^{i\mX_S \alpha} S,
\label{pqsymm}
\end{align}
where $i=1,2,3$ and $r=1,2$. The $\Phi_r$ fields are the Higgs doublets and $S$ is called the PQ scalar. The \textsc{Minimal F-DFSZ} framework enforces three additional conditions that restrict the PQ charges:
\begin{enumerate}
    \item The domain wall number is unity ($\mN_{DW}=1$) in order to circumvent the cosmological domain wall problem for a post-inflationary axion.
    \item The PQ symmetry is \textit{fully horizontal}. Each fermion field is distinguished by its PQ charge.\footnote{A fully horizontal PQ symmetry with $\mN_{DW}=2$ has also been studied~\cite{Bjorkeroth:2018ipq}.}
    \item The resulting Yukawa Lagrangian must be consistent with the measured masses and mixings of the quark sector.
\end{enumerate}
The first condition imposes the following constraint on the colour anomaly equation,
\begin{equation}
    \mN_{DW}=|2N|=\bigg| \sum_{i=1}^{3}\left[2\mX(q_i)-\mX(u_i)-\mX(d_i)\right] \bigg| =1,
    \label{dm1}
\end{equation}
while the second condition requires that each of the charges $\mX(\psi_i)$ be distinct. Notice that $\mN_{DW}$ is invariant under each of the following \textit{permutations} of the PQ charges:
\begin{align}
    \mX(q_k) &\leftrightarrow \mX(q_{\ell}), & \mX(u_k) &\leftrightarrow \mX(u_{\ell}), & \mX(d_k) &\leftrightarrow \mX(d_{\ell}), & \mathcal{X}_1 &\leftrightarrow \mathcal{X}_2, & u &\leftrightarrow d.
    \label{eq:perm_symm}
\end{align}
Although the colour anomaly condition, Eq.~\eqref{dm1}, admits these permutation symmetries, most permutations do not correspond to physically distinct models:
\begin{equation}\label{eq:perm_items}
\begin{minipage}{0.95\linewidth}
\begin{itemize}
    \item Permuting the quark PQ charges $\mX(q_k)$, $\mX(u_k)$, and/or $\mX(d_k)$ is a simple relabelling of the gauge eigenstates and has no physical consequence. 
    \item Permuting the PQ charges of the Higgs doublets, $\mX_1\leftrightarrow \mX_2$, together with $\mX_S~\leftrightarrow~-\mX_S$, corresponds to sending $\tan\beta \to \cot\beta$.
    \item Interchanging the PQ charges of the up- and down-type fermions, $u \leftrightarrow d$, does yield a distinct model, since they carry different hypercharges. Note that such an interchange requires $\mX(q_i) \to -\mX(q_i)$, $\mX(u_i) \to - \mX(d_i)$, and $\mX(d_i) \to -\mX(u_i)$ to allow the required Yukawa couplings.
\end{itemize}
\end{minipage}
\end{equation}

Remarkably, for three generations of quarks, the three requirements above single out a unique set of PQ charges for the quark fields, up to exchanging the charges of the up and down type quarks. We list these SM+PQ charges in Table~\ref{tab:pq_charges}.

\begin{table}[t!]
\centering
\renewcommand{\arraystretch}{1.1}
\setlength{\tabcolsep}{12pt}
\begin{tabular}{ c c c c }
\toprule
\textbf{Field} & \textbf{SM Representation} & $\mX$ ($u$-type model) & $\mX$ ($d$-type model)  \\ 
\midrule
$q_{Li}$   & $(\mathbf{3},\ \mathbf{2},\ +\tfrac{1}{3})$ 
           & $\{1,\ 2,\ 0\}$ & $\{-1,\ -2,\ 0 \}$\\
$u_{Ri}$   & $(\mathbf{3},\ \mathbf{1},\ +\tfrac{4}{3})$ 
           & $\{1+c_\beta^2,\ 2+c_\beta^2,\ -s_\beta^2\}$ & $\{-s_{\beta}^2,\ -(1+s_{\beta}^2),\ c_{\beta}^2\}$\\
$d_{Ri}$   & $(\mathbf{3},\ \mathbf{1},\ -\tfrac{2}{3})$ 
           & $\{s_\beta^2,\ 1+s_\beta^2,\ -c_\beta^2\}$ &$\{-(1+c_{\beta}^2),\ -(2+c_{\beta}^2),\ s_{\beta}^2\}$\\
\midrule
$\Phi_1$   & $(\mathbf{1},\ \mathbf{2},\ +1)$ 
           & $-s_{\beta}^2$ & $-s_{\beta}^2$\\
$\Phi_2$   & $(\mathbf{1},\ \mathbf{2},\ +1)$ 
           & $c_{\beta}^2$ & $c_{\beta}^2$\\
$S$        & $(\mathbf{1},\ \mathbf{1},\ 0)$ 
           & $1$ &$1$\\
\bottomrule
\end{tabular}
\caption{Field content and PQ charge assignments of the up-type and down-type \textsc{Minimal F-DFSZ} models. PQ charges are denoted by $\mX$ and represented in the basis where the PQ and hypercharge currents are orthogonal and $\mX(q_3)=0$. Fermion PQ charges are generation-dependent. Note that $\tan\beta = v_2/v_1$.}
\label{tab:pq_charges}
\end{table}

The PQ symmetry structure has direct phenomenological consequences. In particular, it forbids specific Yukawa operators, leading to a predictive set of texture-zero Yukawa matrices in the quark sector which are stable under renormalisation group evolution. The general Yukawa Lagrangian is
\begin{align}
\label{eq:LYgauge}
\mathcal{L}_Y &=
 -\bar{Q}_L\bigl( \mY_d^1\,\Phi_1 + \mY_d^2\,\Phi_2 \bigr)d_R
  -
  \bar{Q}_L\bigl( \mY_u^1\,\tilde{\Phi}_1 + \mY_u^2\,\tilde{\Phi}_2 \bigr)u_R
  \;+\; \text{h.c.},
\end{align}
where $\tilde{\Phi}_r \equiv i\sigma_2\,\Phi_r^*$. The PQ charges for the up-type model in Table~\ref{tab:pq_charges} enforce the following textures for the Yukawa couplings:
\NiceMatrixOptions{columns-width=5mm}
\begin{align}
    \mY_u^1 = \begin{pNiceMatrix}
        0 &0 &0 \\
        \ast & 0 &0 \\
        0&0&\times
    \end{pNiceMatrix}, &\qquad \mY_u^2 = \begin{pNiceMatrix}
        \times & 0 &0 \\
        0 & \times & 0\\
        0 & 0 & 0
    \end{pNiceMatrix},\nonumber \\ 
    \mY_d^1= \begin{pNiceMatrix}
        0&\times&0 \\
        0&0&0 \\
        \times &0 & 0
    \end{pNiceMatrix}, &\qquad
    \mY_d^2 =  \begin{pNiceMatrix}
        \times & 0& 0\\
        0 & \times & 0 \\
        0 & 0 & \times
    \end{pNiceMatrix},
\label{eq:Yu1Yu2}
\end{align}
where $\times$ $(\ast)$ denotes a real (complex) coupling after absorption of unphysical phases. Without loss of generality, we place the single physical phase in the $(2,1)$-element of $\mY_u^1$.

After electroweak symmetry breaking (EWSB), the corresponding texture-zero mass matrices are given by
\begin{align}
\mM_{q} &= \dfrac{v}{\sqrt{2}}\left(c_{\beta}\mY_{q}^1+s_{\beta}\mY_{q}^2\right), 
\label{eq:massyukreln}
\end{align}
\NiceMatrixOptions{columns-width=5mm}
\begin{align}
    \mM_u\, \sim\, \begin{pNiceMatrix}
\times & 0 & 0 \\
\ast & \times & 0 \\
0 & 0 & \times
\end{pNiceMatrix}, \quad \mM_d\, \sim\, \begin{pNiceMatrix}
\times & \times & 0 \\
0 & \times & 0 \\
\times & 0 & \times
\end{pNiceMatrix}.
\label{eq:Mass_Struc}
\end{align}
There are two important points regarding the Yukawa and mass textures in Eqs.~\eqref{eq:Yu1Yu2} and~\eqref{eq:Mass_Struc}:
\begin{enumerate}
    \item The mass matrix textures in Eq.~\eqref{eq:Mass_Struc} contain exactly ten free parameters (nine magnitudes and one phase) matching the six quark masses and four CKM parameters.
    \label{point1}
    \item Each quark bilinear can couple to only one of the Higgs doublets (i.e. simultaneous non-zero elements in $\mY_{q}^1$ and $\mY_{q}^2$ are forbidden) as can be seen in Eq.~\eqref{eq:Yu1Yu2}. This is a direct consequence of the PQ symmetry.
    \label{point2}
\end{enumerate}
These imply that the Yukawa matrices---and hence the full flavour structure of the new physics---can be expressed entirely in terms of measured observables as well as $\tan\beta$. This predictivity is reminiscent of the \textit{nearest-neighbour interaction} (NNI) form~\cite{Branco:1988iq}, of which Fritzsch textures are a restrictive subclass, as well as BGL models. However, unlike generic NNI and BGL structures, the \textsc{Minimal F-DFSZ} mass matrices feature FCNCs in both sectors as well as a ``\textit{decoupled quark}''.

The decoupled quark follows from the block-diagonal form of one of the mass matrices in Eq.~\eqref{eq:Mass_Struc}: $\mM_u$ for the displayed up-type charge assignment, or $\mM_d$ for the down-type PQ charge assignment. The isolated $(3,3)$ state does not participate in intergenerational mixing, so its gauge and mass eigenstates are aligned. Identifying this state with any one of the six quarks, $\{u,d,s,c,b,t\}$, defines a physically distinct model with quantitatively different flavour phenomenology.

The \textsc{Minimal F-DFSZ} framework encompasses not a single model but a family of six predictive scenarios, each defined by which quark is decoupled. At the same time, the axion---predominantly arising from the phase of the scalar singlet $S$---inherits this flavour structure, coupling to SM fermions in a flavour-violating pattern entirely fixed by the PQ charges. The combination of a predictive quark sector with flavour-violating Higgs couplings and an intrinsically flavoured axion makes the \textsc{Minimal F-DFSZ} framework a compelling and testable scenario for both collider searches and precision flavour experiments.

%============================================================
\section{Reconstruction of the Mass Matrices}
\label{sec:Recons}
%============================================================

In this section, we analytically reconstruct the mass matrices of the \textsc{Minimal F-DFSZ} models in terms of the quark masses and CKM parameters. As discussed in the previous section, these matrices contain exactly ten free parameters. Consequently, the Yukawa couplings can be expressed entirely in terms of physical observables, leaving no additional freedom beyond the assignment of the decoupled quark.

A complex mass matrix $\mM$ can be diagonalised via a singular value decomposition. Introducing the left- and right-handed unitary rotation matrices $\mU_L$ and $\mU_R$, we have
\begin{align}
\mU_L^\dagger\,\mM\,\mU_R=\mathcal{D},
\end{align}
where $\mathcal{D}=\textrm{Diag}\{m_1,m_2,m_3\}$ contains the quark masses, conventionally ordered in ascending magnitude.

The relative misalignment of the left-handed rotations in the up and down sectors defines the CKM matrix,
\NiceMatrixOptions{columns-width=3mm}
\begin{align}
V=\mU_{uL}^\dagger\mU_{dL}&=
\begin{pNiceMatrix}
        V_{ud}&V_{us}&V_{ub}\\
        V_{cd}&V_{cs}&V_{cb}\\
        V_{td}&V_{ts}&V_{tb}
    \end{pNiceMatrix},
\label{eq:standardparam} 
\end{align}
which is usually parameterised in terms of three mixing angles $\theta_{ij}$ and one CP violating phase $\delta_{\rm CP}$. The numerical inputs for these parameters as well as the quark masses are collected in Table~\ref{tab:quarkparams_mz} in App.~\ref{app:MassMatrixFits}. For our analytic reconstruction, it will be more convenient to work directly with the rephasing-independent observables, $|V_{ij}|$ and the CP-violating Jarlskog invariant $\mathcal{J}$ defined by
\begin{align}
    \textrm{Im}(V_{\alpha i}V_{\beta j}V_{\alpha j}^*V_{\beta i}^*)=\mathcal{J}\sum_{\gamma}\epsilon_{\alpha \beta \gamma}\sum_k\epsilon_{ijk}.
\end{align}

To analytically reconstruct the CKM matrix, it is convenient to consider Hermitian combinations of mass matrices $\mH_i = \mM_i \mM_i^{\dagger}$. For \emph{up-type models} (i.e. if the decoupled quark is an up-type quark), these can be parameterised as
\begin{align}
\mH_u=\mM_u\mM_u^{\dagger}&= 
\begin{pmatrix}
E_u & D_ue^{-i\varphi} & 0 \\[1mm]
D_ue^{i\varphi} & C_u & 0 \\[1mm]
0 & 0 & A_u
\end{pmatrix},
\qquad
\mH_d=\mM_d\mM_d^\dagger =
\begin{pmatrix}
E_d & D_d & B_d \\[1mm]
D_d & C_d & 0 \\[1mm]
B_d & 0 & A_d
\end{pmatrix},
\label{eq:H2}
\end{align}
where all parameters are taken to be positive without loss of generality.\footnote{One can always rotate the quark fields such that each parameter in Eq.~\eqref{eq:H2} is positive.} For \emph{down-type models}, the textures are swapped (i.e. replace $u \leftrightarrow d$ in the above expressions).

The eigenvalues of $\mH_{u,d}$ are the squared quark masses, which can be expressed in terms of the matrix invariants:
\begin{align}
    \det(\mH_i) &= \prod_{k=1}^{3}m_k^2, \quad \mathrm{Tr}(\mH_i) = \sum_{k=1}^3m_k^2, \qquad\tfrac{1}{2}\big[\mathrm{Tr}(\mH_i)^2 - \mathrm{Tr}(\mH_i^2)\big] = \sum_{j<k}^3m_j^2 m_k^2.
    \label{eq:matrixinvariants}
\end{align}
Considering both the up- and down-sectors, Eq.~\eqref{eq:matrixinvariants} can be used to fix the parameters $\{A_u, C_u, D_u\}$ and $\{B_d, D_d, E_d\}$ in terms of the quark masses. The identification of the parameter $A_u$ with $\{m_u^2,m_c^2,m_t^2\}$ (or with $\{m_d^2,m_s^2,m_b^2\}$ for the $u \leftrightarrow d$ permutation) defines the $q$-decoupled model. The remaining four parameters $\{A_d, C_d, E_u, \varphi\}$ are to be fixed in terms of CKM matrix elements.

%============================================================
\subsection{Example: $t$-decoupled Model}

For the remainder of this section, we focus on the specific case where $A_u=m_t^2$, known as the ``$t$-decoupled'' model, for illustration. We denote the mass matrices for this model as $\mM_{u}^{(t)}$ and $\mM_d^{(t)}$. The generalisation to ``$q$-decoupled'' is straightforward. 

First, Eq.~\eqref{eq:matrixinvariants} can be used to derive
\begin{align}
    D_u &=  \sqrt{E_u - m_u^2} \sqrt{m_c^2 - E_u},&\qquad E_d = m_d^2 + m_s^2 + m_b^2 - A_d - C_d, \nonumber \\
C_u &= m_u^2 + m_c^2 - E_u, &\qquad D_d = \dfrac{ \sqrt{C_d - m_d^2} \sqrt{m_s^2 - C_d} \sqrt{m_b^2 - C_d} }{ \sqrt{A_d - C_d} }, \nonumber \\
A_u &= m_t^2, &\qquad B_d = \dfrac{ \sqrt{A_d - m_d^2} \sqrt{A_d - m_s^2} \sqrt{m_b^2 - A_d} }{ \sqrt{A_d - C_d} },
\label{eq:inv_soln}
\end{align}
Since all the parameters of $\mH_{u,d}$ are real and positive, Eq.~\eqref{eq:inv_soln} implies the following bounds,
\begin{align}
    m_u^2 &\leq E_u \leq m_c^2, &\qquad m_d^2 &\leq C_d \leq m_s^2, &\qquad m_s^2 &\leq A_d \leq m_b^2.
    \label{eq:bounds}
\end{align}

Diagonalising $\mH_{u,d}$ in Eq.~\eqref{eq:H2} leads to the following expressions for the elements of $\mU_{uL}$ and $\mU_{dL}$ in terms of fermion mass ratios and the parameters $\{A_d,C_d,E_u,\varphi\}$:
\begin{align}
    [\mU_{uL}]_{11}&=\left(1+\left(\frac{m_u^2-E_u}{D_u}\right)^2\right)^{-1/2} \,, &\quad
    [\mU_{uL}]_{22}&=\exp(i\varphi)\left(1+\left(\frac{D_u}{m_c^2-E_u}\right)^2\right)^{-1/2} \,, \nonumber \\ \nonumber \\
    [\mU_{uL}]_{12}&=\exp(-i\varphi)\frac{D_u}{m_c^2-E_u}[\mU_{uL}]_{22} \,, &\quad 
    [\mU_{uL}]_{21}&=\exp(i\varphi)\left(\frac{m_u^2-E_u}{D_u}\right)[\mU_{uL}]_{11} \,, \nonumber\\ \nonumber \\
    [\mU_{uL}]_{3i}&=[\mU_{uL}]_{i3}=\delta_{3i} \,,
    \label{eq:O1}
\end{align}
and
\begin{align}
[\mU_{dL}]_{11} &= \left( 1 + \left( \frac{D_d}{m_d^2 - C_d} \right)^2 + \left( \frac{B_d}{m_d^2 - A_d} \right)^2 \right)^{-1/2} \,,
&\quad
[\mU_{dL}]_{2i} &= \frac{D_d}{m_{d_i}^2 - C_d} [\mU_{dL}]_{1i} \,, \nonumber \\
[\mU_{dL}]_{22} &= \left( 1 + \left( \frac{m_s^2 - C_d}{D_d} \right)^2 + \left( \frac{B_d}{D_d} \cdot \frac{m_s^2 - C_d}{m_s^2 - A_d} \right)^2 \right)^{-1/2} \,,
&\quad
[\mU_{dL}]_{3i} &= \frac{B_d}{m_{d_i}^2 - A_d} [\mU_{dL}]_{1i} \,, \nonumber \\
[\mU_{dL}]_{33} &= \left( 1 + \left( \frac{m_b^2 - A_d}{B_d} \right)^2 + \left( \frac{D_d}{B_d} \cdot \frac{m_b^2 - A_d}{m_b^2 - C_d} \right)^2 \right)^{-1/2} \,.
&&
\label{eq:O2}
\end{align}

For convenience, we define the following down-sector parameters,
\begin{equation}
    r_j \equiv \frac{D_d}{m_{d_j}^2 - C_d}\,, \qquad
    t_j \equiv \frac{B_d}{m_{d_j}^2 - A_d}\,, \qquad
    N_j \equiv \left(1 + r_j^2 + t_j^2\right)^{-1/2}
    = \sqrt{\frac{(C_d - m_{d_j}^2)(A_d - m_{d_j}^2)}{\prod_{l\neq j}(m_{d_j}^2 - m_{d_l}^2)}}\,,
    \label{eq:CKMbuildingblocks}
\end{equation}
together with the up-sector mixing angles
\begin{equation}
    s_u \equiv \sin\theta_u = \sqrt{\frac{E_u-m_u^2}{m_c^2-m_u^2}}\,,
    \qquad
    c_u \equiv \cos\theta_u = \sqrt{\frac{m_c^2-E_u}{m_c^2-m_u^2}}\,.
    \label{eq:upmixing}
\end{equation}
For the $t$-decoupled model, the CKM elements then take the compact form
\begin{equation}
    V_{1j} = N_j\left(c_u - s_u\, e^{-i\varphi}\, r_j\right), \qquad
    V_{2j} = N_j\left(s_u + c_u\, e^{-i\varphi}\, r_j\right), \qquad
    V_{3j} = N_j\, t_j\,.
    \label{eq:CKMANALYTIC}
\end{equation}
Notice that the third row is fixed entirely by the down-sector parameters $A_d$ and $C_d$, and the down-type masses. This is a direct consequence of the decoupled top quark which implies that the rotation matrices $\mU_{uL}$ and $\mU_{uR}$ have a block diagonal form,
\begin{align}
    \mU_{(uL/uR)}=\begin{pNiceMatrix}
        \times & \times & 0\\
        \ast & \ast & 0\\
        0 &0 & 1
    \end{pNiceMatrix}.
    \label{eq:UuL}
\end{align}
Inverting the definition of the CKM $V=\mU_{uL}^\dagger\mU_{dL}$ then implies
\begin{align}
     \mU_{dL}=\begin{pNiceMatrix}
         \times & \times & \times \\
         \times & \times & \times \\
         V_{td} & V_{ts} & V_{tb}
     \end{pNiceMatrix}.
     \label{eq:UdL}
\end{align}

We can now derive analytic solutions for the remaining free parameters $\{A_d,C_d,E_u,\varphi\}$ in terms of CKM matrix elements. We begin with the down-sector parameters $A_d$ and $C_d$. Since the third row of $\mU_{dL}$ \emph{is} the third row of the CKM matrix, comparing Eq.~\eqref{eq:UdL} with its analytic form in Eq.~\eqref{eq:O2} gives
\begin{align}
|V_{td}|&=\frac{\sqrt{(1-\tA_d)(\tA_d-\ve_{sb}^2)}}{\sqrt{(1-\ve_{db}^2)(\ve_{sb}^2-\ve_{db}^2)}} \, \frac{\sqrt{\ve_{sb}^2\tC_d-\ve_{db}^2}}{\sqrt{\tA_d-\ve_{sb}^2\tC_d}}
  \label{eq:V31},\\
|V_{tb}| &= \frac{\sqrt{(\tA_d-\ve_{db}^2)(\tA_d-\ve_{sb}^2)}}{\sqrt{(1-\ve_{db}^2)(1-\ve_{sb}^2)}} \, \frac{\sqrt{1-\ve_{sb}^2\tC_d}}{\sqrt{\tA_d-\ve_{sb}^2\tC_d}},
\label{eq:V33}
\end{align}
where here we have defined the dimensionless parameters,
\begin{align} \label{eq:tilde-params}
    \tE_u \equiv E_u/m_c^2,\quad  \tC_d\equiv C_d/m_s^2, \quad \tA_d \equiv A_d/m_b^2, \quad \ve_{ij}\equiv m_i/m_j,\,
\end{align}
such that $\tE_u,\tC_d$ and $\tA_d$ have upper bounds of unity (for the top-decoupled model).
Equations~\eqref{eq:V31} and \eqref{eq:V33} can be inverted to obtain an exact solution for $\tA_d$ and $\tC_d$ in terms of CKM elements and fermion mass ratios. We find:
\begin{align}
    \tA_d&=|V_{tb}|^2+|V_{ts}|^2\,\ve_{sb}^2+|V_{td}|^2\,\ve_{db}^2, \\
    \tC_d&=\tfrac{|V_{td}|^2}{(1-|V_{tb}|^2)}\,+\,\tfrac{|V_{ts}|^2\,|V_{td}|^2\,(1+|V_{tb}|^2)}{|V_{tb}|^2\,(1-|V_{tb}|^2)^2}\ve_{sb}^2\,+\,\tfrac{|V_{ts}|^2}{(1-|V_{tb}|^2)}\ve_{ds}^2 \, + \, \mO(\ve_{db}^2),
\end{align}
where we have used the unitarity of the CKM matrix and expanded the solution for $\tC_d$ to $\mO(\ve_{ij}^2)$.

Substituting the above solutions into the analytic expression for the CKM matrix then allows us to solve for $\tE_u$. We find to leading order in $\ve_{ij}$ that
\begin{align}
    \tE_u\simeq\frac{|V_{cb}|^2}{1-|V_{tb}|^2}\left(1+\mO(\ve_{sb}^2,\,\ve_{uc}^2)\right).
\end{align}
Similarly, the CP-violating phase $\varphi$ can be expressed in terms of the Jarlskog invariant $\mathcal{J}$, with the leading order solution given by
\begin{align}
    \sin \varphi \simeq \dfrac{\mathcal{J}\left(1-|V_{tb}|^2\right)}{|V_{ub}||V_{cb}||V_{td}||V_{ts}||V_{tb}|} + \mO(\ve_{sb}^2).
\end{align}

In terms of measured observables, the mass matrices for the $t$-decoupled model are then given by
{\setlength{\abovedisplayskip}{6pt}
 \setlength{\belowdisplayskip}{6pt}
 \setlength{\abovedisplayshortskip}{4pt}
 \setlength{\belowdisplayshortskip}{4pt}
\NiceMatrixOptions{columns-width=15mm}
\begin{align}
\mM_u^{(t)}& \simeq m_t \begin{pNiceMatrix}
 \dfrac{|V_{cb}|}{\sqrt{1-|V_{tb}|^2}}\,\ve_{ct} & 0 & 0 \\
\dfrac{ e^{i\varphi} \, |V_{ub}| } { \, \sqrt{1-|V_{tb}|^2}}\ve_{ct} & \dfrac{ \sqrt{1-|V_{tb}|^2}}{|V_{cb}| }\ve_{ut} & 0 \\
0 & 0 & 1 
\end{pNiceMatrix}, \\
\mM_d^{(t)}& \simeq m_b \begin{pNiceMatrix}
\sqrt{1-|V_{tb}|^2}  & \dfrac{|V_{ts}|}{|V_{tb}| \sqrt{1-|V_{tb}|^2}}\ve_{sb} & 0 \\
0 & \dfrac{|V_{td}|}{\sqrt{1-|V_{tb}|^2}}\ve_{sb}& 0 \\
|V_{tb}| & 0 & \dfrac{\ve_{db}}{|V_{td}|}
\end{pNiceMatrix},
\end{align}
}
where we have truncated the series to $\mO(\ve_{ij})$ for illustration.

%============================================================
\subsection{General solution}
\label{sec:General_Solution}
The above analysis for the $t$-decoupled model can be straightforwardly generalised to the other $q$-decoupled models. 

\paragraph{$u,c$--decoupled:}
For the remaining up-type models, the solutions for $\{A_u,C_u,D_u\}$ in Eq.~\eqref{eq:inv_soln} generalise via the replacements
\begin{equation}
\begin{aligned}
u\text{-decoupled: }
(m_u,m_c,m_t) &\rightarrow (m_c,m_t,m_u), \\
c\text{-decoupled: }
(m_u,m_c,m_t) &\rightarrow (m_u,m_t,m_c).
\label{eq:permreplacements}
\end{aligned}
\end{equation}
Note that the ranges for the parameters $\{E_u,C_d,A_d\}$ in Eq.~\eqref{eq:bounds} are also modified. The solutions for the down-sector parameters $\{B_d,D_d,E_d\}$ in Eq.~\eqref{eq:inv_soln} are the same for all up-type decoupled models.

The diagonalisation matrices $\mU_{(uL/uR)}$ for any up-type decoupled model take the general form
\begin{align}
    \mU_{(uL/uR)}=\begin{pNiceMatrix}
        \times & \times & 0\\
        \ast & \ast & 0\\
        0 &0 & 1
    \end{pNiceMatrix} \mathcal{P}_q\qquad \mU_{dL}=\begin{pNiceMatrix}
         \times & \times & \times \\
         \times & \times & \times \\
         V_{qd} & V_{qs} & V_{qb}
     \end{pNiceMatrix},
\end{align}
where the elements denoted by $\times$ and $\ast$ are given by Eqs.~\eqref{eq:O1} and \eqref{eq:O2}, but with the quark mass replacements in Eq.~\eqref{eq:permreplacements}. The matrix $\mathcal{P}_q$ is a permutation matrix:
\begingroup
\setlength{\arraycolsep}{6pt}
\renewcommand{\arraystretch}{1.0}
\begin{align}
\mathcal{P}_u &= \begin{pmatrix}
0 & 1 & 0 \\
0 & 0 & 1 \\
1 & 0 & 0
\end{pmatrix},
&
\mathcal{P}_c &= \begin{pmatrix}
1 & 0 & 0 \\
0 & 0 & 1 \\
0 & 1 & 0
\end{pmatrix},
&
\mathcal{P}_t &= \begin{pmatrix}
1 & 0 & 0 \\
0 & 1 & 0 \\
0 & 0 & 1
\end{pmatrix},
\end{align}
\endgroup
which permutes the columns of the block-diagonal rotation matrix so that the diagonal matrix $\mathcal{D}$ has the conventional mass ordering $\mathcal{D}=\textrm{Diag}\{m_1,m_2,m_3\}$. It then follows from the definition of the CKM matrix, $V=\mU_{uL}^{\dagger}\mU_{dL}$, that $\mathcal{P}_q$ maps the third row of $\mU_{dL}$, $(V_{qd},V_{qs},V_{qb})$, onto the row of $V$ corresponding to the decoupled quark $q$.

Following the same derivation as the $t$-decoupled model, the solutions for $\{\tA_d,\tC_d\}$ for any up-type decoupled model are 
{\setlength{\abovedisplayskip}{6pt}
 \setlength{\belowdisplayskip}{6pt}
 \setlength{\abovedisplayshortskip}{4pt}
 \setlength{\belowdisplayshortskip}{4pt}
\begin{align}
\tA_d&=|V_{qb}|^2+|V_{qs}|^2\,\ve_{sb}^2+|V_{qd}|^2\,\ve_{db}^2, \nonumber\\
    \tC_d&=\tfrac{|V_{qd}|^2}{(1-|V_{qb}|^2)}\,+\,\tfrac{|V_{qs}|^2\,|V_{qd}|^2\,(1+|V_{qb}|^2)}{|V_{qb}|^2\,(1-|V_{qb}|^2)^2}\ve_{sb}^2\,+\,\tfrac{|V_{qs}|^2}{(1-|V_{qb}|^2)}\ve_{ds}^2 \, + \, \mO(\ve_{ij}^4),
    \label{eq:ecusoln}
\end{align}
}
where $q \in \{u,c,t\}$ denotes the decoupled quark. 

In the $(u,c)$-decoupled models we define $\tE_u \equiv E_u/m_t^2$ (note this differs from the definition in Eq.~\eqref{eq:tilde-params}, but is consistent with the replacements in Eq.~\eqref{eq:permreplacements} and ensures $\tE_u<1$), which has the solution
\begin{equation}
    \tE_u \simeq \dfrac{|V_{tb}|^2}{1-|V_{qb}|^2},
\end{equation}
with $q \in \{u,c\}$. Finally, the CP-violating phase can be similarly expressed in terms of the Jarlskog invariant with the leading order solution given by
\begin{align}
    \sin \varphi = \pm \dfrac{\mathcal{J}(1-|V_{qb}|^2)}{|V_{ub}||V_{cb}||V_{qd}||V_{qs}||V_{tb}|},
\end{align}
where $q \in \{u,c,t\}$ is the decoupled quark and  $\sin \varphi$ is positive for the $u,t$-decoupled models and negative for the $c$-decoupled model. 

Putting everything together, the mass matrices for the $u,c$ decoupled models are
\begin{subequations}\label{eq:massmat_cases}
\begin{align}
\text{(i) } &\text{$u$-decoupled:} \nonumber \\[2mm]
\mM_u^{(u)} &\simeq
m_t\begin{pNiceMatrix}
 \,\dfrac{|V_{tb}|}{\sqrt{1-|V_{ub}|^2}} & 0 & 0 \\
e^{i\varphi} \dfrac{|V_{cb}|}{\sqrt{1-|V_{ub}|^2}}  & \dfrac{\sqrt{1-|V_{ub}|^2}}{|V_{tb}|}\ve_{ct} & 0 \\
0 & 0 & \ve_{ut}
\end{pNiceMatrix},
\\
\mM_d^{(u)} &\simeq
m_b\begin{pNiceMatrix}
\sqrt{1-|V_{ub}|^2}  & \dfrac{ |V_{us}|}{|V_{ub}|\sqrt{1-|V_{ub}|^2}} \ve_{sb}& 0 \\
0 & \dfrac{ |V_{ud}|}{\sqrt{1-|V_{ub}|^2}}\ve_{sb} & 0 \\
 |V_{ub}| & 0 & \dfrac{\ve_{db}}{|V_{ud}|}
\end{pNiceMatrix},
\label{eq:massmat_u_decoupled} \\[4mm]
\text{(ii) } &\text{$c$-decoupled:} \nonumber \\[2mm]
\mM_u^{(c)} &\simeq
m_t\begin{pNiceMatrix}
 \,\dfrac{|V_{tb}|}{\sqrt{1-|V_{cb}|^2}} & 0 & 0 \\
e^{i\varphi} \dfrac{|V_{ub}|}{\sqrt{1-|V_{cb}|^2}}  & \dfrac{\sqrt{1-|V_{cb}|^2}}{|V_{tb}|}\ve_{ut} & 0 \\
0 & 0 & \ve_{ct}
\end{pNiceMatrix},
\\
\mM_d^{(c)} &\simeq
m_b\begin{pNiceMatrix}
\sqrt{1-|V_{cb}|^2}  & \dfrac{ |V_{cs}|}{|V_{cb}|\sqrt{1-|V_{cb}|^2}} \ve_{sb}& 0 \\
0 & \dfrac{ |V_{cd}|}{\sqrt{1-|V_{cb}|^2}}\ve_{sb} & 0 \\
 |V_{cb}| & 0 & \dfrac{\ve_{db}}{|V_{cd}|}
\end{pNiceMatrix}.
\label{eq:massmat_c_decoupled}
\end{align}
\end{subequations}
Equation~\eqref{eq:massmat_cases} displays the reconstruction at leading order in $\ve_{ij}$, which suffices to exhibit the flavour structure of each model; the numerical results in the remainder of this work are obtained from the full solutions, without truncation in the mass ratios. 

\paragraph{Down-type decoupled:}
The solutions for the down-type decoupled models are obtained from those above by interchanging the sector labels $(u,c,t) \leftrightarrow (d,s,b)$ throughout and transposing the CKM moduli $|V_{ij}| \leftrightarrow |V_{ji}|$. Moreover, $\sin \varphi$ is negative for the $d,b$-decoupled models and positive for the $s$-decoupled model.

To reiterate, the above reconstruction leaves no free flavour parameters: the mass matrices, and the rotation matrices that diagonalise them, are determined entirely by the quark masses and the CKM moduli and phase. The flavour-changing couplings of the Higgs bosons and the axion are therefore fixed predictions of each model, to which we now turn.

%============================================================
\section{Flavour-Changing Neutral Currents: Higgs and Axion Sectors}
\label{sec:FCNC}
%============================================================

In flavoured DFSZ axion models, both Higgs- and axion-mediated FCNCs are controlled by the unitary rotations that diagonalise the fermion mass matrices. In the Higgs sector, flavour violation arises from the misalignment of the Yukawa structures associated with $\Phi_1$ and $\Phi_2$, while in the axion sector it originates from the generation-dependent Peccei--Quinn charges. In this section, we derive the form of the flavour-changing couplings of both the Higgs and axion fields in the \textsc{Minimal F-DFSZ} model, using the results obtained in the previous section. This makes the flavour structure of the different $q$-decoupled models transparent and clarifies their relation to other controlled flavour-violating 2HDMs, such as the BGL and generalised BGL frameworks~\cite{Botella:2009pq,Alves:2017xmk} which we discuss in Sec.~\ref{sec:comparison}.

%------------------------------------------------------------
\subsection{Flavour-Changing Higgs Couplings}
\label{sec:FCNCHiggs}

After electroweak symmetry breaking, the Yukawa Lagrangian of each \textsc{Minimal F-DFSZ} model in the gauge basis takes the form given in Eq.~\eqref{eq:LYgauge}. In unitary gauge, the physical fields can be embedded into the doublets as,
\begin{align}
\Phi_1 &=
\begin{pmatrix}
-s_{\beta}\,H^+ \\
\dfrac{1}{\sqrt{2}}\left(c_{\beta}v-s_{\alpha}h+c_{\alpha}H-is_{\beta}A\right) 
\end{pmatrix},
 \quad \Phi_2 =
\begin{pmatrix}
c_{\beta}H^+ \\
\dfrac{1}{\sqrt{2}}\left(s_{\beta}v+c_{\alpha}h+s_{\alpha}H+ic_{\beta}A\right) 
\end{pmatrix},
\end{align}
where $v^2 = v_1^2 + v_2^2 \simeq (246~\text{GeV})^2$ with $v_i$ the VEV of $\Phi_i$, and we abbreviate $c_x \equiv \cos x$, $s_x \equiv \sin x$, and $t_x \equiv \tan x$. The angle $\beta$, set by the VEV ratio $t_\beta = v_2/v_1$, rotates the CP-odd and charged fields to the mass eigenstates $A$ and $H^{\pm}$, while $\alpha$, determined by the parameters of the scalar potential, does the same for the CP-even states $h$ and $H$. The potential and the expression for $\alpha$ are given in App.~\ref{app:scalarpotential}.

After electroweak symmetry breaking, the Yukawa Lagrangian can be written in the fermion mass basis as
\begin{align}
\label{eq:Lyuk}
\mathcal{L}_Y^{\text{Higgs}} &\supset
\bar u_{Li} C^{h_u}_{ij} u_{Rj} h
+ \bar u_{Li} C^{H_u}_{ij} u_{Rj} H
+ \bar u_{Li} C^{A_u}_{ij} u_{Rj} A \nonumber\\
&+ \bar d_{Li} C^{h_d}_{ij} d_{Rj} h
+ \bar d_{Li} C^{H_d}_{ij} d_{Rj} H
+ \bar d_{Li} C^{A_d}_{ij} d_{Rj} A \nonumber\\
&+\sqrt{2}\Bigl(
    \bar d_{Li} V^*_{ki} C^{H^-_u}_{kj} u_{Rj} H^-  
   +\bar u_{Li} V_{ik} C^{H^+_d}_{kj} d_{Rj} H^+
 \Bigr) \;+\;\text{h.c.}\, ,
\end{align}
where $V_{ij}$ denotes the CKM matrix as usual. The couplings to the physical Higgs fields $C^X_{ij}$ ($X=h,H,A,H^\pm$) are then given by
\begin{align}
C^{h_u}_{ij} &= \frac{m_{u_i}}{v s_\beta}\,c_\alpha\,\delta_{ij}
   -\frac{m_t}{v}\,c_{\alpha-\beta}\,(t_\beta+t_\beta^{-1})\,\xi^u_{ij}, \qquad
C^{h_d}_{ij} = -\frac{m_{d_i}}{v c_\beta}\,s_\alpha\,\delta_{ij}
   +\frac{m_b}{v}\,c_{\alpha-\beta}\,(t_\beta+t_\beta^{-1})\,\xi^{d}_{ij},
\label{eq:Cuplings_h}\\[3pt]
C^{H_u}_{ij} &= \frac{m_{u_i}}{v s_\beta}\,s_\alpha\,\delta_{ij}
   -\frac{m_t}{v}\,s_{\alpha-\beta}\,(t_\beta+t_\beta^{-1})\,\xi^u_{ij}, \qquad
C^{H_d}_{ij} = \frac{m_{d_i}}{v c_\beta}\,c_\alpha\,\delta_{ij}
   +\frac{m_b}{v}\,s_{\alpha-\beta}\,(t_\beta+t_\beta^{-1})\,\xi^{d}_{ij}, \\[3pt]
C^{A_u}_{ij} &= i\,\frac{m_{u_i}}{v\,t_\beta}\,\delta_{ij}
   -i\,\frac{m_t}{v}\,(t_\beta+t_\beta^{-1})\,\xi^u_{ij}, \qquad
C^{A_d}_{ij} = i\,\frac{m_{d_i}\,t_\beta}{v}\,\delta_{ij}
   -i\,\frac{m_b}{v}\,(t_\beta+t_\beta^{-1})\,\xi^{d}_{ij}, \\[3pt]
C^{H^-_u}_{ij} &= \frac{m_{u_i}}{v\,t_\beta}\,\delta_{ij}
   -\frac{m_t}{v}\,(t_\beta+t_\beta^{-1})\,\xi^u_{ij}, \qquad
C^{H^+_d}_{ij} = \frac{m_{d_i}\,t_\beta}{v}\,\delta_{ij}
   -\frac{m_b}{v}\,(t_\beta+t_\beta^{-1})\,\xi^{d}_{ij},
\label{eq:ChargedHiggsCouplings}
\end{align}
with\footnote{Equivalently, $\xi^u_{ij}$ may be written in terms of
$\mY_2$ and $\xi^d_{ij}$ in terms of $\mY_1$, with a corresponding
change of prefactor in $C^X_{ij}$.}
\begin{align}
\label{eq:EpsilonCouplings}
\xi^u_{ij} = \frac{v\,c_\beta}{\sqrt{2}\,m_t}
  \left(\mU^\dag_{uL}\,\mY_u^1\,\mU_{uR}\right)_{ij}, \qquad
\xi^d_{ij} = \frac{v\,s_\beta}{\sqrt{2}\,m_b}
  \left(\mU^\dag_{dL}\,\mY_d^2\,\mU_{dR}\right)_{ij}.
\end{align}
Despite superficial appearances, $\xi_{ij}^q$ depends only on the fermion mass ratios $\ve_{ij}$ and the CKM magnitudes $|V_{ij}|$ (once $\mY_q^i$ is rewritten in terms of the reconstructed $\mM_q$). It follows that the flavour-violating effects are fixed up to the parameters $\{c_{\alpha-\beta},t_{\beta},m_{\phi_i}\}$, where $m_{\phi_i}$ are the Higgs masses. Furthermore, in the \textsc{Minimal F-DFSZ} models $\xi_{ij}^q$ are real\footnote{This is a consequence of the fact that $\mU_{uL}$ in Eq.~\eqref{eq:O1} (and $\mU_{uR}$) can be factorised in terms of an orthogonal matrix $\mO_{uL}$ and a diagonal matrix of phases $P$: $\mU_{uL}=\mO_{uL}P$. The matrix $P$ is trivially unitary and cancels in the Higgs and axion flavour couplings.}, so CP-violation is mediated entirely by the $W^\pm$ interactions.

The $\xi_{ij}^q$ can be expressed in terms of the quark masses and CKM moduli using the expressions for the left-handed rotation matrices in Eqs.~\eqref{eq:O1} and \eqref{eq:O2}; the right-handed rotation matrix can be obtained using $\mU_R=\mM^{-1}\mU_L\mathcal{D}$. For the sector containing the decoupled quark, the form of $\xi_{ij}^q$ can be computed straightforwardly. Here, we generically find that the off-diagonal couplings for the two non-decoupled generations satisfy $|\xi_{ij}^q| \gg |\xi_{ji}^q|$ for $i < j$ (i.e. there is a hierarchy between the couplings with different chirality). 

For an \emph{up-type decoupled} model, the couplings controlling the tree-level FCNCs in the decoupled sector have the form (to leading order in the quark mass ratios)
\begin{align}
|(\xi_{ij}^u)|=\ve_{jt}\dfrac{|V_{ib}||V_{jb}|}{1-|V_{rb}|^2}, \quad i<j, 
\label{eq:decoupledsectorFCNCUp}
\end{align}
where $r$ indicates the decoupled quark. Similarly, for a \emph{down-type decoupled} model they take the form
\begin{align}
|(\xi_{ij}^d)|=\ve_{jb}\dfrac{|V_{ti}||V_{tj}|}{1-|V_{tr}|^2},\quad i<j.
\label{eq:decoupledsectorFCNCDown}
\end{align}

The form of $\xi_{ij}^q$ for the sector without the decoupled quark is generally more complicated. Approximate expressions in terms of $|V_{ij}|$ and $\ve_{ij}$ are collected for all six models in App.~\ref{app:analyticFCNCXI}. Here, for the sake of illustration, we present the expressions for the $t$-decoupled model, expanded to leading order in the Wolfenstein parameter $\lambda = 0.225$. Given that the CKM elements scale with $\lambda$ according to~\cite{Wolfenstein:1983yz,Ahn:2011fg}
\begin{align}
\label{eq:WolfCKM}
    |V_{ij}| \sim \begin{pmatrix}
        1 & \lambda & \lambda^3 \\
        \lambda & 1 & \lambda^2 \\
        \lambda^3 & \lambda^2 & 1
    \end{pmatrix},
\end{align}
while the quark mass ratios approximately scale as,
\begin{align}
\label{eq:Wolfeps}
\begin{pmatrix}
        \ve_{ut} & \ve_{ct} \\
        \ve_{db} &\ve_{sb}
    \end{pmatrix} \sim \begin{pmatrix}
        \lambda^8 & \lambda^4 \\
        \lambda^5 & \lambda^3
    \end{pmatrix},
\end{align}
the elements of $\xi_{ij}^d$ (for the $t$-decoupled model) are given by
\begin{align}
|\xi_{ds}^d| &\simeq \ve_{sb}\dfrac{|V_{td}||V_{ts}|}{1-|V_{tb}|^2}, &
|\xi_{sd}^d| &\simeq 2\,\ve_{db}\,\dfrac{|V_{ts}|}{|V_{td}|}, &
|\xi_{bd}^d| &\simeq \ve_{db}\,\dfrac{|V_{td}|}{|V_{td}|^2+|V_{ts}|^2\ve_{ds}^2},
\\[8pt]
|\xi_{db}^d| &\simeq |V_{td}||V_{tb}|, &
|\xi_{sb}^d| &\simeq |V_{ts}||V_{tb}|, &
|\xi_{bs}^d| &\simeq \ve_{sb}\,\dfrac{|V_{ts}|\,\ve_{ds}^2}{|V_{td}|^2+|V_{ts}|^2\ve_{ds}^2},
\end{align}
to leading order in $\lambda$. Note that unlike the couplings in the decoupled sector ($\xi_{ij}^u$ for the $t$-decoupled model), the couplings in the non-decoupled sector do not have a universal chiral hierarchy. 

Table~\ref{tab:Orderinlambda} shows the order of magnitude of the matrices $\xi^q$ for all six models. This illustrates the relative strength of the couplings and, in particular, allows for the identification of the most relevant flavour-violating couplings in each model. 
\begin{table}[htpb]
    \centering
    \begin{tabular}{||c|cc||}
    \hline
       Model  &$|\xi_{ij}^u|$ & $|\xi_{ij}^d|$  \\ \hline \hline
        $t$-decoupled & $\sim\begin{pmatrix}
10^{-7} & 10^{-3} & 0 \\
10^{-8} & 10^{-5} & 0 \\
0 & 0 & 1
\end{pmatrix}$ & $\sim\begin{pmatrix}
10^{-3} & 10^{-2} & 10^{-2} \\
10^{-2} & 10^{-3} & 10^{-1} \\
10^{-1} & 10^{-2} & 10^{-2}
\end{pmatrix}$ \\[2pt]
        $c$-decoupled & $\sim\begin{pmatrix}
10^{-10} & 0 & 10^{-2} \\
0 & 10^{-2} & 0 \\
10^{-12} & 0 & 10^{-5}
\end{pmatrix}$ & $\sim\begin{pmatrix}
10^{-3} & 10^{-2} & 10^{-2} \\
10^{-2} & 10^{-2} & 10^{-1} \\
10^{-1} & 1 & 1
\end{pmatrix}$ \\[2pt]
        $u$-decoupled & $\sim\begin{pmatrix}
10^{-5} & 0 & 0 \\
0 & 10^{-5} & 10^{-1} \\
0 & 10^{-6} & 10^{-3}
\end{pmatrix}$ & $\sim\begin{pmatrix}
10^{-3} & 10^{-2} & 10^{-3} \\
10^{-4} & 10^{-2} & 10^{-2} \\
10^{-2} & 1 & 1
\end{pmatrix}$ \\[2pt]
        $b$-decoupled & $\sim\begin{pmatrix}
10^{-5} & 10^{-3} & 10^{-2} \\
10^{-4} & 10^{-2} & 10^{-1} \\
10^{-3} & 10^{-4} & 1
\end{pmatrix}$ & $\sim\begin{pmatrix}
10^{-3} & 10^{-2} & 0 \\
10^{-5} & 10^{-2} & 0 \\
0 & 0 & 0
\end{pmatrix}$ \\[2pt]
        $s$-decoupled & $\sim\begin{pmatrix}
10^{-5} & 10^{-3} & 10^{-2} \\
10^{-4} & 10^{-2} & 10^{-1} \\
10^{-3} & 10^{-1} & 10^{-2}
\end{pmatrix}$ & $\sim\begin{pmatrix}
10^{-3} & 0 & 10^{-2} \\
0 & 0 & 0 \\
10^{-9} & 0 & 1
\end{pmatrix}$ \\[2pt]
        $d$-decoupled & $\sim\begin{pmatrix}
10^{-6} & 10^{-3} & 10^{-2} \\
10^{-7} & 10^{-3} & 10^{-3} \\
10^{-4} & 10^{-1} & 10^{-2}
\end{pmatrix}$ & $\sim\begin{pmatrix}
0 & 0 & 0 \\
0 & 10^{-2} & 10^{-1} \\
0 & 10^{-6} & 1
\end{pmatrix}$ \\[2pt]
\hline
    \end{tabular}
    \caption{Order of magnitudes of the entries of the flavour-changing matrices $|\xi_{ij}^u|$ and $|\xi_{ij}^d|$, defined in Eq.~\eqref{eq:EpsilonCouplings}, for each $q$-decoupled model.}
    \label{tab:Orderinlambda}
\end{table}

%------------------------------------------------------------
\subsection{Flavour-Changing Axion--Quark Couplings}
\label{subsec:FCNCaxion}

Below the electroweak scale, the effective Lagrangian for the axion field can be written as
\begin{align}
    \mathcal{L}_a \supset \dfrac{g_s^2}{32\pi^2}\dfrac{a}{f_a}G\tilde{G}+\dfrac{E}{N}\dfrac{e^2}{32\pi^2}\dfrac{a}{f_a}F\tilde{F}+\sum_{q=u,d,e}\dfrac{\partial_{\mu}a}{2f_a}\bar{q}_i\gamma^{\mu}\left((C^V_{q_iq_j})+(C^A_{q_iq_j})\gamma_5\right)q_j,
\label{eq:axionefflag}
\end{align}
where $E/N$ is the ratio of the electromagnetic and QCD anomalies, and the axion decay constant is defined by $f_a =v_{\rm PQ}/(2N)$, where $\langle S \rangle =v_{\rm PQ}/\sqrt{2}$.

The vector and axial couplings of the axion are, respectively,
\begin{align}
    C^{V,A}_{q_iq_j} = C^L_{q_iq_j} \pm C^R_{q_iq_j}, \qquad C^{L,R}_{q_iq_j}=\dfrac{1}{2N}\,\left(\mU_{q_{L,R}}^\dagger\,\mX_{f_{L,R}}\,\mU_{q_{L,R}}\right)_{ij},
    \label{eq:axfermC}
\end{align}
where $\mX_{q_{L,R}}$ are the diagonal matrices of PQ charges given in Table~\ref{tab:pq_charges} and $\mU_{q_{L,R}}$ are the unitary rotation matrices. 

The flavour-dependent PQ charges of the quarks imply that $C_q^{V,A}$ contain flavour off-diagonal terms\footnote{It is interesting to note that since both sources of flavour violation originate from the Yukawa Lagrangian, an explicit expression can be derived relating $\xi_{ij}^q$ to $C_{q_iq_j}^{L,R}$~\cite{DiLuzio:2023ndz}.}, unlike in the usual, flavour-universal DFSZ model. These give rise to FCNCs mediated by the axion. We find that while the diagonal\footnote{Note that the flavour-diagonal vector couplings are unphysical.} entries of $C_q^{V,A}$ depend on $\tan\beta$, the off-diagonal entries do not. The flavour-violating axion couplings are therefore entirely fixed, up to the overall scaling with $f_a$. These are given in Table~\ref{tab:fitandfvcouplings} for each $q$-decoupled model.

\begin{table}[t]
\centering
\scriptsize
\setlength{\tabcolsep}{4pt}
\renewcommand{\arraystretch}{1.12}

\begin{tabular}{c ccc ccc}
\toprule
\multicolumn{7}{c}{\textit{(a) Off-diagonal vector couplings, $C_{ij}^V$}}\\
\midrule
& \multicolumn{3}{c}{Up sector}
& \multicolumn{3}{c}{Down sector}\\
\cmidrule(lr){2-4}\cmidrule(lr){5-7}
Model
& $C^V_{uc}$ & $C^V_{ut}$ & $C^V_{ct}$
& $C^V_{ds}$ & $C^V_{db}$ & $C^V_{sb}$\\
\midrule
$t$
& $-0.09$ & $0$ & $0$
& $-0.65$ & $-0.11$ & $-0.07$\\

$c$
& $0$ & $4.79\times10^{-3}$ & $0$
& $-0.82$ & $-0.07$ & $0.39$\\

$u$
& $0$ & $0$ & $-0.05$
& $-0.46$ & $6.73\times10^{-3}$ & $-0.52$\\
\addlinespace

$b$
& $-0.13$ & $5.68\times10^{-3}$ & $-0.04$
& $-0.21$ & $0$ & $0$\\

$s$
& $-0.46$ & $8.44\times10^{-3}$ & $0.13$
& $0$ & $8.66\times10^{-3}$ & $0$\\

$d$
& $-0.44$ & $8.24\times10^{-3}$ & $0.10$
& $0$ & $0$ & $0.04$\\
\bottomrule
\end{tabular}

\vspace{2ex}

\begin{tabular}{c ccc ccc}
\toprule
\multicolumn{7}{c}{\textit{(b) Diagonal axial couplings, $C_{ii}^A$}}\\
\midrule
& \multicolumn{3}{c}{Up sector}
& \multicolumn{3}{c}{Down sector}\\
\cmidrule(lr){2-4}\cmidrule(lr){5-7}
Model
& $C^A_{uu}$ & $C^A_{cc}$ & $C^A_{tt}$
& $C^A_{dd}$ & $C^A_{ss}$ & $C^A_{bb}$\\
\midrule
$t$
& $-0.01-c_\beta^2$
& $0.01-c_\beta^2$
& $1.00-c_\beta^2$
& $1.84+c_\beta^2$
& $-0.85+c_\beta^2$
& $-0.99+c_\beta^2$\\

$c$
& $-c_\beta^2$
& $1.00-c_\beta^2$
& $-c_\beta^2$
& $1.82+c_\beta^2$
& $-1.67+c_\beta^2$
& $-0.15+c_\beta^2$\\

$u$
& $1.00-c_\beta^2$
& $-c_\beta^2$
& $-c_\beta^2$
& $0.10+c_\beta^2$
& $0.46+c_\beta^2$
& $-0.56+c_\beta^2$\\
\addlinespace

$b$
& $2+c_\beta^2$
& $-1+c_\beta^2$
& $-1+c_\beta^2$
& $-0.04-c_\beta^2$
& $0.04-c_\beta^2$
& $1.00-c_\beta^2$\\

$s$
& $1.90+c_\beta^2$
& $-1.89+c_\beta^2$
& $-0.01+c_\beta^2$
& $-c_\beta^2$
& $1-c_\beta^2$
& $-c_\beta^2$\\

$d$
& $0.10+c_\beta^2$
& $-0.09+c_\beta^2$
& $-0.01+c_\beta^2$
& $1-c_\beta^2$
& $-c_\beta^2$
& $-c_\beta^2$\\
\bottomrule
\end{tabular}

\caption{Axion--quark couplings in the six $q$-decoupled models. Panel~(a) gives the $\tan\beta$-independent, flavour-violating vector
couplings relevant for rare meson decay bounds. Omitted entries are fixed by hermiticity. Panel~(b) gives the diagonal axial couplings relevant for astrophysical bounds.}
\label{tab:fitandfvcouplings}
\end{table}

%------------------------------------------------------------
\subsection{Comparison with Other Flavour Frameworks}
\label{sec:comparison}
%------------------------------------------------------------
The \textsc{Minimal F-DFSZ} models belong to a broader class of 2HDMs in which FCNCs are controlled by an underlying flavour structure, rather than by the additional free Yukawa parameters present in the general type-III 2HDM.\footnote{See Ref.~\cite{Crivellin:2013wna} for a review of the flavour phenomenology of the general type-III 2HDM.} Their closest symmetry-based counterparts are BGL models and their generalisations. 

\paragraph{BGL and generalised BGL models:}
In BGL models~\cite{Branco:1996bq,Botella:2009pq}, an Abelian symmetry fixes the neutral-scalar FCNCs in terms of quark masses and CKM elements, while restricting them to a single quark sector. Their couplings have the schematic form~\cite{Botella:2015hoa}
\begin{align}
\text{up-type BGL:}\qquad
|\xi^d_{ij}(\widetilde u_r)|
&=
\ve_{jb}\,
|V_{ri}^{}V_{rj}|,
\nonumber\\
\text{down-type BGL:}\qquad
|\xi^u_{ij}(\widetilde d_r)|
&=
\ve_{jt}\,
|V_{ir}V_{jr}|,
\qquad i\neq j,
\label{eq:BGL}
\end{align}
where $\widetilde u_r$ or $\widetilde d_r$ denotes the generation singled out by the BGL symmetry, with FCNCs arising in the opposite quark sector.

The FCNCs of the \textsc{Minimal F-DFSZ} models exhibit a dependence on quark-mass ratios and CKM elements closely analogous to that of BGL models, as shown in Eq.~\eqref{eq:BGL}. In particular, both frameworks predict the chiral hierarchy $|\xi^q_{ij}|\gg|\xi^q_{ji}|$ for $i<j$.

There are, however, two important structural differences. First, the \textsc{Minimal F-DFSZ} models contain tree-level FCNCs in both the up and down quark sectors, whereas ordinary BGL models restrict them to one sector. This leads to very different flavour phenomenology. Generalised BGL constructions can generate FCNCs in both sectors, but introduce additional flavour parameters beyond the quark masses and CKM matrix~\cite{Alves:2017xmk}. Second, the flavour symmetry of the \textsc{Minimal F-DFSZ} framework is identified with an anomalous PQ symmetry. The same symmetry therefore controls the heavy-scalar flavour structure, solves the strong CP problem, and predicts flavour-violating axion couplings correlated with those of the Higgs sector~\cite{DiLuzio:2023ndz}. It is also worth noting that there are (non-minimal) flavoured DFSZ models that yield the coupling structures in Eq.~\eqref{eq:BGL} if the scalar sector is extended to three Higgs doublets, as studied in Ref.~\cite{Celis:2014jua}.

%============================================================
\section{Axion Phenomenology}
\label{sec:axionpheno}
%============================================================

The axion phenomenology of the \textsc{Minimal F-DFSZ} models involves two complementary classes of observables: flavour-violating meson decays, which probe the off-diagonal axion couplings, and astrophysical observations, which constrain the flavour-diagonal couplings to quarks and nucleons. Since both sets of couplings are fixed by the flavour structure of the model (up to $t_{\beta}$ for the diagonal couplings), the resulting limits on the axion decay constant $f_a$ become sharply model-dependent. Using the QCD axion relation~\cite{Weinberg:1977ma}
\begin{align}
    m_a \simeq 5.7\,\mu\mathrm{eV}
    \left(\frac{10^{12}\,\mathrm{GeV}}{f_a}\right),
    \label{eq:ma_fa_relation}
\end{align}
we translate these constraints into upper bounds on $m_a$ for each $q$-decoupled model. We also compare these constraints with the post-inflationary axion dark matter window. We then examine the axion-photon coupling in simple lepton-sector extensions of the model, and comment on the possibility of a nucleophobic axion.

%------------------------------------------------------------
\subsection{Rare Meson Decays and Astrophysical Bounds}
\paragraph{Rare meson decays:}

Measurements of meson decays involving neutrinos can be recast as bounds on axion couplings~\cite{BaBar:2004xlo,BaBar:2013npw,CLEO:2008ffk,E949:2007xyy,NA62:2025upx}. In Table~\ref{tab:mesondecayslimits}, we quote the corresponding $90\%$ CL lower bounds on $f_a/|C^V_{ij}|$~\cite{MartinCamalich:2020dfe}. The decay $K^+ \to \pi^+ a$ is the most tightly constrained experimentally and provides the strongest constraint on $f_a$ in all models where the $s \to d$ transition occurs at tree-level. The exceptions are the $d$- and $s$-decoupled models. For these two models, the leading meson-decay bounds arise instead from the tree-level $c \to u$ transition that produces $D \to \pi a$ decays and from the $t \to c$ transition that leads to $B \to K a$ decays at one-loop. The resulting bounds on $f_a$ are therefore several orders of magnitude weaker in these models. 

\begin{table}[t]
    \centering
    \begin{tabular}{||c|c|c|c||}
        \hline
        Process & Flavour-Violating Coupling & Bound on $f_a$ (GeV)&Ref. \\
        \hline\hline
        $K^+ \rightarrow \pi^+ a$ & $s \to d$ 
        & $f_a > 5.3 \times 10^{11}\,|C_{sd}^V|$&\cite{NA62:2025upx} \\
        $B \rightarrow K a$ & $b \to s$ & $f_a > 1.65 \times 10^{8}\,|C_{bs}^V|$&\cite{BaBar:2013npw} \\
        $B \rightarrow \pi a$ & $b \to d$ & $f_a > 1.1 \times 10^{8}\,|C_{db}^V|$&\cite{BaBar:2004xlo} \\
        $D \to \pi a$ & $c \to u$ & $f_a>4.85 \times 10^7|C_{cu}^V|$&\cite{CLEO:2008ffk} \\
        $K^+ \rightarrow \pi^+ a$ (loop) & $t \to u$ & $f_a > 1.5 \times 10^8 |C_{ut}^V|$ & \cite{E949:2007xyy}\\
        $B \to Ka$ (loop) &$t \to c$ &$f_a>3.5 \times 10^8|C_{tc}^V|$&\cite{E949:2007xyy} \\
        \hline\hline
    \end{tabular}
    \caption{Strongest constraints from flavour-violating meson decays on each flavour transition~\cite{MartinCamalich:2020dfe}.}
    \label{tab:mesondecayslimits}
\end{table}

\paragraph{Astrophysical bounds:}
The flavour-diagonal axial couplings are constrained by energy loss in SN1987A~\cite{Carenza:2019pxu} and neutron-star cooling~\cite{Beznogov:2018fda}. The dimensionless axion couplings to nucleons, $C_p$ and $C_n$, are given in terms of the diagonal axial quark couplings defined in Eq.~\eqref{eq:axfermC} by~\cite{GrillidiCortona:2015jxo}
\begin{align}
C_p + C_n &= 
0.50(5)\left( C_{uu}^A + C_{dd}^A - 1 \right) - 2\delta \, ,\label{eq:CppCn} \\
C_p - C_n &= 1.273(2)\left( C_{uu}^A - C_{dd}^A - \frac{1-z}{1+z} \right) ,
\label{eq:CpmCn}
\end{align}
where $z = m_u/m_d = 0.48(3)$, and
\begin{align}
    \delta \equiv 0.038(5) C_{ss}^A + 0.012(5) C_{cc}^A + 0.009(2) C_{bb}^A + 0.0035(4) C_{tt}^A,
\end{align}
collects the contributions from the heavy-quark axial couplings to the nucleon couplings. The effective axion-nucleon couplings are then given by $g_{ai} = C_i m_i/f_a$, for $i=p,n$. In terms of these quantities, the SN1987A bound is~\cite{Carenza:2019pxu}
\begin{equation}
0.61\,g_{ap}^2 + g_{an}^2 + 0.53\, g_{an} g_{ap} < 8.26\times 10^{-19} .
\label{eq:SN1987A}
\end{equation}
The strongest neutron star constraint arises from observations of HESS~J1731--347 and is~\cite{Beznogov:2018fda}
\begin{align}
    g_{an} \lesssim 2.8 \times 10^{-10}.
    \label{eq:HESS-J1731-347}
\end{align}
Which of these provides a stronger constraint on $f_a$ depends on the specific $q$-decoupled model under consideration. Furthermore, since the flavour-diagonal couplings depend on $t_\beta$, so too do the astrophysical lower bounds on $f_a$. However, this dependence is mild, typically modifying the bound by a $\mO(1)$ factor over the relevant range of $t_\beta$.

\paragraph{Combined bounds on the axion mass:}
For models in which flavour-violating rare-meson decays provide the dominant constraint, the bounds in Table~\ref{tab:mesondecayslimits} translate directly into upper limits on the axion mass. By contrast, converting the astrophysical limits in Eqs.~\eqref{eq:SN1987A}--\eqref{eq:HESS-J1731-347} into upper bounds on $m_a$ via Eq.~\eqref{eq:ma_fa_relation} requires scanning across a viable range of $t_{\beta}$ and choosing the most conservative upper limit on $m_a$.

The viable $t_{\beta}$ range is set by the requirement that the Yukawa couplings, which scale as either $1/s_\beta$ or $1/c_\beta$ depending on the Higgs-doublet, satisfy the perturbativity condition $|(\mathcal{Y}_{1,2}^q)_{ij}|\leq\sqrt{4\pi}$. This condition bounds $t_{\beta}$ from above and below. The resulting intervals, obtained using the numerical mass-matrix fits in Table~\ref{tab:MuMdNum} in App.~\ref{app:MassMatrixFits}, are shown in Table~\ref{tab:axion_bounds}, together with the upper bound on $m_a$ for each $q$-decoupled model. 

For the $u$-, $c$-, $t$-, and $b$-decoupled models the dominant constraint is from $K^+\to\pi^+a$, with similar upper bounds on $m_a$ across these models. This directly follows from the comparable values for the $C^V_{sd}$ couplings in Table~\ref{tab:fitandfvcouplings}. The $d$- and $s$-decoupled models evade this constraint, leaving neutron-star cooling and SN1987A as the dominant probes. Since the flavour-diagonal axion couplings depend on $t_\beta$, we evaluate Eqs.~\eqref{eq:SN1987A}--\eqref{eq:HESS-J1731-347} across the perturbatively viable range and identify, at each point, the stronger constraint. We then select the value $t_\beta^\star$ that yields the most conservative upper bound on $m_a$. For the $d$- and $s$-decoupled models, the dominant constraint arises from SN1987A and neutron star cooling, respectively.

\begin{table}[t!]
\centering
\begin{tabular}{c c c c c}
\hline\hline
$q$ & Perturbative $t_{\beta}$ range & $m_a$ upper bound $[\mu\mathrm{eV}]$
  & Leading constraint & $t_{\beta}^{\star}$ \\
\hline
$t$ & $0.001 \lesssim t_{\beta} \lesssim 3.5$  & $16.5$ & $K^+\to \pi^+a$ & -- \\
$c$ & $0.29 \lesssim t_{\beta} \lesssim 554$  & $13.1$ & $K^+\to \pi^+a$ & -- \\
$u$ & $0.29 \lesssim t_{\beta} \lesssim 67.9$ & $23.6$ & $K^+\to \pi^+a$ & -- \\
$b$ & $0.01 \lesssim t_{\beta} \lesssim 3.5$  & $51.1$ & $K^+\to \pi^+a$ & -- \\
$s$ & $0.28 \lesssim t_{\beta} \lesssim 42$   & $2.18\times10^3$ & HESS J1731-347 & $42$ \\
$d$ & $0.28 \lesssim t_{\beta} \lesssim 37.6$ & $8.22\times 10^{4}$ & SN1987A & $0.74$ \\
\hline\hline
\end{tabular}
\caption{Perturbative $t_{\beta}$ ranges and upper bounds on the axion mass in each $q$-decoupled model, together with the leading constraint. For the $d$- and $s$-decoupled models, the leading (astrophysical) constraints depend on $t_\beta$ and the quoted bound is the most conservative over the perturbative range, with $t_\beta^\star$ denoting the value at which this bound is attained.}
\label{tab:axion_bounds}
\end{table}

\paragraph{Post-inflationary axion dark matter:}
It is interesting to compare the limits on $m_a$ in Table~\ref{tab:axion_bounds} with the axion mass required to achieve the observed dark matter abundance in the post-inflationary PQ-breaking scenario. While this value remains subject to significant theoretical uncertainty, a representative post-inflationary dark matter window is $40\,\mu\textrm{eV}\leq m_a\leq500\,\mu\rm eV$~\cite{Gorghetto:2020qws,Hoof:2021jft,Buschmann:2021sdq,Sopov:2022bog}\footnote{Recent work shows that previously unaccounted for friction between the thermal bath and axion-string network may increase the upper bound of this window by several orders of magnitude~\cite{Hook:2026grn}.} which, along with the model-dependent upper bounds on the axion mass, are displayed in Fig.~\ref{fig:axionphotoncoupling}. The $u$-, $c$- and $t$-decoupled models already lie in serious tension with saturating the observed relic abundance, as they over-produce the dark matter density. In this sense, the flavour structure may directly determine which realisations of the model contain a viable dark matter candidate, at least in the post-inflationary PQ-breaking cosmology.

%------------------------------------------------------------
\subsection{Nucleophobia}

In \(\mN_{\rm DW}=1\) axion models it is possible to suppress the axion--nucleon couplings, rendering the axion effectively nucleophobic~\cite{DiLuzio:2017ogq} and thereby relaxing the astrophysical bounds. This can be achieved by requiring the expressions in Eqs.~\eqref{eq:CppCn} and~\eqref{eq:CpmCn} to vanish, implying
\begin{align}
    C_{uu}^A + C_{dd}^A &\simeq 1,\\
    C_{uu}^A - C_{dd}^A &\simeq \frac{1-z}{1+z}.
\end{align}
In the limit where there is no flavour mixing ($\mU_{L/R}=\mathds{1})$, the first condition can be satisfied by requiring that the light-quark contribution saturates the colour anomaly
\begin{align}
    C_{uu}^A + C_{dd}^A &=\dfrac{1}{2N}\left(2\mX(q_1)-\mX(u_1)-\mX(d_1)\right)=1,  
    \label{eq:nuceq1}
\end{align}
while the second condition requires a tuning of $\tan\beta$ such that,
\begin{align}
    C_{uu}^A-C_{dd}^A &=\dfrac{1}{2N}\left(\mX(d_1)-\mX(u_1)\right)\simeq \dfrac{1-z}{1+z}=1/3,
    \label{eq:nuceq2}
\end{align}
up to sub-leading corrections from the sea quarks. The first condition holds only for the $u$- and $d$- decoupled models and the second condition is then satisfied for $\tan\beta=\sqrt{2},\,1/\sqrt{2}$, respectively. Although the astrophysical constraints are weakened at these parameter points, flavour observables still provide a stringent constraint on $m_a$. In the $u$-decoupled model, the bound on $m_a$ from $K^+ \to \pi^+a$ is unaffected, while in the $d$-decoupled model the loop-level $B \to Ka$ transition sets the upper bound, $m_a \lesssim \mO(0.1)\,\rm eV$.

%------------------------------------------------------------
\subsection{Axion-Photon Coupling and the Lepton Sector}
\label{sec:axion_photon}
\begin{table}[t!]
    \centering
    \begin{tabular}{c c c c}
        \hline\hline
        Model & Leptons couple to & Decoupled sector & $E/N$ \\
        \hline
        \multirow{2}{*}{\textsc{F-DFSZ-I}}
            & \multirow{2}{*}{$\Phi_1$}
            & up-type & $-10/3$ \\
            &
            & down-type & $20/3$ \\
        \hline
        \multirow{2}{*}{\textsc{F-DFSZ-II}}
            & \multirow{2}{*}{$\Phi_2$}
            & up-type & $8/3$ \\
            &
            & down-type & $2/3$ \\
        \hline\hline
    \end{tabular}
    \caption{$E/N$ values for the F-DFSZ models. The two possible $E/N$ values for \textsc{F-DFSZ-I} and \textsc{F-DFSZ-II} arise from whether the decoupled quark is an up-type or down-type quark.}
    \label{tab:E/N}
\end{table}
A central target of the axion experimental program is the axion-photon interaction,
\begin{align}
\mathcal{L} \supset -\frac{1}{4}\, g_{a\gamma\gamma}\, a\,F_{\mu\nu}\tilde F^{\mu\nu},
\end{align}
where~\cite{GrillidiCortona:2015jxo},
\begin{align}
g_{a\gamma\gamma}
=
\frac{\alpha_{\rm em}}{2\pi f_a}
\left(
\frac{E}{N}
-
1.92(4)
\right) .
\label{eq:gagg}
\end{align}
The model dependence enters through the ratio \(E/N\) (see~\cite{DiLuzio:2017pfr} for typical values in DFSZ-type models), where
\begin{align}
    E = \sum_{f}N_C^f\,\left[\mX(f_{L})-\mX(f_{R})\right]\,Q_f^2
    \label{eq:EManom}
\end{align}
is the electromagnetic anomaly, $N_C^f$ denotes the number of colours for fermion $f$, $\mX(f_{L/R})$ denotes the PQ charge of the chiral fermion, as given in Table~\ref{tab:pq_charges}, and $Q_f$ denotes the electric charge. The colour anomaly $N$ is defined in the usual way via Eq.~\ref{dm1} and we note that for up-type and down-type decoupled models, $2N=1\text{ and}\,-1$, respectively.

To evaluate Eq.~\eqref{eq:EManom}, the PQ charges of the charged leptons must also be specified. The two simplest extensions of the F-DFSZ framework to incorporate the lepton sector are:
\begin{align}
    \textrm{\textsc{F-DFSZ-I}: \textrm{All leptons couple to $\Phi_1$}, } &\Rightarrow \mX(e_i)=s_{\beta}^2,\nonumber\\
    \textrm{\textsc{F-DFSZ-II}: \textrm{All leptons couple to $\Phi_2$}, } &\Rightarrow \mX(e_i)=-c_{\beta}^2.
\end{align}
Pairing these two choices with either an up-type or down-type decoupled model leads to four values of $E/N$ which we collect in Table~\ref{tab:E/N}. 

\begin{figure}[t]
    \centering
    \includegraphics[width=\linewidth]{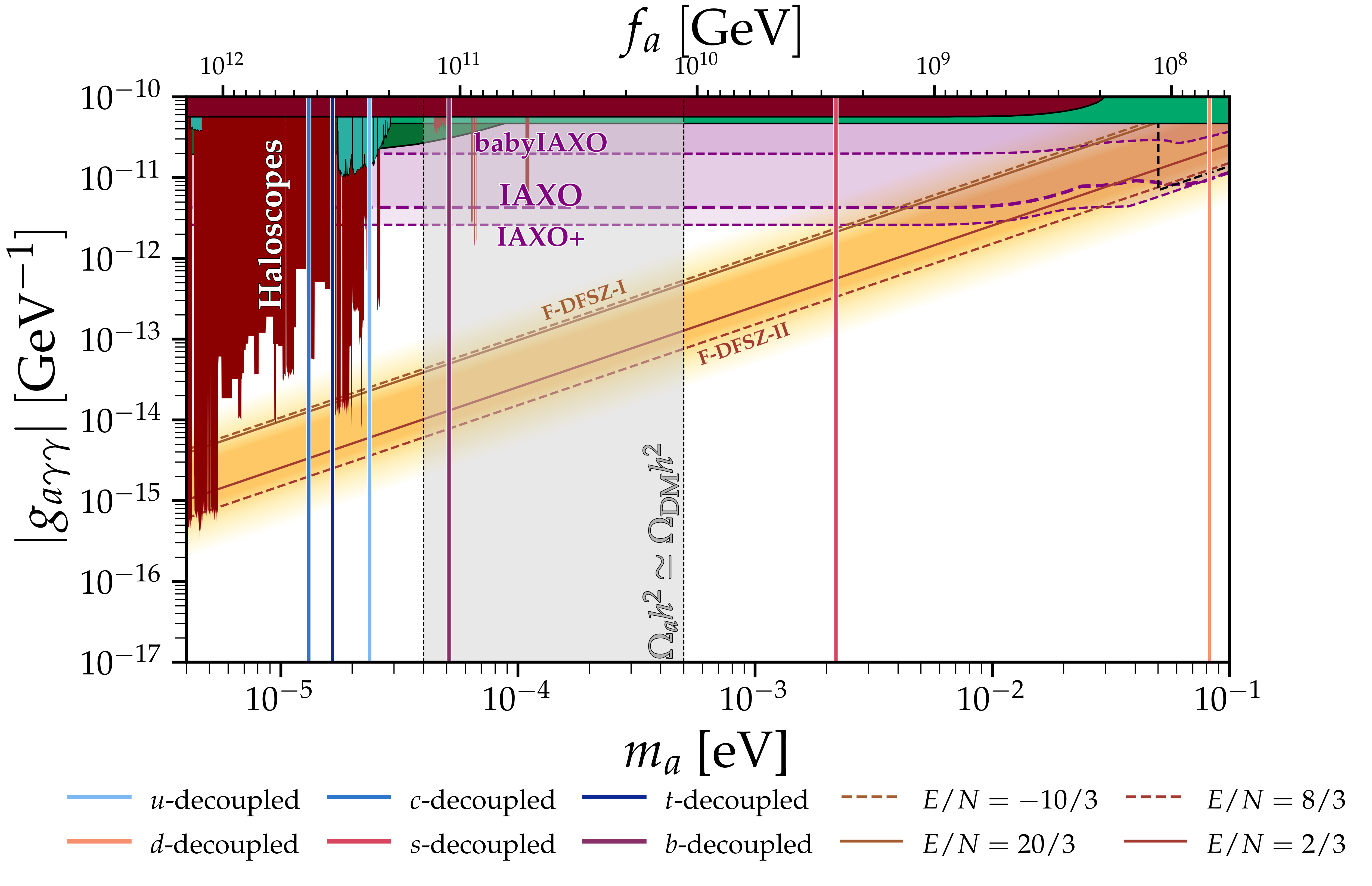}
    \caption{The vertical coloured lines show the upper bounds on the axion mass, $m_a$, for each $q$-decoupled model, derived from flavour-violating meson decays and astrophysical observations, as summarised in Table~\ref{tab:axion_bounds}. The grey band marks the mass range in which post-inflationary axions can reproduce the observed dark-matter abundance. The bounds in the $u$-, $c$-, and $t$-decoupled models lie below this range, implying an overabundance of axion dark matter in the standard post-inflationary scenario. Also shown are the axion--photon couplings for the two simplest lepton-sector extensions: \textsc{F-DFSZ-I}, in which all leptons couple to $\Phi_1$, and \textsc{F-DFSZ-II}, in which all leptons couple to $\Phi_2$. Adapted from Ref.~\cite{OHare:2020AxionLimits}.}
    \label{fig:axionphotoncoupling}
\end{figure}

The corresponding predictions for the axion--photon coupling, $g_{a\gamma\gamma}$, are shown in Fig.~\ref{fig:axionphotoncoupling}, together with existing experimental bounds. The microwave cavity haloscopes ADMX~\cite{ADMX:2025vom}, HAYSTAC~\cite{HAYSTAC:2024jch}, QUAX~\cite{QUAX:2024fut}, TASEH~\cite{TASEH:2022vvu}, and CAPP~\cite{Yi:2022fmn} are starting to probe both the \textsc{F-DFSZ-I} and \textsc{F-DFSZ-II} models in the $\sim\mu\rm eV$ mass range, assuming pre-inflationary PQ-breaking. The future helioscope experiment IAXO is projected to probe new parameter space in the $d$-decoupled \textsc{F-DFSZ-I} model, while the enhanced IAXO+ sensitivity would extend this reach to the $s$-decoupled \textsc{F-DFSZ-I} model. In both cases, the accessible parameter space again lies outside the post-inflationary dark-matter window.

One could also consider more general, flavour-dependent charge assignments in the lepton sector. The primary constraint on their PQ charge assignments is that they lead to Yukawa textures that can fit the observed charged-lepton masses. Flavour-dependent charges can give different $E/N$ values and hence different axion-photon couplings, see for example~\cite{DiLuzio:2017pfr, Rocha:2025ade}.

The lepton sector can be further extended to include either Dirac or Majorana neutrino masses, in which case right-handed neutrinos may also carry PQ charges, subject to the requirement that viable neutrino mass textures are generated and that the resulting Pontecorvo--Maki--Nakagawa--Sakata (PMNS) matrix is consistent with experiment.

In scenarios with Majorana neutrinos, the complex scalar field $S$ responsible for PQ breaking can also break lepton number and generate heavy Majorana masses. Such constructions have been explored, for example in Ref.~\cite{Rocha:2025ade} where a flavour-dependent PQ symmetry leads to a maximally constrained set of texture zeros in the lepton sector.

%============================================================
\section{Higgs and Top Phenomenology}
\label{sec:higgspheno}
%============================================================

A characteristic feature of the \textsc{Minimal F-DFSZ} framework is the presence of tree-level flavour-changing neutral Higgs couplings. In this section, we confront the scalar sector of these models with collider data from Higgs and top-quark physics.

LHC measurements of the $h(125)$ resonance indicate that its production rates and decay branching fractions are consistent with SM predictions within current experimental uncertainties. Interpreted in the context of 2HDMs, this implies that we must lie close to the \emph{alignment limit} ($c_{\alpha-\beta} \to 0$)~\cite{Bernon:2015qea}, where the tree-level couplings of the SM-like Higgs $h$ to gauge bosons and fermions reduce to their SM values. ATLAS measurements place a robust upper bound on departures from alignment, $c_{\alpha-\beta} \lesssim 0.3\,(0.1)$ for the type-I (II) 2HDM~\cite{ATLAS:2021vrm}.\footnote{The type-I 2HDM has universal couplings of all RH quarks to one of the Higgs doublets. In the type-II 2HDM, the RH up- and down-type quarks couple to different Higgs doublets. The original, flavour-independent, DFSZ model is of type-II~\cite{Dine:1981rt}.} 

Alignment may be realised either with decoupling, \(\Lambda \equiv m_H\sim m_A\sim m_{H^\pm} \gg v\), for which \(c_{\alpha-\beta}^2\sim v^2/\Lambda^2\ll1\), or without decoupling through a suppression of the scalar-potential parameter \(\Lambda_6\)~\cite{Gunion:2002zf,Carena:2013ooa} (see App.~\ref{app:alignment}).\footnote{In the alignment-without-decoupling scenario, either $c_{\alpha-\beta} \to 0$~\cite{Bernon:2015qea} or $s_{\alpha-\beta} \to 0$~\cite{Bernon:2015wef}, in which case $h$ or $H$ represents the SM-like Higgs, respectively. We only consider the former branch, so $h$ always denotes the SM-like Higgs.}

In a general (type-III) 2HDM, the standard collider probes of misalignment are the Higgs signal strengths (Sec.~\ref{sec:HiggsSignalStrength}), flavour-violating Higgs decays $h \to b\bar{s}$ and $h \to b\bar{d}$ (Sec.~\ref{sec:FVHiggsDecays}), and the rare top decays $t \to qh$ (Sec.~\ref{sec:RareTopDecays})~\cite{Herrero-Garcia:2019mcy,Botella:2015hoa,Chiang:2015cba,Abbas:2015cua}. Whereas in a general 2HDM these flavour-violating rates are controlled by free parameters in the Yukawa sector, in the \textsc{Minimal F-DFSZ} framework they are fixed entirely by measured quark masses and CKM elements, up to the universal factor $c_{\alpha-\beta}^2(t_{\beta}+t_{\beta}^{-1})^2$, as can be seen from Eq.~\eqref{eq:Cuplings_h}. We exploit this to translate current/projected upper bounds on the branching ratios of $h \to b\bar{q}$ and $t \to qh$ into current/projected constraints on $|c_{\alpha-\beta}(t_{\beta}+t_{\beta}^{-1})|$ in each of the six $q$-decoupled realisations. Interestingly, this common factor cancels in ratios of flavour-violating decay rates, which therefore have sharp predictions in these models. However, the absolute rates of these processes are suppressed due to bounds from neutral meson mixing (to be discussed in Sec.~\ref{sec:flavourpheno}) which enforce alignment with decoupling in most $q$-decoupled models. The exception is the $d$-decoupled model, as we show in Sec.~\ref{sec:results}.

%------------------------------------------------------------
\subsection{Higgs Signal Strengths}
\label{sec:HiggsSignalStrength}

To confront the \textsc{Minimal F-DFSZ} models with Higgs signal-strength data, we employ the $\kappa$-framework in which deviations of Higgs couplings from their SM values are parameterised by rescaling factors $\kappa_i=g_{hii}^{\rm Model}/g_{hii}^{\rm SM}$. In the \textsc{Minimal F-DFSZ} models, the deviations in the fermion and $W$/$Z$-boson couplings are given by 
\begin{align}
    \kappa_f = \frac{v}{m_f}\,C_{ff}^{h_f}, \qquad \kappa_W = \kappa_Z =\sin(\beta-\alpha),
    \label{eq:kappaFermionGauge}
\end{align}
where $C_{ff}^{h_f}$ denotes the coupling of the Higgs $h$ to the fermion $f$ (see Eq.~\eqref{eq:Cuplings_h}). The current experimental constraints on the $\kappa_i$ from the ATLAS experiment~\cite{ATLAS:2021vrm} are summarised in Table~\ref{tab:kappameasurements}. 
\begin{table}[t!]
    \centering
    \begin{tabular}{c c c c c c}
        \hline \hline
        $\kappa_Z$ & $\kappa_W$ & $\kappa_b$ & $\kappa_t$ & $\kappa_{\tau}$ & $\kappa_{\mu}$ \\
        \hline
        $0.99 \pm 0.06$ & $1.03 \pm 0.05$ & $0.88 \pm 0.11$ & $0.92 \pm 0.06$ & $0.92 \pm 0.07$ & $1.07^{+0.25}_{-0.31}$ \\
        \hline \hline
    \end{tabular}
    \caption{Measurements of the $\kappa$ parameters from Run-II of ATLAS with $139\,\rm fb^{-1}$~\cite{ATLAS:2021vrm}.}
    \label{tab:kappameasurements}
\end{table}
Assuming a \textsc{F-DFSZ-I} leptonic structure\footnote{Such a structure implies that $\kappa_{\tau}=-s_{\alpha}/c_{\beta}$.}, we perform a $\chi^2$ fit, treating each $\kappa_i$ independently, and obtain the upper bounds on $|c_{\alpha-\beta}(t_{\beta}+t_{\beta}^{-1})|$ in Table~\ref{tab:SignalStrengthBounds}.

There are also contributions to $\kappa_{\gamma}$ that depend on the charged Higgs mass~\cite{Posch:2010hx}. A full analysis of flavour observables is performed in Sec.~\ref{sec:flavourpheno} and Sec.~\ref{sec:results}, and we have checked that for the parameter space considered, the constraint from $\kappa_{\gamma}$ is sub-leading relative to that from the global fit of the fermion and $W$/$Z$ couplings above. Interestingly, we find that the constraint from the global fit of the $\kappa_i$ is stronger than the constraints from flavour-violating top decays across much of the parameter space---a feature of the natural CKM-like suppression of flavour off-diagonal processes which we now discuss in detail.

\begin{table}[t]
\centering
\begin{tabular}{|c||c|c|c|c|c|c|}
\hline\hline
Model
& $u$ & $c$ & $t$ & $d$ & $s$ & $b$
\\
\hline
$|c_{\alpha-\beta}(t_\beta+t_\beta^{-1})|$
& $\lesssim 0.87$
& $\lesssim 0.46$
& $\lesssim 0.23$
& $\lesssim 0.39$
& $\lesssim 0.39$
& $\lesssim 0.23$
\\
\hline\hline
\end{tabular}
\caption{Upper bounds on $|c_{\alpha-\beta}(t_\beta+t_\beta^{-1})|$ from Higgs signal strength $\chi^2$ fit at $95\%$~CL for each $q$-decoupled model.}
\label{tab:SignalStrengthBounds}
\end{table}

%------------------------------------------------------------
\subsection{Rare Top Decays}
\label{sec:RareTopDecays}

Currently, the best direct probes of flavour-changing Higgs couplings are the rare top quark decays \(t\to ch\) and \(t\to uh\). In the SM, these transitions arise only at loop level and are strongly GIM-suppressed, with \(\mB(t\to ch)_{\rm SM}\sim \mO(10^{-15})\) and \(\mB(t\to uh)_{\rm SM}\sim \mO(10^{-17})\)~\cite{Altmannshofer:2019ogm}, far below the foreseeable collider sensitivity. By contrast, in a general (type-III) 2HDM, these decays can occur at tree level, and have been widely studied both in that setting~\cite{Baum:2008qm,Chen:2013qta,Altunkaynak:2015twa} and in related \(\mN_{\rm DW}=1\) flavoured DFSZ constructions~\cite{Chiang:2015cba,Chiang:2017fjr}, where the rates are controlled by additional free parameters in the Yukawa sector.

In the \textsc{Minimal F-DFSZ} framework, however, the rates for \(t\to qh\) are fixed by SM quark-sector observables up to the two scalar-sector parameters $(c_{\alpha-\beta},t_\beta)$:
\begin{align}
    \Gamma(t\rightarrow qh)
    &=
    \frac{m_t}{32\pi}\,
    \left(|C^{h_u}_{tq}|^2+|C^{h_u}_{qt}|^2\right)
    \left(1-\frac{m_h^2}{m_t^2}\right)^2, \nonumber\\
    &=\dfrac{m_t^3c_{\alpha-\beta}^2}{32\pi\, v^2}(t_{\beta}+t_{\beta}^{-1})^2\left(|\xi^u_{tq}|^2+|\xi^u_{qt}|^2 \right)\left(1-\dfrac{m_h^2}{m_t^2} \right)^2,
    \label{eq:tdecaypartialwidth}
\end{align}
where $q=u,c$ and we have neglected the light-quark masses. In each up-type-decoupled model, flavour-changing couplings involving the decoupled quark vanish. Consequently, both $t\to uh$ and $t\to ch$ are absent at tree level in the $t$-decoupled model, while $t\to uh$ ($t\to ch$) is absent in the $u$-decoupled ($c$-decoupled) model.

Since $\mB(t\to bW)\simeq 1$ and using
\begin{align}
\Gamma(t \to bW)
=
\frac{G_F}{8\pi \sqrt{2}}
|V_{tb}|^2 m_t^3
\left(1-\frac{m_W^2}{m_t^2}\right)^2
\left(1+2\frac{m_W^2}{m_t^2}\right),
\end{align}
we obtain
\begin{align}
    \mB(t \to qh)&\simeq0.1279 \times c_{\alpha-\beta}^2(t_{\beta}+t_{\beta}^{-1})^2\,(|\xi_{tq}|^2+|\xi_{qt}|^2).
\end{align}
We summarise the relevant model dependent factors $(|\xi_{ij}^q|^2+|\xi_{ji}^q|^2)$ (which were derived in Sec.~\ref{sec:FCNC} and App.~\ref{app:analyticFCNCXI}) for both $\mB(t \to uh)$ and $\mB(t \to ch)$ in Table~\ref{tab:top_ckm_factors}. In particular, notice that for all models the decay $t \to uh$ is suppressed by $\lambda^6$, whereas $t \to ch$ is suppressed by $\lambda^4$, with $\lambda=0.225$ the Wolfenstein parameter. Consequently, whenever the $t \to ch$ transition is present, it provides the dominant bound on $|c_{\alpha-\beta}(t_{\beta}+t_{\beta}^{-1})|$ from top decays.

The current ATLAS bounds on the branching ratios are~\cite{ATLAS:2024mih}
\begin{align}
\mB(t \rightarrow ch)_{\rm exp} < 3.3 \times 10^{-4}, \qquad
\mB(t \rightarrow uh)_{\rm exp} < 2.8 \times 10^{-4},
\end{align}
at $95\%$~CL. In Table~\ref{tab:top_ckm_factors}, we translate these into a bound on $|c_{\alpha-\beta}(t_{\beta}+t_{\beta}^{-1})|$ in each of the $q$-decoupled models. Moreover, for the $u$- and $c$-decoupled models, the chiral hierarchy in the $\xi^u_{ij}$ (as discussed in Sec.~\ref{sec:FCNCHiggs}) implies that the final-state light quark is predominantly left-handed. Given sufficient event yields, this prediction could be tested using helicity-sensitive observables, as previously explored for flavoured $\mN_{\rm DW}=1$ DFSZ models~\cite{Chiang:2017fjr,Chiang:2015cba}. 
In practice, however, the stringent meson-mixing constraints discussed in Sec.~\ref{sec:flavourpheno} force the scalar sector close to alignment, strongly suppressing the rare-top-decay branching ratios.

\begin{table}[t!]
\centering
\begin{tabular}{|c||c|c||c|c|}
\hline\hline
& \multicolumn{2}{c||}{$\mB(t\to uh)$} & \multicolumn{2}{c|}{$\mB(t\to ch)$} \\
\hline
Model
& $|\xi_{ut}^u|^2+|\xi_{tu}^u|^2$
& $|c_{\alpha-\beta}(t_{\beta}+t_{\beta}^{-1})|$
& $|\xi_{ct}^u|^2+|\xi_{tc}^u|^2$
& $|c_{\alpha-\beta}(t_{\beta}+t_{\beta}^{-1})|$
\\
\hline
$t$
&
0
&
$-$
&
0
&
$-$
\\[6pt]

$c$
&
$\displaystyle
|V_{ub}|^2 \sim \mathcal O(\lambda^6)$
&
$<9.8$
&
0
&
$-$
\\[8pt]

$u$
&
0
&
$-$
&
$\displaystyle
|V_{cb}|^2 \sim \mathcal O(\lambda^4)$
&
$<0.95$
\\[8pt]

$b$
&
$\displaystyle
\frac{|V_{ub}|^2|V_{cb}|^2}{1-|V_{tb}|^2}
\sim \mathcal O(\lambda^6)$
&
$<11.15$
&
$\displaystyle
\frac{|V_{cb}|^4}{1-|V_{tb}|^2}
\sim \mathcal O(\lambda^4)$
&
$<1.22$
\\[10pt]

$s$
&
$\displaystyle
|V_{ts}|^2|V_{us}|^2
\sim \mathcal O(\lambda^6)$
&
$<5.1$
&
$\displaystyle
\left(\frac{m_c}{m_t}\right)^2\frac{|V_{cs}|^2}{|V_{ts}|^2}
\sim \mathcal O(\lambda^4)$
&
$<0.54$
\\[10pt]

$d$
&
$\displaystyle
|V_{td}|^2|V_{ud}|^2
\sim \mathcal O(\lambda^6)$
&
$<5.67$
&
$\displaystyle
\left(\frac{m_c}{m_t}\right)^2\frac{|V_{cd}|^2}{|V_{td}|^2}
\sim \mathcal O(\lambda^4)$
&
$<0.53$
\\
\hline\hline
\end{tabular}
\caption{Suppression factors entering the flavour-violating top decays $t\to uh$ and $t\to ch$, together with the corresponding constraint on $|c_{\alpha-\beta}(t_{\beta}+t_{\beta}^{-1})|$ from the experimental bounds on $\mB(t \to qh)$.}
\label{tab:top_ckm_factors}
\end{table}

%------------------------------------------------------------
\subsection{Flavour-Violating Higgs Decays}
\label{sec:FVHiggsDecays}

The \textsc{Minimal F-DFSZ} models also predict the flavour-violating decays $h \to b\bar{s}$ and $h \to b\bar{d}$. There is currently no direct bound on these decay modes from the LHC; however, the proposed FCC-ee collider could achieve sensitivities of $\mB(h \to b\bar{s}) \sim 10^{-4}$~\cite{Kamenik:2023hvi,Arroyo-Urena:2025mju}. We therefore enumerate the predictions for these decays in the \textsc{Minimal F-DFSZ} framework. The branching ratio of $h \to b\bar{q}$ is given by
\begin{align}
    \mB(h \to b\bar{q})+\mB(h \to \bar{b}q) = \dfrac{3\,m_h}{8\pi\Gamma_h}\left(|C_{qb}^{h_d}|^2+|C_{bq}^{h_d}|^2\right).
    \label{eq:brhtobq}
\end{align}
Taking $\Gamma_h=\Gamma_h^{\rm SM} \simeq 4.1\,\rm MeV$, we find that
\begin{align}
    \mB(h \to b\bar{q})+\mB(h \to \bar{b}q)=0.4857 \times c_{\alpha-\beta}^2(t_{\beta}+t_{\beta}^{-1})^2\left[ |\xi_{qb}|^2+|\xi_{bq}|^2\right].
\end{align}
The leading contributions to the combination $|\xi_{qb}|^2+|\xi_{bq}|^2$ are listed in Table~\ref{tab:higgs_ckm_factors} for each $q$-decoupled model. 

As previously mentioned, deviations from alignment must be extremely suppressed due to constraints from meson mixing observables. We shall see in Sec.~\ref{sec:results} that the least constrained model is the $d$-decoupled model with an upper bound $|c_{\alpha-\beta}(t_{\beta}+t_{\beta}^{-1})|<0.038$. Even in this scenario, the allowed deviations from alignment result in a maximum branching ratio for $h \to\bar{b}s$ of $\sim 10^{-6}$, well below the potential sensitivity of even FCC-ee.

\begin{table}[htbp]
\centering
\begin{tabular}{|c||c||c|}
\hline\hline 
& $\mB(h\to b\bar d)+\mB(h \to \bar b d)$ & $\mB(h\to b\bar s)+\mB(h \to \bar b s)$ \\
\hline
Model
& $|\xi_{db}|^2+|\xi_{bd}|^2$
& $|\xi_{sb}|^2+|\xi_{bs}|^2$
\\
\hline
$t$
&
$\displaystyle
\ve_{db}^2\dfrac{|V_{td}|^2}{(|V_{td}|^2+|V_{ts}|^2\ve_{ds}^2)^2}\sim \mO(\lambda^2)$
&
$\displaystyle
|V_{ts}|^2|V_{tb}|^2
\sim \mO(\lambda^4)$
\\[8pt]
$c$
&
$\displaystyle
\dfrac{\ve_{db}^2}{|V_{cd}|^2|V_{cb}|^2}\sim \mO(\lambda^4)
$
&
$\displaystyle
\ve_{sb}^2\dfrac{|V_{cs}|^2}{|V_{cb}|^2} \sim \mO(\lambda^2)
$
\\[10pt]
$u$
&
$\displaystyle
\ve_{db}^2\dfrac{|V_{us}|^{4}|V_{ub}|^2|V_{ud}|^2}{(|V_{ub}|^{2}+|V_{us}|^{2}\ve_{sb}^{2})^2} \sim \mO(\lambda^8)
$
&
$\displaystyle
\ve_{sb}^2\,\dfrac{|V_{us}|^2\,|V_{ub}|^2}{(|V_{ub}|^{2}+|V_{us}|^{2}\ve_{sb}^{2})^2} \sim \mO(\lambda^2)
$
\\[8pt]
$b$
&
0
&
0
\\[6pt]
$s$
&
$\displaystyle
\dfrac{|V_{tb}|^2|V_{td}|^2}{1-|V_{ts}|^2}
\sim \mathcal O(\lambda^6)$
&
0
\\[10pt]
$d$
&
0
&
$\displaystyle
\frac{|V_{tb}|^2|V_{ts}|^2}{(1-|V_{td}|^2)^2} \sim \mO(\lambda^4)
$
\\
\hline\hline
\end{tabular}
\caption{Approximate suppression factors entering the flavour-violating Higgs decays $h\to b\bar d$ and $h\to b\bar s$.}
\label{tab:higgs_ckm_factors}
\end{table}

%============================================================
\section{Flavour Observables}
\label{sec:flavourpheno}
%============================================================

Having discussed the constraints from Higgs and top quark decays in Sec.~\ref{sec:higgspheno}, we now turn to the constraints from precision flavour observables: neutral meson mixing in the $B_q$, $K$, and $D$ systems, the CP-violating parameter $\ve_K$, the radiative decay $B \to X_s \gamma$, as well as the electroweak precision observable $Z \to b\bar{b}$. Among these, neutral-meson mixing usually provides the leading constraints on the \textsc{Minimal F-DFSZ} models for most of the relevant parameter space.

We perform our analysis in the Weak Effective Theory, integrating out the Higgs fields and electroweak gauge bosons at the common matching scale $\mu=m_W$ at one-loop. The Wilson coefficients are subsequently evolved to the hadronic scale appropriate for each meson system using one-loop QCD renormalisation-group evolution.

We restrict our analysis to observables whose predictions are fixed entirely by the quark-sector structure. We therefore do not include rare leptonic decays such as $B_s\to\mu^+\mu^-$, which additionally depend on the choice of lepton-sector completion. Nevertheless, we have verified that for the two simplest completions, \textsc{F-DFSZ-I} and \textsc{F-DFSZ-II} (discussed in Sec.~\ref{sec:axion_photon}) the constraint from $B_s\to\mu^+\mu^-$ is subdominant to the leading quark-sector bounds or is comparable to the bound from $\bar{B} \to X_s \gamma$.

The remainder of this section is organised as follows. In Sec.~\ref{sec:deltaf}, we set up the $\Delta F=2$ effective Hamiltonian formalism underlying the mixing observables. Our analyses of $B_q$-, $K$-, and $D$-meson mixing are presented in Secs.~\ref{sec:Bqmixing},~\ref{sec:Kmixing}, and~\ref{sec:Dmixing}, respectively, followed by the CP-violating parameter $\ve_K$ in Sec.~\ref{sec:KCP}. We then discuss the loop-induced constraints from $Z \to b\bar{b}$ in Sec.~\ref{sec:Zbb} and from the radiative decay $\bar{B} \to X_s \gamma$ in Sec.~\ref{sec:bsgamma}. The relevant experimental inputs are collected in Table~\ref{tab:FlavObsExp} in App.~\ref{app:params}. The combined results of our analysis are presented in Sec.~\ref{sec:results}.
		\begin{table}[t!]
			\renewcommand*{\arraystretch}{1.2}
			\centering
			\begin{tabular}{lccc}
				\hline \hline
				Observable \ & Constraint & Ref.  \\
				\hline
				$\Delta M_{B_s}$   & $17.765 \pm 0.00565\, \rm ps^{-1}$ & HFLAV~\cite{HFLAV:2022esi}  
				\\
				$\Delta M_{B_d}$  & $0.5065\pm0.0019 \, \rm ps^{-1}$ & HFLAV~\cite{HFLAV:2022esi} 
				\\
				$\Delta M_K$   & $3.484(6) \times 10^{-12}\,\,\rm MeV$ & PDG~\cite{ParticleDataGroup:2024cfk}
				\\
                $\Delta M_D$ & $(6.56 \pm 0.76)\times 10^{-15}\, \rm GeV$ & HFLAV~\cite{HFLAV:2022esi} \\
				|$\ve_K$|   & $2.228(11)\times10^{-3}$ & PDG~\cite{ParticleDataGroup:2024cfk} 
				\\
				$\mB(\bar{B} \rightarrow X_s\gamma)$   & $(3.49 \pm 0.19) \times 10^{-4}$ & HFLAV~\cite{HeavyFlavorAveragingGroupHFLAV:2024ctg} 
				\\
				\hline
				\hline
			\end{tabular}
			\caption{Current experimental bounds on meson mixing observables and radiative meson decays.} 
			\label{tab:FlavObsExp}
		\end{table}

%------------------------------------------------------------
\subsection{$\Delta F = 2$ Observables}
\label{sec:deltaf}

For a neutral meson system $\{X,\bar{X}\}$, the mass difference between the mass eigenstates, $\Delta M_X$, is determined by the dispersive part of the mixing amplitude, $M_{12}$, and is highly sensitive to heavy virtual particles. The short-distance contribution to $M_{12}$ is described below the electroweak scale by the $\Delta F=2$ effective Hamiltonian~\cite{Gabbiani:1996hi}
\begin{align}
\mathcal{H}_{\rm eff}^{\Delta F=2}
=
\sum_{i=1}^{5} \mC_i(\mu)\,O_i
+
\sum_{i=1}^{3} \mC_i'(\mu)\,O_i',
\end{align}
where,
\begin{align}
O_1 &= (\bar q'^\alpha \gamma_\mu P_L q^\alpha)(\bar q'^\beta \gamma^\mu P_L q^\beta), &
O_2 &= (\bar q'^\alpha P_L q^\alpha)(\bar q'^\beta P_L q^\beta), \nonumber\\
O_3 &= (\bar q'^\alpha P_L q^\beta)(\bar q'^\beta P_L q^\alpha), &
O_4 &= (\bar q'^\alpha P_L q^\alpha)(\bar q'^\beta P_R q^\beta), \nonumber \\
O_5 &= (\bar q'^\alpha P_L q^\beta)(\bar q'^\beta P_R q^\alpha).
\label{eqn:effectiveoperators}
\end{align}
Here, $\alpha$ and $\beta$ denote colour indices, with the $SU(3)_C$ contractions shown explicitly, and the operators $O_i'$ are given by the replacement $P_L \rightleftarrows P_R$.\footnote{Note that only $O_1,O_2$ and $O_3$ map to a different $O_i'$.}

\begin{figure}[t]
    \centering

    % ---------- Row 1 ----------
    % Standard Model W-W box
    \begin{subfigure}[b]{0.45\textwidth}
        \centering
        \includegraphics[]{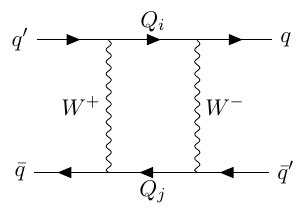}
        \caption{Standard Model \(W^+\)--\(W^-\) box}
        \label{fig:ww-box}
    \end{subfigure}
    \hfill
    % Tree-level neutral-scalar exchange
    \begin{subfigure}[b]{0.45\textwidth}
        \centering
        \includegraphics[]{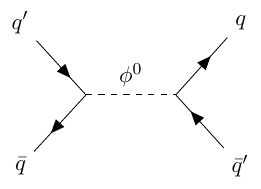}
        \caption{Tree-level neutral-scalar exchange}
        \label{fig:neutral-scalar-exchange}
    \end{subfigure}

    \vspace{1em}

    % ---------- Row 2 ----------
    % Neutral-scalar box
    \begin{subfigure}[b]{0.45\textwidth}
        \centering
        \includegraphics[]{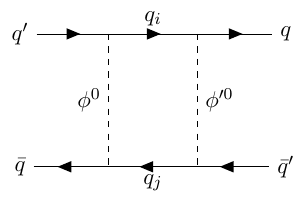}
        \caption{Neutral--scalar box}
        \label{fig:neutral-scalar-box}
    \end{subfigure}
    \hfill
    % Charged-Higgs box
    \begin{subfigure}[b]{0.45\textwidth}
        \centering
        \includegraphics[]{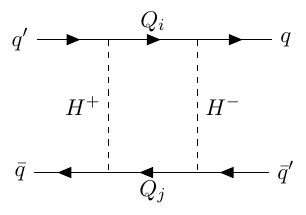}
        \caption{Charged-Higgs \(H^+\)--\(H^-\) box}
        \label{fig:hh-box}
    \end{subfigure}

    \vspace{1em}

    % ---------- Row 3 ----------
    % Mixed W-H box
    \begin{subfigure}[b]{0.45\textwidth}
        \centering
        \includegraphics[]{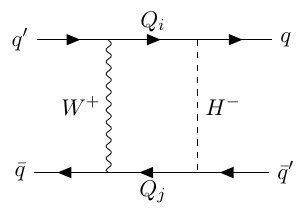}
        \caption{Mixed \(W^\pm\) --\(H^\mp\) box}
        \label{fig:wh-box}
    \end{subfigure}

    \caption{Tree- and one-loop diagrams contributing to neutral-meson mixing. Here, \(q\) and \(q'\) denote distinct external quark flavours, \(q_i\) and \(q_j\) denote internal quarks with the same electric charge as the external quarks, and \(Q_i\) and \(Q_j\) denote their weak-isospin partners. The neutral Higgs are denoted by \(\phi^0,\phi'^0\in\{h,H,A\}\). Only a single diagram has been shown for each class.}
    \label{fig:boxdiagrams}
\end{figure}

The SM contributes only to $O_1$ at loop-level via box diagrams such as that in Fig.~\ref{fig:ww-box}, with the corresponding Wilson coefficient denoted by $\mC_1^{\textrm{SM}}(\mu_X)$. In the \textsc{Minimal F-DFSZ} models, tree-level contributions from neutral scalar exchange, such as displayed in Fig.~\ref{fig:neutral-scalar-exchange}, give the following contributions to the Wilson coefficients for down-type mesons
\begin{align}
\mC_2(\mu) &= -\sum_{\phi=h,H,A}\frac{1}{2\,m_{\phi}^2}\,\bigl(C^{\phi_d}_{qq'}\bigr)^{\!*2},\qquad
\mC_2'(\mu) = -\sum_{\phi=h,H,A}\frac{1}{2\,m_{\phi}^2}\,\bigl(C^{\phi_d}_{q'q}\bigr)^{2}, \nonumber \\
\mC_4(\mu) &= -\sum_{\phi=h,H,A}\frac{1}{m_{\phi}^2}\,\bigl(C^{\phi_d}_{q'q}\,C^{\phi_d*}_{qq'}\bigr).
\label{eq:tree-levelWC}
\end{align}
The analogous tree-level contributions to $D$-mixing can be found by replacing $\phi_d \to \phi_u$. 

In addition to the tree-level contributions, there can be important one-loop contributions. For a given neutral meson system, the tree and one-loop diagrams are proportional to different products of Higgs couplings. Therefore, depending on the flavour structure of $\xi^u$ and $\xi^d$, there can be cases in which one-loop contributions dominate.

The tree-level contributions to $B_d$, $B_s$ and $K$ mixing are proportional to different products of off-diagonal elements of $\xi^d$. For this reason, loop-level contributions proportional to $\xi_{ij}^d$ are never relevant in determining the leading bound from meson mixing. We therefore only include one-loop contributions that are proportional to $\xi_{ij}^u$. For $B_q$ and $K$ mixing, these arise from box diagrams involving the charged Higgs (e.g. Figs.~\ref{fig:hh-box},~\ref{fig:wh-box}). For $D$-meson mixing, the tree-level contributions are proportional to $\xi^u_{uc}$ and/or $\xi^u_{cu}$. The potentially relevant one-loop contributions are the neutral Higgs boxes with an internal top quark, which dominate over the charged Higgs boxes due to a top mass enhancement. However, we find that in all the \textsc{Minimal F-DFSZ} models they are sub-leading with respect to the tree-level contributions to $D$-mixing, giving percent level corrections.

The neutral and charged Higgs box diagrams contribute to $O_1,...,O_5$. The corresponding Wilson coefficients are taken from Ref.~\cite{Crivellin:2013wna} and collected in App.~\ref{app:PVfuncs}. The inclusion of these one-loop contributions distinguishes our analysis from that of Ref.~\cite{Rocha:2024twm}.

All Wilson coefficients are matched at the electroweak scale $\mu_{\rm high}=m_W$. They are then evolved to the hadronic scale, $\mu_X\sim M_X$, according to
\begin{align}
\mC_i(\mu_{\rm low})
=
U_{ij}(\mu_{\rm low},\mu_{\rm high})
\mC_j(\mu_{\rm high}),
\end{align}
using the leading-order QCD evolution matrix $U_{ij}$ given in App.~\ref{app:params}. The renormalisation-group evolution mixes the operator pairs $(O_2,O_3)$ and $(O_4,O_5)$.

The short-distance contribution to $M_{12}$ is then computed via
\begin{align}
\label{eq:dispersiveamplitude}
M_{12}^{X} 
= \frac{\langle X|\mathcal{H}_{\rm eff}^{\Delta F=2}|\bar X\rangle}{2M_{X}}.
\end{align}
The hadronic matrix elements are parameterised as~\cite{FermilabLattice:2016ipl},
\begin{align}
\langle O_1 \rangle 
&\equiv \bra{X}O_1(\mu_X)\ket{\bar X}
= b_1\,M_{X}^2 f_{X}^2\, B^{(1)}(\mu_X),\\
\langle O_i \rangle
&= b_i\,\left[\left(\dfrac{M_{X}}{m_q(\mu_X)\,+\,m_{q'}(\mu_X)} \right)^2+d_i\right]\,M_{X}^2 f_{X}^2\, B^{{(i)}}(\mu_X), \qquad i=2,\dots,5,
\end{align}
with $b_i = (2/3,\,-5/12,\,1/12,\,1/2,\,1/6)$. The $B^{(i)}$ are bag parameters encoding the deviation of each matrix element from its vacuum saturation approximation (VSA) value. The constants $d_i$ for the chirality-flipping operators $O_4$ and $O_5$ depend on the convention in which the bag parameters are defined; we adopt the full-VSA convention for $B_q$ mixing ($d_4=1/6$,
$d_5=3/2$)~\cite{FermilabLattice:2016ipl,Greljo:2022jac} and the usual lattice convention for $K$ and $D$ mixing ($d_i=0$)~\cite{Garron:2016mva,FlavourLatticeAveragingGroupFLAG:2024oxs}, with $d_2=d_3=0$ throughout. All relevant meson masses and decay constants are listed in Table~\ref{tab:mesonmixinginputs1} in App.~\ref{app:params}. We now discuss the specific meson systems.

\subsubsection{$B_q-\bar{B}_q$ mixing}
\label{sec:Bqmixing}

For $B_q^0$ with $q=d,s$, the mass difference is dominated by short distance physics such that $\Delta M_{B_q} \simeq 2 |M_{12}^{B_q}|$. The theoretical prediction for $\Delta M_{B_q}$ can then be straightforwardly computed using the $\Delta F=2$ effective Hamiltonian in Eq.~\eqref{eqn:effectiveoperators} and compared with experiment.

The SM contribution to $O_1$ is given by the Wilson coefficient
\begin{equation}
\mC_1^{\rm SM}(\mu_{B_s})=\frac{G_F^2 m_W^2}{4\pi^2}\,\bigl(V_{tq}^*V_{tb}\bigr)^2\,\hat\eta_B(\mu_{B_{s}})\,S_0(x_t),\quad x_t=\tfrac{m_t^2}{m_W^2},
\end{equation}
where $\hat{\eta}_B\approx0.84$ encodes the NLO running from $\mu_W$ to $\mu_{B_s}$ and $S_0(x_t)$ is the Inami--Lim function defined as~\cite{Inami:1980fz},
\begin{align}
S_0(x)
&=
\frac{x}{(1-x)^2}
\left(
1
- \frac{11}{4}x
+ \frac{x^2}{4}
- \frac{3x^2}{2(1-x)}\ln x
\right).
\end{align}
The remaining Wilson coefficients are evolved down to $\mu = M_{B_s}$ using the evolution matrix given in Table~\ref{tab:mesonmixinginputs1} in App.~\ref{app:params}.

It is known that the dominant contribution to the theoretical uncertainties is the lattice-determined bag factors, $B^{(i)}$, while the other uncertainties arising from experimental inputs such as CKM parameters are negligible by comparison~\cite{Albrecht:2024oyn}. In our analysis, we construct a multivariate distribution of the $B_q$ bag factors given in Table~\ref{tab:bag_parameters}, with correlations given in~\cite{Greljo:2022jac}. The theoretical uncertainty is then obtained by random sampling of this distribution and added in quadrature with the experimental uncertainty given in Table~\ref{tab:FlavObsExp}.
\begin{table}[t]
\centering
\renewcommand{\arraystretch}{1.3}
\begin{tabular}{c|ccccc}
\hline\hline
 $B_q$ bag factor& $B^{(1)}$ & $B^{(2)}$ & $B^{(3)}$ & $B^{(4)}$ & $B^{(5)}$ \\
\hline
$B_d$ &
$0.837 \pm 0.030$ &
$0.808 \pm 0.037$ &
$0.761 \pm 0.053$ &
$1.054 \pm 0.042$ &
$0.986 \pm 0.037$ \\

$B_s$ &
$0.826 \pm 0.029$ &
$0.824 \pm 0.037$ &
$0.830 \pm 0.053$ &
$1.040 \pm 0.041$ &
$0.964 \pm 0.034$ \\
\hline\hline
\end{tabular}
\caption{Combined HPQCD~\cite{Dowdall:2019bea} and KLR~\cite{King:2019lal} determinations of the $B_d$ and $B_s$ bag parameters, taken from Ref.~\cite{Greljo:2022jac}. The corresponding correlation matrix is also provided therein.}
\label{tab:bag_parameters}
\end{table}

\subsubsection{$K-\bar{K}$ mixing}
\label{sec:Kmixing}
The neutral kaon system receives
non-negligible long-distance contributions that are not under full theoretical control. As a result, only the short-distance component of the mass difference $\Delta M_K$ can be reliably computed. 

The SM value for the short-distance contribution to $\Delta M_K$ is given by~\cite{Brod:2011ty}
\begin{align}
   \left( \Delta M_K^{\textrm{SD}}\right)_{\textrm{SM}}=3.1(1.2)\times10^{-15}\,\rm GeV.
\end{align}
Due to the poorly known long-distance effects, we impose the commonly used requirement (e.g.~\cite{Rocha:2024twm,Botella:2014ska,Ferreira:2019aps}) that the new physics short-distance contribution does not exceed the SM short-distance contribution by more than $10\%$:
\begin{align}
   \Bigg| \dfrac{\left(\Delta M_K^{\rm SD}\right)_{\rm NP}}{\left(\Delta M_K^{\rm SD}\right)_{\rm SM}} \Bigg |\leq 0.1 ,
\end{align}
where we conservatively take the SM value to be $2\sigma$ above the central theoretical value. The bag factors, evaluated at $\mu \sim 3\,\rm GeV$, are given in Table~\ref{tab:KaonBags}.

\begin{table}[t]
    \centering
    \renewcommand{\arraystretch}{1.2}
    \begin{tabular}{c c c c c c}
        \hline\hline
        Kaon Bag Factor & $B_1$ & $B_2$ & $B_3$ & $B_4$ & $B_5$ \\
        \hline
        Value & $0.507(13)(11)$ & $0.46(1)(3)$ & $0.79(2)(5)$ & $0.78(2)(4)$ & $0.49(3)(3)$ \\
        \hline\hline
    \end{tabular}
    \caption{Kaon bag factors used to determine $\Delta M_K$ and $\ve_K$, computed in the $N_f=4$ RI/MOM scheme and converted to the $\overline{\mathrm{MS}}$ scheme at $\mu=3\,\mathrm{GeV}$~\cite{FlavourLatticeAveragingGroupFLAG:2024oxs,Carrasco:2015pra}.}
    \label{tab:KaonBags}
\end{table}

\subsubsection{$D^0-\bar{D}^0$ mixing}
\label{sec:Dmixing}
The $D^0 - \bar D^0$ system differs qualitatively from both the $B$ and $K$ systems. In the SM, the short-distance contribution to $\Delta M_D$ is highly GIM suppressed and is estimated to lie four orders of magnitude below the experimentally measured value~\cite{Golowich:2005pt}. The observed mass difference is therefore believed to be dominated by long-distance effects, which are difficult to calculate theoretically.\footnote{Although some progress has been made recently~\cite{Dulibic:2025emg,DiCarlo:2025mnm}.} In our analysis, we therefore constrain only the new-physics contribution to the dispersive mixing amplitude,
\[
\Delta M_D^{\rm NP} \;\simeq\; 2\,\bigl| (M_{12}^D)_{\rm NP} \bigr|,
\]
and require that it does not exceed the experimental value $\bigl|\Delta M_D^{\rm NP}\bigr|\;\le\;\bigl|\Delta M_D^{\rm exp}\bigr|$. The lattice-determined hadronic matrix elements are given in Table~\ref{tab:DMesMatElts}.
\begin{table}[t]
    \centering
    \renewcommand{\arraystretch}{1.2}
\begin{tabular}{c c c c c}
    \hline\hline
    $\langle O_1 \rangle$ &
    $\langle O_2 \rangle$ &
    $\langle O_3 \rangle$ &
    $\langle O_4 \rangle$ &
    $\langle O_5 \rangle$ \\
    \hline
    $0.0805(55)(16)$ &
    $-0.1561(70)$ &
    $0.0464(31)(9)$ &
    $0.2747(129)$ &
    $0.1035(71)(21)$ \\
    \hline\hline
\end{tabular}
    \caption{Hadronic matrix elements for $D$-meson mixing, computed using lattice QCD in the $\overline{\textrm{MS}}$-NDR scheme at $\mu=3\,\rm GeV$~\cite{Bazavov:2017weg}.}
    \label{tab:DMesMatElts}
\end{table}

%------------------------------------------------------------
\subsubsection{CP Violation: $\ve_K$}
\label{sec:KCP}

Another important constraint from $K - \bar K$ mixing arises from the CP-violating observable $\ve_K$ \cite{Buras:2008nn}:
\begin{align}
    \ve_K = \kappa_{\ve}\,\frac{e^{i\phi_{\ve}}}{\sqrt{2}}\,
    \frac{\textrm{Im}\!\left(M_{12}^K\,\lambda_c^{\ast 2}\right)}
    {\left[\Delta M_K\right]_{\rm exp}\,|\lambda_c|^2}\,,
\end{align}
where $\lambda_c=V_{cs}^*V_{cd}$, $\kappa_{\ve}\simeq0.94$ \cite{Brod:2019rzc}, and $\phi_{\ve}\simeq43.52^\circ$ \cite{FlavourLatticeAveragingGroupFLAG:2024oxs}.\footnote{The factors of $\lambda_c$ ensure that our computation of $\ve_K$ is phase-convention independent.} Unlike $\Delta M_K$ and $\Delta M_D$, $\ve_K$ is not dominated by incalculable long-distance contributions, and can therefore be used as a direct constraint on new physics. Using the $\Delta F=2$ Hamiltonian in Eq.~\eqref{eqn:effectiveoperators}, we compute the dispersive mixing amplitude $M_{12}^K$ through Eq.~\eqref{eq:dispersiveamplitude}, including both tree- and loop-level contributions as in Sec.~\ref{sec:Kmixing}.

For the SM contribution, we use $|\ve_K|_{\rm SM}= 2.16(6)(8)(15)\times 10^{-3}$ \cite{Brod:2019rzc} and impose a $2\sigma$ bound on the new-physics contribution by comparing $|\ve_K|_{\rm SM+NP}$ to the experimental value, combining experimental, SM, and bag parameter uncertainties in quadrature, where the uncertainty on the new physics contribution is obtained via Monte Carlo propagation of the bag-factor uncertainties in Table~\ref{tab:KaonBags}.

%------------------------------------------------------------
\subsection{$Z \rightarrow \bar{b}b$}
\label{sec:Zbb}

Loop corrections involving charged and neutral scalars can induce sizeable modifications to the effective $Z\bar{b}b$ couplings, making
this observable a sensitive probe of extended Higgs sectors. We follow Refs.~\cite{Rocha:2024twm,Ferreira:2019aps} and impose the $2\sigma$ constraint derived from LEP and SLC measurements,
\begin{equation}
    2(\bar{g}_b^L)^2 + 2(\bar{g}_b^R)^2
    = 0.36782 \pm  0.00143 \,,
\end{equation}
where $\bar{g}_b^{L,R}$ denote the effective left- and right-handed $Z\bar{b}b$ couplings.

These couplings can be written as
\begin{align}
    \bar{g}_b^L &= -0.42112 + \delta\bar{g}_b^L \,, \qquad
    \bar{g}_b^R = 0.07744 + \delta\bar{g}_b^R \,,
\end{align}
where the numerical values are the SM contributions and the dominant one-loop corrections arise from charged-Higgs loops. The leading contributions are given by~\cite{Haber:1999zh,Ferreira:2019aps,Hernandez-Sanchez:2012vxa}
\begin{align}
    \delta\bar{g}_b^L &=
    \frac{\sqrt{2}G_Fv^2}{16\pi^2}
    \left| (C^{H_u^-\dagger} V)_{33} \right|^2
    f_1\!\left(\frac{m_t^2}{m_{H^\pm}^2}\right) \,, \\
    \delta\bar{g}_b^R &=-
    \frac{\sqrt{2}G_Fv^2}{16\pi^2}
    \left| (V C^{H_d^+})_{33} \right|^2
    f_1\!\left(\frac{m_t^2}{m_{H^\pm}^2}\right) \,,
\end{align}
where the loop function is
\begin{equation}
    f_1(x) = \frac{x}{x-1}
    \left(1 - \frac{\ln x}{x-1}\right) \,.
\end{equation}

%------------------------------------------------------------
\subsection{Radiative Meson Decay \texorpdfstring{$\bar{B}\to X_s\gamma$}{b -> s gamma}}
\label{sec:bsgamma}

\begin{figure}[t]
\centering
 
\begin{subfigure}[b]{0.47\textwidth}
\centering
\includegraphics[]{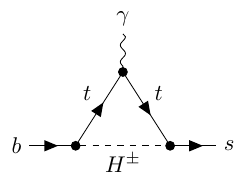}
\caption{\(b \to s\gamma\): 2HDM contribution to $O_7$}
\end{subfigure}
\hfill
\begin{subfigure}[b]{0.47\textwidth}
\centering
\includegraphics[]{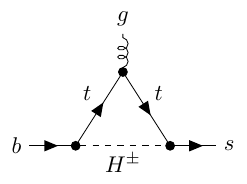}
\caption{\(b \to sg\): 2HDM contribution to $O_8$}
\end{subfigure}
\caption{Charged-Higgs penguin contributions to $b \to s \gamma$ and $b \to sg$.}
\label{fig:bsgamma_bsg_hpm}
\end{figure}

At the parton level, the $\bar{B}\to X_s\gamma$ decay is dominated by dipole operators, to which both the charged- and neutral-scalars contribute at one loop in the \textsc{Minimal F-DFSZ}, as shown in Fig.~\ref{fig:bsgamma_bsg_hpm} for the charged Higgs case. 
 
At the parton level, the $b\to s\gamma$ transition is described by the standard effective Hamiltonian,
\begin{equation}
\mathcal H_{\rm eff}^{b\to s\gamma}
= -\frac{4G_F}{\sqrt{2}}\,V_{tb}V_{ts}^\ast
\sum_{i=7,8}\Bigl[
\mC_i(\mu_{B_s})\,O_i + \mC_i'(\mu_{B_s})\,O_i'
\Bigr],
\end{equation}
where $\mu_{B_s}= M_{B_s}$. The dipole operators are
\begin{align}
 O_7 &= \frac{e}{16\pi^2}\,m_b\,
\bar s_{\alpha}\sigma^{\mu\nu}P_R b_{\alpha}\,F_{\mu\nu},
&
O_8 &= \frac{g_s}{16\pi^2}\,m_b\,
\bar s_{\alpha}\sigma^{\mu\nu}P_R \Bigl(\frac{\lambda^a}{2}\Bigr)_{\alpha\beta} b_{\beta}\,G^a_{\mu\nu},
\\
O_7' &= \frac{e}{16\pi^2}\,m_b\,
\bar s_{\alpha}\sigma^{\mu\nu}P_L b_{\alpha}\,F_{\mu\nu},
&
O_8' &= \frac{g_s}{16\pi^2}\,m_b\,
\bar s_{\alpha}\sigma^{\mu\nu}P_L \Bigl(\frac{\lambda^a}{2}\Bigr)_{\alpha\beta} b_{\beta}\,G^a_{\mu\nu}.
\end{align}
The inclusive branching ratio is then given by~\cite{Misiak:2006ab,Buras:2011zb,Blanke:2011ry,Blanke:2012tv},
\begin{equation}
\mB\!\left(\bar B\to X_s\gamma\right)
=\mB_{\rm SM}
+0.00247\Bigl[
|\Delta \mC_7(\mu_{B_s})|^2+|\Delta \mC_7'(\mu_{B_s})|^2
-0.706\,\mathrm{Re}\,\Delta \mC_7(\mu_{B_s})
\Bigr],
\label{eq:bsg_master}
\end{equation}
where $\mB_{\rm SM}=(3.40\pm0.17)\times10^{-4}$ denotes the NNLO SM prediction~\cite{Misiak:2020vlo} and where $\Delta \mC_i$ are defined in the following way:
\begin{equation}
\mC_{i}^{(\prime)}(\mu)=\mC_{i,\rm SM}^{(\prime)}(\mu)+\Delta \mC_i^{(\prime)}(\mu),
\qquad i=7,8.
\end{equation}

At the heavy scale $\Lambda$, the new-physics contributions can be written as a sum of charged- and neutral-scalar pieces,
\begin{equation}
\Delta \mC_i^{(\prime)}(\mu)=
\Delta \mC_{i}^{(\prime),H^\pm}(\mu)
+\sum_{\phi=h,H,A}\Delta \mC_{i}^{(\prime),\phi}(\mu),
\qquad i=7,8.
\label{eq:bsg_decomposition}
\end{equation}
The explicit matching expressions and the relevant loop functions are collected in App.~\ref{app:bsgamma}. Here, the matching is performed at the scale of the mediating scalar, $\mu_{\phi}$, and evolved to $\mu_{B_s}$ using leading-logarithmic QCD running,
\begin{equation}
\Delta \mC_7^{(\prime)}(\mu_{B_s})\simeq
\eta^{16/23}\,\Delta \mC_7^{(\prime)}(\mu_\phi)
+\frac{8}{3}\Bigl(\eta^{14/23}-\eta^{16/23}\Bigr)\Delta \mC_8^{(\prime)}(\mu_\phi),
\qquad
\eta=\frac{\alpha_s(\mu_{\phi})}{\alpha_s(\mu_{B_s})}.
\label{eq:bsg_running}
\end{equation}
The RG running induces operator mixing that necessitates the inclusion of the $b \to sg$ diagram in Fig.~\ref{fig:bsgamma_bsg_hpm}. 

The new physics contributions are constrained by comparing Eq.~\eqref{eq:bsg_master} to the experimental world average
\cite{HeavyFlavorAveragingGroupHFLAV:2024ctg},
\begin{equation}
\mB_{\rm exp}\equiv \mB(\bar B\to X_s\gamma)_{\rm exp}
=\left(3.49\pm0.19\right)\times10^{-4}.
\end{equation}

%============================================================
\section{Results and Discussion}
\label{sec:results}
%============================================================
In this section, we discuss the full flavour analysis of all $q$-decoupled models. We impose all flavour constraints at $2\sigma$ and, for simplicity, we take the heavy scalars to be degenerate, $\Lambda \equiv m_H = m_A = m_{H^{\pm}}$. While scalar mass splittings are automatically suppressed in the decoupling limit $\Lambda \gg v$, we impose this degeneracy across the full scan range $\Lambda \sim \mO(1 \text{--} 10^5\,\rm TeV)$ and verify in App.~\ref{app:scalarpotential} that it is compatible with the theoretical and phenomenological requirements of the scalar potential. The parameter space of each model is then spanned by $(c_{\alpha-\beta}, t_{\beta}, \Lambda)$, of which we study two complementary slices. 

\begin{figure}[!p]
    \centering
    \includegraphics[width=\textwidth]{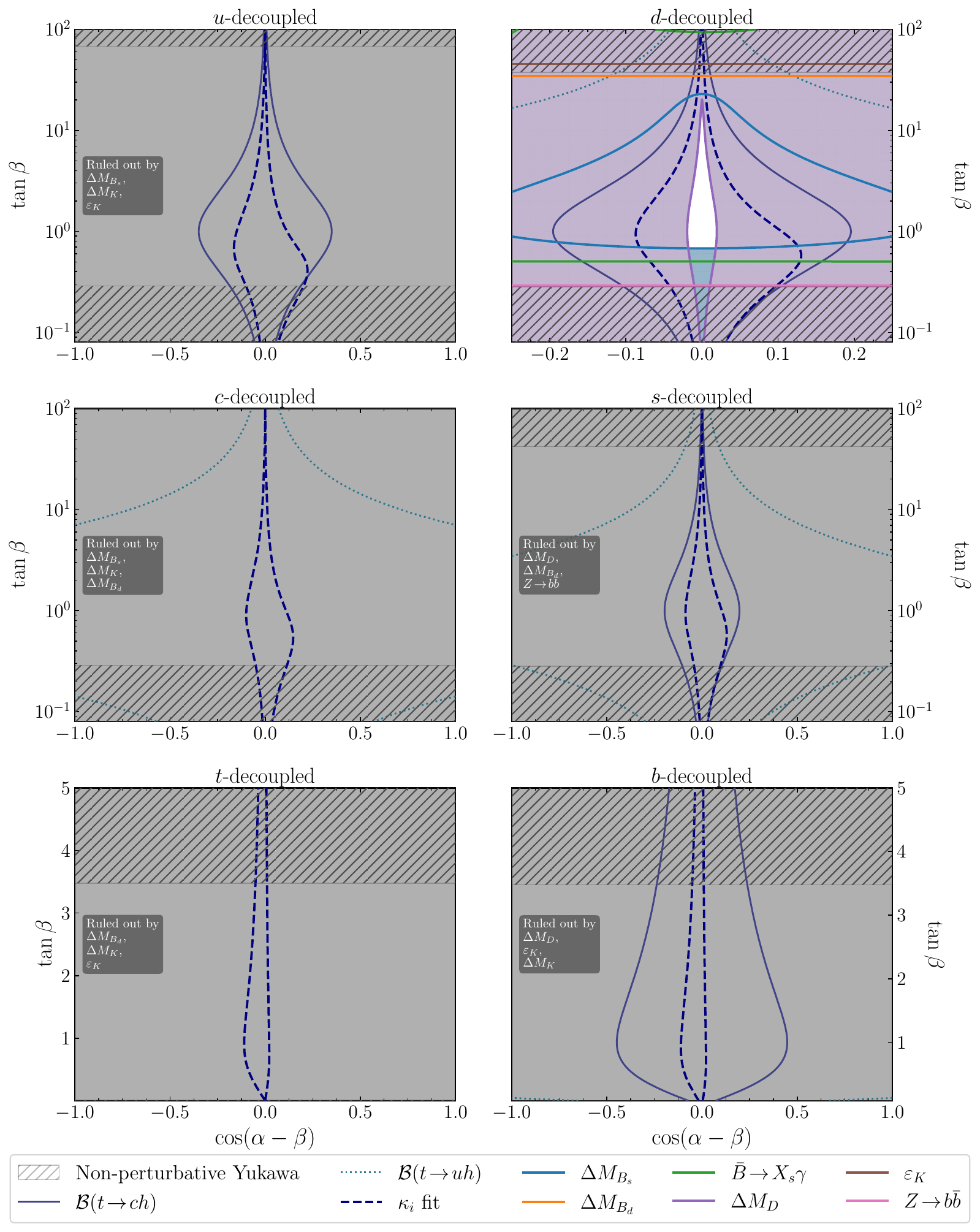}
    \caption{Flavour constraints on the six $q$-decoupled models in the $(c_{\alpha-\beta},\,t_\beta)$ plane, assuming degenerate non-SM scalars with a common mass $\Lambda=2\,\mathrm{TeV}$. Grey shading denotes regions excluded by meson mixing observables. At this benchmark scale, the entire displayed parameter space is excluded for every realisation except the $d$-decoupled model for which we show a smaller $c_{\alpha-\beta}$ range. The allowed region for the $d$-decoupled model is indicated by the white region.}
    \label{fig:grid_all}
\end{figure}
\begin{figure}[!p]
    \centering
    \includegraphics[width=\textwidth]{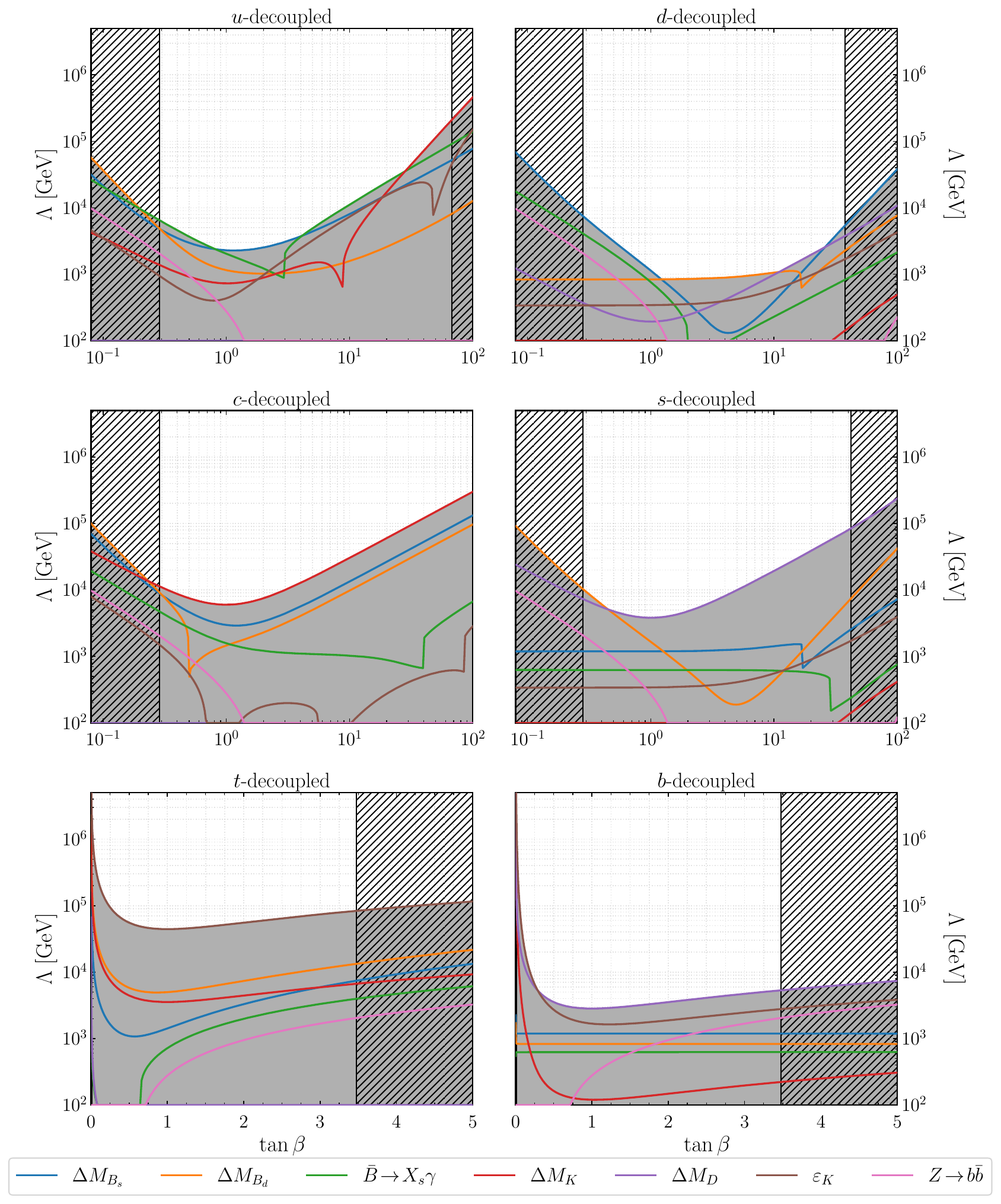}
    \caption{Flavour constraints on all $q$-decoupled models in the alignment limit, $c_{\alpha-\beta} = 0$. We have assumed all non-SM scalar masses are degenerate $\Lambda = m_H = m_A = m_{H^\pm}$. The grey shaded regions are experimentally excluded, while the coloured contours indicate the $2\sigma$ upper bounds on $\Lambda$ from each observable. The black hatching indicates the region in which Yukawa couplings become non-perturbative.}
    \label{fig:flavour_constraints_all}
\end{figure}

\paragraph{Higgs Misalignment:}
We first consider departures from the alignment limit. Appreciable misalignment requires the additional Higgs states to remain relatively light, with $\Lambda$ of order a few TeV or below. We adopt the benchmark $\Lambda=2\,\rm TeV$ which, as shown in App.~\ref{app:benchmark}, permits values up to $|c_{\alpha-\beta}|\sim0.2$ while the scalar sector parameters remain perturbative.\footnote{Note that we nevertheless show a larger range of $c_{\alpha-\beta}$ in Fig.~\ref{fig:grid_all} to illustrate the constraints from $t \to qh$ and the Higgs signal strengths, as these are independent of $\Lambda$.} 

The resulting constraints in the $(c_{\alpha-\beta},t_\beta)$ plane are displayed in Fig.~\ref{fig:grid_all}. Among the Higgs- and top-sector observables discussed in Sec.~\ref{sec:higgspheno}, the leading bounds arise from the global fit on the Higgs coupling modifiers, $\kappa_i$, described in Sec.~\ref{sec:HiggsSignalStrength}, since the CKM-like suppression of the flavour-violating couplings weakens the constraints from rare top decays. Once meson-mixing constraints are included, however, the entire displayed parameter region is excluded at $\Lambda=2~\mathrm{TeV}$ for every realisation except the $d$-decoupled model.

\paragraph{Higgs Misalignment in the $d$-decoupled model:}
The $d$-decoupled model forbids tree-level contributions to $\Delta M_K$ and $\Delta M_{B_d}$, while the residual loop-level contribution to $\Delta M_{B_d}$ imposes approximately $t_\beta\lesssim20$. The remaining constraints from $\Delta M_{B_s}$ and $\Delta M_D$ are weakened by the small products $|\xi_{sb} \, \xi_{bs}|$ and $|\xi_{uc}\,\xi_{cu}|$ in the $b \leftrightarrow s$ and $c \leftrightarrow u$ sectors, respectively, as can be seen in Table~\ref{tab:Orderinlambda}. 

Furthermore, the numerical analysis, as displayed in Fig.~\ref{fig:grid_all}, shows that for $t_\beta\lesssim1$, the loop-induced $\Delta M_{B_s}$ constraint dominates, while $\Delta M_D$ is the dominant bound for $t_{\beta} \gtrsim 1$. Taken together, these observables restrict the departure from alignment to approximately $|c_{\alpha-\beta}(t_\beta+t_\beta^{-1})|\lesssim0.038$, permitting values of $|c_{\alpha-\beta}|$ as large as $\sim0.02$ near $t_\beta\sim1$. This maximal misalignment allows $\mB(t \to ch) \sim 10^{-6}$, which under optimistic assumptions could be probed at FCC-hh~\cite{Papaefstathiou:2017xuv}.

\paragraph{Alignment limit:}
For the benchmark scalar mass scale $\Lambda=2\,\rm TeV$, meson-mixing constraints exclude appreciable misalignment in all realisations except the $d$-decoupled model. We therefore turn to the alignment limit, $c_{\alpha-\beta}=0$, where the SM-like Higgs $h$ has no tree-level FCNCs, and derive lower bounds on the common heavy scalar mass scale in each model as a function of $t_{\beta}$. The results are shown in Fig.~\ref{fig:flavour_constraints_all}. 

Neutral meson mixing continues to set the leading constraint in most cases. The relative importance of the various meson mixing constraints can be estimated from the flavour-changing matrices in Table~\ref{tab:Orderinlambda}. When $H$ and $A$ have degenerate masses, the tree-level contributions to both $O_2$ and $O_2'$ exactly cancel, as can be seen from Eq.~\eqref{eq:tree-levelWC} and Eq.~\eqref{eq:Lyuk}. The mixed-chirality combination $|\xi_{ij}^q\xi_{ji}^q|$, which generates $O_4$ at tree-level, therefore remains as the most relevant diagnostic. 

\begin{enumerate}
    \item \textbf{Top decoupled:} This is the most strongly constrained realisation. Decoupling the top quark suppresses up-sector flavour violation, with $|\xi^u_{uc}\,\xi^u_{cu}|\sim10^{-11}$, but leaves sizeable and comparatively symmetric flavour-changing couplings in the down sector ($\xi_{ij} \sim \xi_{ji}$). In particular, $|\xi^d_{ds}\,\xi^d_{sd}|\sim10^{-4}$, so the extreme sensitivity of the kaon system makes $\ve_K$ the dominant constraint. The fact that $\ve_K$, rather than $\Delta M_K$, provides the stronger bound reflects the phase structure of the couplings. Yukawa perturbativity restricts $0.01\lesssim t_\beta\lesssim3.5$, while $\ve_K$ constrains $\Lambda$ to be at least several tens of TeV near $t_\beta\sim\mO(1)$ and drives this constraint towards $\mO(10^3\,\rm TeV)$ at very small $t_\beta$.
  
    \item \textbf{Charm decoupled:} The charm-decoupled model also contains unsuppressed down-sector flavour violation. Although the $b \leftrightarrow s$ flavour violation is particularly large, $|\xi^d_{sb}\,\xi^d_{bs}| \sim 10^{-1}$, the much greater sensitivity of the kaon system to new physics allows the more modest combination $|\xi^d_{ds}\,\xi^d_{sd}| \sim 10^{-4}$ to set the leading constraint in the perturbative Yukawa region, $\tan\beta \gtrsim 0.3$. Masses of $\mO(10\,\rm TeV)$ are permitted near $t_{\beta} \sim 1$, while this lower bound rises to $\mO(100\,\rm TeV)$ for $\tan\beta \sim \mO(10)$.
    
    \item \textbf{Up decoupled:} Decoupling the up quark eliminates tree-level contributions to $D$-mixing. By contrast, the down-sector hierarchy $|\xi^d_{sb}\,\xi^d_{bs}|\sim 10^{-2} \gg|\xi^d_{db}\,\xi^d_{bd}|\sim 10^{-5}>|\xi^d_{ds}\,\xi^d_{sd}| \sim 10^{-6}$ singles out the $b \leftrightarrow s$ transition. Consequently, the dominant bound is from either $\Delta M_{B_s}$ or $b \to s \gamma$ for most of the perturbative Yukawa range, $0.3\lesssim t_\beta\lesssim68$, with scalar masses $\Lambda \sim \textrm{a few TeV}$ viable. Towards the upper end of this $t_\beta$ range, the kaon-mixing bounds become competitive and eventually dominate, pushing the bound on $\Lambda$ to $\mO(100\,\rm TeV)$. 
    
    \item \textbf{Bottom decoupled:} Decoupling the bottom quark removes the neutral-scalar couplings responsible for $B_d$- and $B_s$-meson mixing at tree-level. The leading neutral-meson constraints therefore come from $\Delta M_D$ and $\ve_K$, where the corresponding mixed-chirality combinations are $|\xi^u_{uc}\,\xi^u_{cu}|\sim10^{-7}$ and $|\xi^d_{ds}\,\xi^d_{sd}|\sim10^{-7}$; since the Higgs couplings in the up sector are also proportional to $m_t/v$, compared to $m_b/v$ in the down sector, $\Delta M_D$ provides the dominant bound across most of the $t_\beta$ range. The scalar mass scale $\Lambda$ can be as low as a few-TeV near $t_\beta\sim\mO(1)$, with perturbativity requiring $0.01\lesssim t_\beta\lesssim3.5$.
    
    \item \textbf{Strange decoupled:}  Decoupling the strange quark eliminates tree-level neutral-scalar contributions to both $K$- and $B_s$-meson mixing. The remaining down-sector $b\leftrightarrow d$ couplings satisfy $|\xi^d_{db}\,\xi^d_{bd}|\sim10^{-11}$, greatly suppressing the corresponding mixed-chirality tree-level amplitude. The $\Delta M_{B_d}$ bound visible at low $t_\beta$ therefore arises predominantly from loop-level charged-Higgs effects. For larger $t_\beta$, the up-sector combination $|\xi^u_{uc}\,\xi^u_{cu}|\sim10^{-7}$ makes $\Delta M_D$ the dominant constraint. Yukawa perturbativity allows $0.3\lesssim t_\beta\lesssim42$, with the lower bound on the scalar masses in the multi-TeV range near $t_\beta\sim\mO(1)$, but increasing rapidly for larger $t_\beta$.
    
    \item \textbf{Down decoupled:}
    This is the least constrained realisation. Decoupling the down quark eliminates tree-level neutral scalar contributions to both $K$ and $B_d$ mixing. The remaining neutral-meson mixing processes are also strongly suppressed by their mixed chirality products satisfying $|\xi^d_{sb}\,\xi^d_{bs}|\sim10^{-7}$ and $|\xi^u_{uc}\,\xi^u_{cu}|\sim10^{-10}$. At low $t_\beta$, the strongest constraint is set by $\Delta M_{B_s}$, with $\bar{B}\to X_s\gamma$ providing a similar constraint. At intermediate $t_\beta$, the residual loop-level contribution to $\Delta M_{B_d}$ becomes important and sets the dominant bound. Towards larger $t_\beta$, $\Delta M_D$ becomes competitive before $\Delta M_{B_s}$ again sets the strongest bound near the upper end of the perturbative interval, which is $0.3\lesssim t_\beta\lesssim38$. Together, the absence of tree-level $K$- and $B_d$-mixing contributions and the strong chiral suppression of the remaining FCNC amplitudes explain why this model remains compatible with a heavy-scalar scale in the few-TeV range (for $1\lesssim t_\beta \lesssim 10$), including the benchmark $\Lambda=2\,\rm TeV$ used in Fig.~\ref{fig:grid_all}.
\end{enumerate}

\paragraph{Summary of results:}
The \textsc{Minimal F-DFSZ} models are distinguished by the identity of the decoupled quark $q$, which does not participate in FCNCs. Relative to the general type-III 2HDM, they provide a natural suppression of the contributions to the precision flavour observables discussed in Sec.~\ref{sec:flavourpheno}. Tree-level FCNCs occur in both the up and down sectors, with their magnitudes and chiral structures fixed by quark masses and CKM elements in a highly model-dependent manner. The combined constraints from neutral meson mixing and the decay $b \to s\gamma$ provide a stringent and highly model-dependent test of all six $q$-decoupled realisations.

For the benchmark scalar mass scale $\Lambda=2\,\rm TeV$, misalignment is excluded in the entire allowed $t_{\beta}$ parameter space for all $q$-decoupled models except for the $d$-decoupled model. In this latter case, deviations from alignment remain possible, subject to $|c_{\alpha-\beta}(t_{\beta}+t_{\beta}^{-1})|\lesssim 0.038$ which permits $c_{\alpha-\beta} \sim 0.02$ near $t_{\beta} \sim \mO(1)$. While increasing the heavy scalar masses suppresses the flavour amplitudes, potentially re-opening parameter space away from exact alignment, the scalar sector enforces $c_{\alpha-\beta} \sim v^2/\Lambda^2$ for $\Lambda \gg v$, driving $c_{\alpha-\beta} \to 0$. Thus, for the $u,c,t,s,b$-decoupled models, viable parameter space is driven towards alignment either by flavour constraints when $\Lambda$ is at the TeV scale or by scalar sector decoupling at larger $\Lambda$.

The relative importance of the various constraints in each of the six models is controlled primarily by the identity of the decoupled quark. Decoupling an up-type quark generally leaves the highly sensitive down-sector transitions active; most notably, the sizeable $d\leftrightarrow s$ couplings of the $t$-decoupled model make $\ve_K$ the strongest constraint and require heavy-scalar masses of $\mO(100\,\rm TeV)$ across much of its parameter space, making this realisation the most strongly constrained. By contrast, decoupling a down-type quark removes one or more tree-level transitions in the $K$ and $B$ systems and shifts the leading sensitivity towards up-sector observables such as $D$-mixing. The least constrained realisation is the $d$-decoupled model, in which $K$- and $B_d$-meson mixing are absent at tree-level and the mixed chiral products $|\xi^d_{sb}\,\xi_{bs}^d|$ and $|\xi_{uc}^u\,\xi_{cu}^u|$ in the $b\leftrightarrow s$ and $u\leftrightarrow c$ sectors respectively are suppressed.

Consequently, the $d$-decoupled model admits the lowest heavy-scalar scale, with $\Lambda$ of $\mO(1)$ TeV viable near $t_\beta\sim\mO(1)$. This region may be accessible to direct LHC searches for the charged and neutral Higgs. Existing searches can already reach masses above 1\,TeV~\cite{ATLAS:2021upq,ATLAS:2024dkv,CMS:2025dzq} depending on the couplings and branching ratios. Projections for the HL-LHC suggest that masses as large as $2\text{--}2.5\,\mathrm{TeV}$ may also be probed in some 2HDM scenarios~\cite{Cepeda:2019klc,Bahl:2020kwe}. Future improvements in $D$- and $B_{s,d}$-meson mixing, $\bar{B}\to X_s\gamma$, and the corresponding hadronic inputs will provide complementary sensitivity to the remaining multi-TeV parameter space.

%============================================================
\section{Conclusion}
%============================================================

In this work, we have investigated the \textsc{Minimal F-DFSZ} framework: a set of six phenomenologically distinct flavoured DFSZ axion models, each defined by a single \emph{decoupled quark} $q$ and characterised by a domain-wall number $\mN_{DW}=1$ that admits a post-inflationary axion without the formation of stable domain walls. The fully horizontal $U(1)_{\rm PQ}$ symmetry enforces predictive texture-zero Yukawa matrices whose entries are fixed entirely by the measured quark masses and CKM parameters. As shown in Sec.~\ref{sec:FCNC}, this flavour structure in turn determines both the Higgs- and axion-mediated FCNCs in terms of the same Standard Model observables, with closed-form expressions for the former collected in App.~\ref{app:analyticFCNCXI}. The result is a highly predictive construction, similar in spirit to nearest-neighbour-interaction and BGL-type models, but supplemented by a QCD axion whose couplings are tied directly to the quark flavour structure.

The axion phenomenology, studied in Sec.~\ref{sec:axionpheno}, is dominated by rare flavour-violating meson decays, principally $K^+\to\pi^+a$, which translate into model-dependent upper bounds on $m_a$. In the $u$-, $c$- and $t$-decoupled models these bounds fall below the post-inflationary dark matter window, so the axion over-produces the relic abundance in the standard cosmology. The quark flavour structure therefore selects which realisations admit the axion as the dominant dark matter component.

A detailed study of the flavour phenomenology of all $q$-decoupled realisations was presented in Secs.~\ref{sec:flavourpheno} and~\ref{sec:results}. Experimental constraints from neutral-meson mixing and the radiative decay $B\to X_s\gamma$ place stringent bounds on the remaining free parameters $(\tan\beta,\,\cos(\alpha-\beta),\,m_H,m_A,m_{H^\pm})$. For the degenerate mass benchmark $\Lambda\equiv m_H=m_A=m_{H^\pm}=2\,\rm TeV$, only the $d$-decoupled model retains appreciable freedom to depart from alignment, permitting $c_{\alpha-\beta}$ as large as $\sim0.02$ near $t_\beta\sim\mO(1)$. In the alignment limit $c_{\alpha-\beta}=0$, the lower bound on the common scalar mass, $\Lambda$, ranges from a few TeV in the least-constrained $d$-decoupled model up to $\mO(100\,\rm TeV)$ in the most-constrained $t$-decoupled model. The $d$-decoupled realisation is thus singled out on both counts as it uniquely tolerates misalignment at the benchmark scale and admits new scalars in the few-TeV range, potentially within reach of current colliders.

Several extensions merit further study. The most immediate is to extend the same predictive PQ flavour structure to the lepton sector, along the lines discussed in Sec.~\ref{sec:axion_photon}; a full treatment of the resulting leptonic flavour phenomenology would complete the analysis of this set of predictive textures. Such a leptonic \textsc{Minimal F-DFSZ} model is viable for both Dirac and Majorana neutrinos; the latter can be achieved by the PQ scalar, $S$, spontaneously breaking lepton number. Such a model could yield nontrivial predictions for neutrinoless double-beta decay and leptogenesis.

Taken together, these results establish the \textsc{Minimal F-DFSZ} models as a predictive bridge between the strong CP problem, fermion flavour, axion phenomenology, and extended Higgs-sector searches. The combination of $\mN_{DW}=1$, fully reconstructed Yukawa textures leading to calculable FCNCs, and correlated axion and Higgs signatures renders the framework at once theoretically economical and experimentally falsifiable.

\section*{Acknowledgements}

This work was supported in part by the ARC Centre of Excellence for Dark Matter Particle Physics CE200100008, and in part by ARC Discovery Project DP260104148. P.C. is supported by the Australian Research Council Discovery Early Career Researcher Award DE210100446.

\newpage

\appendix

%============================================================
\section{Input Parameters}
\label{app:params}
%============================================================

%------------------------------------------------------------
\subsection{Meson Mixing Parameters}

\begin{table}[h!]
    \centering
    \renewcommand{\arraystretch}{1.15}
    \resizebox{\textwidth}{!}{%
    \begin{tabular}{||c|c||c|c||c@{\hspace{0.75cm}}||c|c||}
    \hline\hline
    \multicolumn{2}{||c||}{$B_d$} & \multicolumn{2}{c||}{$B_s$} & \multicolumn{1}{c@{\hspace{1cm}}||}{} & \multicolumn{2}{c||}{$K$} \\
    \hline
    $M_{B_d}$ & $5.279\,\mathrm{GeV}$ & $M_{B_s}$ & $5.366\,\mathrm{GeV}$ & & $M_K$ & $0.497\,\mathrm{GeV}$ \\
    $f_{B_d}$ & $0.190\,\mathrm{GeV}$ & $f_{B_s}$ & $0.230\,\mathrm{GeV}$ & & $f_K$ & $0.1557\,\mathrm{GeV}$ \\
    $B_{B_d}^{(1)}$ & $0.837$ & $B_{B_s}^{(1)}$ & $0.826$ & & $B_K^{(1)}$ & $0.507$ \\
    $B_{B_d}^{(2)}$ & $0.808$ & $B_{B_s}^{(2)}$ & $0.824$ & & $B_K^{(2)}$ & $0.46$ \\
    $B_{B_d}^{(3)}$ & $0.761$ & $B_{B_s}^{(3)}$ & $0.830$ & & $B_K^{(3)}$ & $0.79$ \\
    $B_{B_d}^{(4)}$ & $1.054$ & $B_{B_s}^{(4)}$ & $1.040$ & & $B_K^{(4)}$ & $0.78$ \\
    $B_{B_d}^{(5)}$ & $0.986$ & $B_{B_s}^{(5)}$ & $0.964$ & & $B_K^{(5)}$ & $0.49$ \\
    \hline\hline
    \multicolumn{4}{||c||}{$B_q$ evolution matrix $U_{ij}^{B_q}$($m_W,  M_{B_s}$)} & \multicolumn{1}{c@{\hspace{1cm}}||}{} & \multicolumn{2}{c||}{$K$ evolution matrix $U_{ij}^K$ $(m_W,  3\,\mathrm{GeV})$} \\
    \hline
    \multicolumn{4}{||c||}{$\begin{pmatrix}
        0.86 & 0 & 0 & 0 & 0 \\
        0 & 1.43 & -0.20 & 0 & 0 \\
        0 & -0.05 & 0.68 & 0 & 0 \\
        0 & 0 & 0 & 1.82 & 0.30 \\
        0 & 0 & 0 & 0 & 0.93
    \end{pmatrix}$}
    & &
    \multicolumn{2}{c||}{$\begin{pmatrix}
        0.82 & 0 & 0 & 0 & 0 \\
        0 & 1.60 & -0.27 & 0 & 0 \\
        0 & -0.07 & 0.60 & 0 & 0 \\
        0 & 0 & 0 & 2.21 & 0.43 \\
        0 & 0 & 0 & 0 & 0.91
    \end{pmatrix}$} \\
    \hline\hline
    \end{tabular}}
    \caption{Input parameters for the $B_d$, $B_s$, and $K^0$ systems, together with the RG evolution matrices for $B$- and $K$-meson mixing. The RG evolution matrices were computed using the package \textsc{Wilson}~\cite{Aebischer:2018bkb}.}
    \label{tab:mesonmixinginputs1}
\end{table}

%============================================================
\section{Scalar Sector: Constraints and Alignment}
\label{app:scalarpotential}
%============================================================

The scalar potential of the underlying 2HDM in the \textsc{Minimal F-DFSZ} (in the PQ-broken phase) is
\begin{align}
    V(\Phi_1,\Phi_2)=m_{11}^2\Phi_1^{\dagger}\Phi_1+m_{22}^2\Phi_2^{\dagger}\Phi_2-(m_{12}^2\,\Phi_1^{\dagger}\Phi_2 + \textrm{h.c.}) 
    +\frac{\lambda_1}{2}(\Phi_1^\dagger\Phi_1)^2+\frac{\lambda_2}{2}(\Phi_2^{\dagger}\Phi_2)^2 \nonumber\\
    +\lambda_3(\Phi_1^{\dagger}\Phi_1)(\Phi_2^{\dagger}\Phi_2)+\lambda_4(\Phi_1^{\dagger}\Phi_2)(\Phi_2^{\dagger}\Phi_1),
\label{eq:scalarpotential}
\end{align}
with $\langle\Phi_a\rangle=(0,\,v_a)^T/\sqrt{2}$ and $ v=\sqrt{v_1^2+v_2^2}\simeq246\,$GeV. The horizontal PQ symmetry forbids the quartic terms $(\Phi_1^\dagger\Phi_2)^2$ and $(\Phi_a^\dagger\Phi_a)(\Phi_1^\dagger\Phi_2)$, while the bilinear $m_{12}^2$ is generated by the PQ-breaking singlet VEV. Since $m_{12}^2$ can always be made real by a field redefinition the vacuum is CP-conserving. 

In the alignment limit, the parameters of the scalar potential can be expressed in terms of the physical scalar masses according to~\cite{Branco:2011iw}
\begin{align}
    m_{12}^2 &= m_A^2\,s_\beta c_\beta, \qquad 
    \lambda_1=\dfrac{m_h^2+(m_H^2-m_A^2)\,t^2_\beta}{v^2}, \qquad 
    \lambda_2= \dfrac{m_h^2+(m_H^2-m_A^2)\,t_\beta^{-2}}{v^2}, \nonumber \\
    \lambda_3&=\dfrac{2m_{H^\pm}^2+m_h^2-m_H^2-m_A^2}{v^2}, \qquad
    \lambda_4=\dfrac{2(m_A^2-m_{H^\pm}^2)}{v^2}\,.
\label{eq:lambdainversion}
\end{align}

%------------------------------------------------------------
\subsection{Theoretical Constraints}
\label{app:theoryconstraints}

Two standard requirements restrict the quartic couplings. First, vacuum stability (boundedness from below) demands~\cite{Deshpande:1977rw,Branco:2011iw}
\begin{align}
    \lambda_1 >0, \qquad \lambda_2>0, \qquad 
    \lambda_3+\sqrt{\lambda_1\lambda_2}>0, \qquad 
    \lambda_3 + \lambda_4 + \sqrt{\lambda_1 \lambda_2} > 0\,,
\label{eq:BFB}
\end{align}
supplemented by~\cite{Barroso:2012mj,Ivanov:2015nea}
\begin{align}
    \dfrac{2m_{12}^2}{v^2 c_{\beta}s_{\beta}} > \lambda_3 + \lambda_4 - \sqrt{\lambda_1 \lambda_2}\,.
\end{align}
Second, tree-level unitarity of the $2\to2$ scalar scattering matrix bounds the eigenvalues of the partial-wave amplitudes~\cite{Lee:1977eg,Lee:1977yc,Branco:2011iw}:
\begin{align}
    |\lambda_1|,  \, |\lambda_2|, \, |\lambda_3|, \, |\lambda_3+\lambda_4|, \, |\lambda_3-\lambda_4|, \, |\lambda_3 + 2 \lambda_4 | &\leq 8\pi, \nonumber\\
    \bigg|\dfrac{3}{2}(\lambda_1+\lambda_2) \pm \sqrt{\dfrac{9}{4}(\lambda_1-\lambda_2)^2+(2\lambda_3+\lambda_4)^2}\bigg| &\leq 8\pi, \nonumber \\
    \bigg|\dfrac{1}{2}(\lambda_1+\lambda_2) \pm \dfrac{1}{2}\sqrt{(\lambda_1-\lambda_2)^2+4\lambda_4^2}\bigg| &\leq 8\pi, \nonumber \\
    |(\mY_{q}^{1,2})_{ij}| &\leq \sqrt{4\pi}.
\label{eq:PertUnit}
\end{align}
A key structural feature of Eq.~\eqref{eq:lambdainversion} is that (in the alignment limit) the quartic couplings depend on $\{m_H,m_A,m_{H^\pm}\}$ only via mass-squared differences. For an exactly degenerate spectrum, $m_H = m_A = m_{H^\pm}$, one finds
\begin{equation}
    \lambda_1=\lambda_2=\lambda_3=\frac{m_h^2}{v^2}\simeq0.26\,,
    \qquad \lambda_4 = 0\,.
    \label{eq:degeneratequartics}
\end{equation}
Both Eq.~\eqref{eq:BFB} and Eq.~\eqref{eq:PertUnit} are then satisfied trivially at any scale $\Lambda=m_H=m_A=m_{H^\pm}$.

%------------------------------------------------------------
\subsection{Alignment with and without Decoupling}
\label{app:alignment}

Here, we discuss how alignment can be achieved in the 2HDM. Rotating the doublets by the angle $\beta$ into the Higgs basis, in which only one doublet has a VEV, the potential takes the general form given in~\cite{Gunion:2002zf,Branco:2011iw} with quartic couplings $\Lambda_{1\text{--}7}$. The mixing angle for the CP-even scalars $\alpha-\beta$ then satisfies the relations~\cite{Gunion:2002zf}
\begin{equation}
    \label{eq:sin2alphabeta}
    s_{2(\alpha-\beta)}=\frac{2\Lambda_6 v^2}{m_H^2-m_h^2},
    \qquad 
    c_{2(\alpha-\beta)}=\frac{(\Lambda_1-\Lambda_5)v^2-m_A^2}{m_H^2-m_h^2}\,,
\end{equation}
\begin{equation}
    \label{eq:cos2alphabeta}
    c^2_{\alpha-\beta}=\frac{\Lambda_1 v^2 -m_h^2}{m_H^2-m_h^2} 
    = \frac{\Lambda^2_6\, v^4}{(m_H^2-m_h^2)(m_H^2-\Lambda_1 v^2)} \,,
\end{equation}
which can be rearranged to give
\begin{align}
    \label{eq:massrelations1}
    \Lambda_1 v^2 & = m_h^2\, s_{\alpha-\beta}^2+m_H^2\, c_{\alpha-\beta}^2 \, , \\
    \label{eq:massrelations2}
    \Lambda_6 v^2 & = ( m_H^2 - m_h^2 )\, s_{\alpha-\beta}\, c_{\alpha-\beta} \, , \\
    \label{eq:massrelations3}
    \Lambda_5 v^2 & = m_H^2\, s_{\alpha-\beta}^2+m_h^2\, c_{\alpha-\beta}^2 - m_A^2\,.
\end{align}

Alignment, $c_{\alpha-\beta}\to0$, can be reached in two distinct ways:
\begin{itemize}
    \item \textbf{Alignment with decoupling:} Taking $m_H^2 \gg v^2$, Eq.~\eqref{eq:cos2alphabeta} enforces $c_{\alpha-\beta}\simeq \Lambda_6 v^2/m_H^2 \to 0$~\cite{Gunion:2002zf,Bernon:2015qea}. Equation~\eqref{eq:massrelations3} then implies $m_A^2 \simeq m_H^2$. A similar relation holds for $m_{H^\pm}$, see Ref.~\cite{Gunion:2002zf}. Therefore, the heavy spectrum becomes \emph{automatically degenerate}, with splittings of $\mathcal{O}(v^2/m_H^2)$.
    
    \item \textbf{Alignment without decoupling:} Setting $\Lambda_6\to0$ directly leads to alignment~\cite{Bernon:2015qea,Carena:2013ooa} without requiring a hierarchy between the additional Higgs masses and $v$ and leaves the $m_H,\,m_A,\,m_{H^\pm}$ spectrum a priori non-degenerate. In our numerical analyses, we nevertheless impose degeneracy in the non-decoupling regime as well, since it immediately ensures that the stability, perturbativity and electroweak-precision constraints are satisfied, as shown in Eq.~\eqref{eq:degeneratequartics} and in App.~\ref{app:EWPO} below.
\end{itemize}

%------------------------------------------------------------
\subsection{Misalignment Window}
\label{app:benchmark}

A degenerate spectrum does not enforce $\Lambda_6\to0$, so deviations from alignment remain possible in the non-decoupling regime. However, perturbativity of the quartic couplings restricts the maximum misalignment that can be achieved for a given heavy mass scale $\Lambda \equiv m_H=m_A=m_{H^\pm}$. This can be seen from Eqs.~\eqref{eq:massrelations1}~--~\eqref{eq:massrelations3}. Expanding to leading order around $c_{\alpha-\beta}=0$, Eq.~\eqref{eq:massrelations2} becomes
\begin{equation}
    \Lambda_6 \simeq c_{\alpha-\beta}\,\frac{\Lambda^2-m_h^2}{v^2}\,,
    \label{eq:Lambda6misalignment}
\end{equation}
showing that for fixed $c_{\alpha-\beta}$, $\Lambda_6$ grows as $\Lambda^2/v^2$. Perturbativity, $|\Lambda_6|\leq 4\pi$, therefore caps the attainable misalignment at
\begin{equation}
    |c_{\alpha-\beta}^{\max}| \simeq
    \frac{4\pi\, v^2}{\Lambda^2 - m_h^2}\,.
    \label{eq:misalignmentcap}
\end{equation}
We adopt $\Lambda=2\,$TeV as the benchmark for Fig.~\ref{fig:grid_all} in Sec.~\ref{sec:results}. Equation~\eqref{eq:misalignmentcap} then gives $|c_{\alpha-\beta}^{\max}|\simeq0.2$.

%------------------------------------------------------------
\subsection{Electroweak Precision Observables}
\label{app:EWPO}
The additional scalars in the 2HDM contribute to the electroweak oblique parameters, $S,T$ and $U$ which are constrained at $95\%$~CL to~\cite{ParticleDataGroup:2024cfk}
\begin{equation}
    S = -0.04 \pm 0.10 \,, \qquad
    T = 0.01 \pm 0.12 \,, \qquad
    U = -0.01 \pm 0.09 \,.
    \label{eqn:obliqueparams}
\end{equation}

We evaluate the charged and neutral Higgs contributions using the general multi-doublet expressions of Refs.~\cite{Grimus:2007if,Grimus:2008nb}. In the alignment limit they reduce to~\cite{Grimus:2007if,Haber:2010bw,Branco:2011iw}
\begin{align}
    T &=
    \frac{1}{16\pi s_W^2 m_W^2}
    \Bigl[
        F(m_{H^\pm}^2,m_A^2)
        +F(m_{H^\pm}^2,m_H^2)
        -F(m_A^2,m_H^2)
    \Bigr] \,,
    \label{eq:Talignment} \\
    S &= \frac{1}{\pi m_Z^2}
    \Bigl[
        \bar{B}_{22}(m_Z^2;m_H^2,m_A^2)
        -\bar{B}_{22}(m_Z^2;m_{H^\pm}^2,m_{H^\pm}^2)
    \Bigr] \,,
    \label{eq:Salignment} \\
    S+U &= \frac{1}{\pi m_W^2}
    \Bigl[
        \bar{B}_{22}(m_W^2;m_{H^\pm}^2,m_H^2)
        +\bar{B}_{22}(m_W^2;m_{H^\pm}^2,m_A^2)
        -2\,\bar{B}_{22}(m_W^2;m_{H^\pm}^2,m_{H^\pm}^2)
    \Bigr] \,,
    \label{eq:SUalignment}
\end{align}
with $\bar{B}_{22}(q^2;x,y)\equiv B_{22}(q^2;x,y)-B_{22}(0;x,y)$ defined in~\cite{Grimus:2008nb,Haber:2010bw} and
$F(x,y) \equiv \frac{x+y}{2}-\frac{xy}{x-y}\ln\frac{x}{y}$. 

For an exactly degenerate mass spectrum, $m_H=m_A=m_{H^\pm}$, the contributions to $S$, $T$ and $U$ vanish identically in the alignment limit. Misalignment corrections enter at $\mathcal{O}(c^2_{\alpha-\beta})$ and remain well within Eq.~\eqref{eqn:obliqueparams} for the benchmark of App.~\ref{app:benchmark}, as verified numerically using the full expressions in Refs.~\cite{Grimus:2007if,Grimus:2008nb}.

%============================================================
\section{Mass Matrix Fits}
\label{app:MassMatrixFits}
%============================================================

\begin{table}[t!]
\centering
\begin{minipage}{\textwidth}
\caption{\label{tab:quarkparams_mz}
Quark masses in the $\overline{\textrm{MS}}$ scheme at $\mu=m_Z$~\cite{Huang:2020hdv} and the CKM mixing parameters.}
\centering
\begin{tabular}{lclc}
\hline
Parameter  & $\mu_i \, \pm \, 1\sigma_i$ & Input  & $\mu_i \, \pm \, 1\sigma_i$ \\ \hline
$m_{u}/\text{MeV}$ & $1.23 \pm 0.32$ & $\theta_{12}^{q}$ & $13.04^{\circ} \pm 0.05^{\circ}$ \\
$m_{d}/\text{MeV}$ & $2.67 \pm 0.38$ & $\theta_{13}^{q}$ & $0.224^{\circ} \pm 0.014^{\circ}$ \\
$m_{s}/\text{MeV}$ & $53.16 \pm 6.33$ & $\theta_{23}^{q}$ & $2.35^{\circ} \pm 0.05^{\circ}$ \\
$m_{c}/\text{GeV}$ & $0.620 \pm 0.021$ & $\delta_{\text{CP}}^{q}$ & $68.9^{\circ} \pm 3.5^{\circ}$ \\
$m_{b}/\text{GeV}$ & $2.839 \pm 0.019$ & & \\
$m_{t}/\text{GeV}$ & $168.26 \pm 0.75$ & & \\
\hline
\end{tabular}
\end{minipage}
\end{table}

In this appendix, we provide the numerical mass matrices for each $q$-decoupled model, which are used in our analysis in Sec.~\ref{sec:results}. These are obtained using the solutions for the mass matrix parameters in Sec.~\ref{sec:General_Solution} (without any expansion in the quark mass ratios). The numerical inputs are the quark masses in the $\overline{\textrm{MS}}$ scheme at $\mu=m_Z$~\cite{Huang:2020hdv} and the CKM parameters (which do not run appreciably below $\mu\sim m_Z$), which are collected in Table~\ref{tab:quarkparams_mz}. The mass matrices for all six $q$-decoupled models are given in Table~\ref{tab:MuMdNum}.\footnote{We note that the recent work in Ref.~\cite{Arkani-Hamed:2026uos} scanned all full-rank quark mass matrices with nine total non-zero entries (``9-link textures''), of which our mass matrices are examples.}
\newpage
\begin{table}[h!]
\centering
\begin{minipage}{\textwidth}
\caption{\label{tab:MuMdNum} Reconstructed mass matrices $\mM_u^{(q)}$ and $\mM_d^{(q)}$ from the CKM fits for all $q$-decoupled models.}
\centering
\renewcommand{\arraystretch}{1.2}
\setlength{\tabcolsep}{10pt}
\begin{tabular}{lclc}
\hline
\textbf{$\mM_u^{(q)}$} & \textbf{Numerical Fit} $(\rm GeV)$ &
\textbf{$\mM_d^{(q)}$} & \textbf{Numerical Fit} $(\rm GeV)$\\
\hline

% --- u-decoupled ---
$\mM_u^{(u)}$ &
$\displaystyle
\begin{pNiceMatrix}
168.02 & 0 & 0 \\[2mm]
9.023\,e^{\mathrm i 0.79} & 0.621 & 0 \\[2mm]
0 & 0 & 0.00123
\end{pNiceMatrix}
$ &
$\mM_d^{(u)}$ &
$\displaystyle
\begin{pNiceMatrix}
1.886 & 2.121 & 0 \\[2mm]
0 & 0.078 & 0 \\[2mm]
0.01596 & 0 & 0.00274
\end{pNiceMatrix}
$
\\[10pt]

% --- c-decoupled ---
$\mM_u^{(c)}$ &
$\displaystyle
\begin{pNiceMatrix}
168.26 & 0 & 0 \\[2mm]
0.806\,e^{-\mathrm i 0.790} & 0.00123 & 0 \\[2mm]
0 & 0 & 0.620
\end{pNiceMatrix}
$ &
$\mM_d^{(c)}$ &
$\displaystyle
\begin{pNiceMatrix}
2.610 & 1.111 & 0 \\[2mm]
0 & 0.01331 & 0 \\[2mm]
0.1290 & 0 & 0.0116
\end{pNiceMatrix}
$
\\[10pt]

% --- t-decoupled ---
$\mM_u^{(t)}$ &
$\displaystyle
\begin{pNiceMatrix}
0.6175 & 0 & 0 \\[2mm]
0.0551\,e^{\mathrm i 1.616} & 0.00123 & 0 \\[2mm]
0 & 0 & 168.26
\end{pNiceMatrix}
$ &
$\mM_d^{(t)}$ &
$\displaystyle
\begin{pNiceMatrix}
0.1199 & 0.05051 & 0 \\[2mm]
0 & 0.01118 & 0 \\[2mm]
2.821 & 0 & 0.3008
\end{pNiceMatrix}
$
\\[10pt]

% --- d-decoupled ---
$\mM_u^{(d)}$ &
$\displaystyle
\begin{pNiceMatrix}
167.47 & 16.17 & 0 \\[2mm]
0 & 0.6069 & 0 \\[2mm]
1.451 & 0 & 0.001262
\end{pNiceMatrix}
$ &
$\mM_d^{(d)}$ &
$\displaystyle
\begin{pNiceMatrix}
2.837 & 0 & 0 \\[2mm]
0.118\,e^{-\mathrm i 0.412} & 0.05321 & 0 \\[2mm]
0 & 0 & 0.002670
\end{pNiceMatrix}
$
\\[10pt]

% --- s-decoupled ---
$\mM_u^{(s)}$ &
$\displaystyle
\begin{pNiceMatrix}
167.48 & 14.64 & 0 \\[2mm]
0 & 0.1402 & 0 \\[2mm]
6.943 & 0 & 0.00547
\end{pNiceMatrix}
$ &
$\mM_d^{(s)}$ &
$\displaystyle
\begin{pNiceMatrix}
2.839 & 0 & 0 \\[2mm]
0.0246\,e^{\mathrm i 0.412} & 0.002670 & 0 \\[2mm]
0 & 0 & 0.05316
\end{pNiceMatrix}
$
\\[10pt]

% --- b-decoupled ---
$\mM_u^{(b)}$ &
$\displaystyle
\begin{pNiceMatrix}
7.066 & 0.6179 & 0 \\[2mm]
0 & 0.05513 & 0 \\[2mm]
168.11 & 0 & 0.3294
\end{pNiceMatrix}
$ &
$\mM_d^{(b)}$ &
$\displaystyle
\begin{pNiceMatrix}
0.05204 & 0 & 0 \\[2mm]
0.0108\,e^{-\mathrm i 1.617} & 0.002727 & 0 \\[2mm]
0 & 0 & 2.839
\end{pNiceMatrix}
$
\\[10pt]

\hline
\end{tabular}
\end{minipage}
\end{table}

%============================================================
\section{Approximate Expressions for the FCNC Couplings}
\label{app:analyticFCNCXI}
%============================================================

The flavour-changing couplings $\xi_{ij}^u$, $\xi_{ij}^d$ are defined in Eq.~\eqref{eq:EpsilonCouplings}. Here, we give approximate analytic expressions for these couplings, obtained using the left- and right-handed rotation matrices in Eqs.~\eqref{eq:O1} and \eqref{eq:O2} and the reconstructed mass matrices discussed in Sec.~\ref{sec:General_Solution}. We expand the expressions for each coupling to leading non-zero order in the Wolfenstein parameter $\lambda$, using the quark mass ratio scaling in Eq.~\eqref{eq:Wolfeps}. (Note that in some cases we have chosen to retain sub-leading terms by not expanding the denominators.) The resulting expressions for each model are:
\newpage

\begin{minipage}{\linewidth}
\vspace{-0.5cm}\underline{\textit{Top decoupled:}}

\[
\xi_{ij}^u
\simeq
\begin{pmatrix}
\ve_{ut}\,\dfrac{|V_{ub}|^2}{1- |V_{tb}|^2}
&
-\,\ve_{ct}\,\dfrac{|V_{ub}| |V_{cb}|}{1- |V_{tb}|^2}
&
0
\\[10pt]
-\,\ve_{ut}\,\dfrac{|V_{ub}|^3}{|V_{cb}|(1 - |V_{tb}|^2)}
&
\ve_{ct}\,\dfrac{|V_{ub}|^2}{1 - |V_{tb}|^2}
&
0
\\[10pt]
0 & 0 & 1
\end{pmatrix}
\]

\[
\xi_{ij}^d
\simeq
\begin{pmatrix}
3\, \ve_{db}
&
-\ve_{sb}\dfrac{ |V_{td}||V_{ts}|}{1-|V_{tb}|^2}
&
 \, |V_{td}||V_{tb}|
\\[12pt]
2\, \ve_{db}\, \dfrac{|V_{ts}|}{|V_{td}|}
&
\ve_{sb}\left[
\dfrac{|V_{td}|^2}{1-|V_{tb}|^2}
+
\dfrac{\bigl(1+|V_{tb}|^2\bigr)\bigl(1-|V_{tb}|^2\bigr)}{|V_{td}|^2}\,\ve_{ds}^2
\right]
&
\, |V_{ts}||V_{tb}|
\\[12pt]
-\ve_{db}\,\dfrac{ |V_{td}|}{|V_{td}|^2+|V_{ts}|^2\ve_{ds}^2}
&
-\ve_{sb}\,\dfrac{\,|V_{ts}|\,\ve_{ds}^2}{|V_{td}|^2+|V_{ts}|^2\ve_{ds}^2}
&
|V_{ts}|^2+\dfrac{\ve_{db}^2}{|V_{td}|^2}
\end{pmatrix} 
\]

\vspace{0.3cm}\underline{\textit{Charm decoupled:}}

\[
\xi_{ij}^u
\simeq
\begin{pmatrix}
-\,\ve_{ut}\,\dfrac{|V_{ub}|^2}{1- |V_{cb}|^{2}}
&
0
&
-\,\dfrac{|V_{ub}||V_{tb}|}{1-|V_{cb}|^{2}}
\\[12pt]
0 & \ve_{ct} & 0
\\[12pt]
\ve_{ut}\,\dfrac{|V_{ub}|^{3}}{|V_{tb}|(1-|V_{cb}|^{2})}
&
0
&
\dfrac{|V_{ub}|^2}{1 - |V_{cb}|^{2}}
\end{pmatrix}
\]
\[
\xi_{ij}^d
\simeq
\begin{pmatrix}
3 \ve_{db}  
&
-2 \ve_{sb} |V_{cd}| |V_{cs}|
&
\dfrac{|V_{cd}| |V_{cb}|^{3}}{|V_{cb}|^{2}+|V_{cs}|^{2}\ve_{sb}^{2}}
\\[12pt]
\dfrac{2 \ve_{db}  }{|V_{cd}|}(1-|V_{cd}|^2)^{3/2}
&
- \ve_{sb}\left(1-\dfrac{2\,\ve_{sb}^{2}}{|V_{cb}|^{2}+|V_{cs}|^{2}\ve_{sb}^{2}}\right)
&
\dfrac{|V_{cs}| |V_{cb}|^{3}}{|V_{cb}|^{2}+|V_{cs}|^{2}\ve_{sb}^{2}}
\\[12pt]
\ve_{db}\dfrac{|V_{cb}|}{|V_{cd}|\left(|V_{cb}|^{2}+|V_{cs}|^{2}\ve_{sb}^{2}\right)}
&
- \ve_{sb}\dfrac{|V_{cs}||V_{cb}|}{|V_{cb}|^{2}+|V_{cs}|^{2}\ve_{sb}^{2}}
&
\dfrac{|V_{cb}|^{2}}{|V_{cb}|^{2}+|V_{cs}|^{2}\ve_{sb}^{2}}
\end{pmatrix} 
\]

\vspace{0.3cm}\underline{\textit{Up decoupled:}}

\[
\xi_{ij}^u \simeq
\begin{pmatrix}
\ve_{ut} & 0 & 0
\\[8pt]
0 &
-\,\ve_{ct}\,\dfrac{|V_{cb}|^2}{1 - |V_{ub}|^2}
&
-\dfrac{|V_{cb}||V_{tb}|}{1 - |V_{ub}|^2}
\\[10pt]
0 &
\ve_{ct}\,
\dfrac{|V_{cb}|^3}{|V_{tb}|(1 - |V_{ub}|^2)}
&
\dfrac{|V_{cb}|^2}{1 - |V_{ub}|^2}
\end{pmatrix}
\]

\[
\xi_{ij}^d \simeq
\begin{pmatrix}
\ve_{db}
&
-2\,\ve_{sb}|V_{us}||V_{ud}|
&
\dfrac{|V_{ub}|^{2}-|V_{us}|^{2}\ve_{sb}^{2}}{|V_{ub}|}
\\[1.2em]
-\ve_{db}\dfrac{|V_{us}|^{3}\,\ve_{sb}^{2}}{|V_{ub}|^{2}+|V_{us}|^{2}\ve_{sb}^{2}}
&
\ve_{sb}\,\dfrac{|V_{ub}|^{2}+2|V_{us}|^{2}\ve_{sb}^{2}}{|V_{ub}|^{2}+|V_{us}|^{2}\ve_{sb}^{2}}
&
-\ve_{sb}\dfrac{|V_{us}|^{3}\,\ve_{sb}^{3}}{|V_{ub}|\left(|V_{ub}|^{2}+|V_{us}|^{2}\ve_{sb}^{2}\right)}
\\[1.2em]
\ve_{db}\dfrac{|V_{us}|^{2}|V_{ub}||V_{ud}|}{|V_{ub}|^{2}+|V_{us}|^{2}\ve_{sb}^{2}}
&
\dfrac{|V_{us}|\,\ve_{sb}\,|V_{ub}|}{|V_{ub}|^{2}+|V_{us}|^{2}\ve_{sb}^{2}}
&
\dfrac{|V_{ub}|^{2}}{|V_{ub}|^{2}+|V_{us}|^{2}\ve_{sb}^{2}}
\end{pmatrix}
\]
\end{minipage}

\underline{\textit{Bottom decoupled:}}

\[
\xi_{ij}^u
\simeq
\begin{pmatrix}
-2\, \ve_{ut}
&
\ve_{ct}\dfrac{\,|V_{ub}||V_{cb}|}{1-|V_{tb}|^2}
&
- \,\dfrac{|V_{ub}||V_{cb}|}{\sqrt{1-|V_{tb}|^2}}
\\[12pt]
-2\, \ve_{ut}\, \dfrac{|V_{cb}|}{|V_{ub}|}
&
\ve_{ct} 
&
-\dfrac{|V_{cb}|^2}{\sqrt{1-|V_{tb}|^2}}
\\[12pt]
\ve_{ut}\dfrac{ \sqrt{1-|V_{tb}|^2}}{|V_{ub}||V_{cb}|}
&
\ve_{ut}\,\dfrac{|V_{cb}|\ve_{uc}}{|V_{ub}|^2}\,
&
1
\end{pmatrix}
\]

\[
\xi_{ij}^d
\simeq
\begin{pmatrix}
\ve_{db}
&
\ve_{sb}\,\dfrac{|V_{ts}|\,|V_{td}|}{1 - |V_{tb}|^{2}}
&
0
\\[12pt]
-\,\ve_{db}\,\dfrac{|V_{td}|^{3}}{|V_{ts}|\,\bigl(1 - |V_{tb}|^{2}\bigr)}
&
\ve_{sb}\,\dfrac{|V_{ts}|^{2}}{1 - |V_{tb}|^{2}}
&
0
\\[12pt]
0 & 0 & 0
\end{pmatrix}
\]

\underline{\textit{Strange decoupled:}}

\[
\xi_{ij}^u
\simeq
\begin{pmatrix}
-2\, \ve_{ut}\,|V_{cs}|^{2}
&
2\ve_{ct} |V_{us}|\,|V_{cs}|
&
-|V_{ts}|\,|V_{us}|
\\[12pt]
-\dfrac{2\ve_{ut}}{|V_{us}|}
|V_{cs}|
&
-2\,\ve_{ct}\,|V_{cs}|^{2}
&
-|V_{ts}||V_{cs}|
\\[12pt]
-\ve_{ut}\dfrac{1}{|V_{ts}|\,|V_{us}|}
&
\ve_{ct}\,\dfrac{|V_{cs}|}{|V_{ts}|}
&
|V_{ts}|^2+\dfrac{\ve_{ct}^2}{|V_{ts}|^2}
\end{pmatrix}
\]

\[
\xi_{ij}^d
\simeq
\begin{pmatrix}
\displaystyle
\ve_{db}
&
0
&
\displaystyle
\frac{\,|V_{tb}|\,|V_{td}|}{1 - |V_{ts}|^2}
\\[1.2em]
0 & 0 & 0
\\[1.2em]
\displaystyle
-\,\ve_{db}\frac{\,|V_{td}|^3}
{|V_{tb}|\,\bigl(1 - |V_{ts}|^2\bigr)}
&
0
&
\displaystyle
\frac{\,|V_{tb}|^2}{1 - |V_{ts}|^2}
\end{pmatrix}
\]

\underline{\textit{Down decoupled:}}

\[
\xi_{ij}^u \simeq
\begin{pmatrix}
-2\,\ve_{ut}\,|V_{cd}|^2
&
2\,\ve_{ct} |V_{cd}||V_{ud}|
&
-|V_{td}|\,|V_{ud}|
\\[12pt]
-2\,\ve_{ut}\,|V_{cd}|^3
&
2\,\ve_{ct}\,|V_{cd}|^2
&
-|V_{td}||V_{cd}|
\\[12pt]
-\ve_{ut}\dfrac{ |V_{cd}|^2}{|V_{td}||V_{ud}|}
&
\ve_{ct}\dfrac{ |V_{cd}|}
{|V_{td}|}
&
\ve_{ct}^2\dfrac{|V_{cd}|^2}{|V_{td}|^2}
\end{pmatrix}
\]

\[
\xi_{ij}^d
\simeq
\begin{pmatrix}
0 & 0 & 0 \\[1.0em]
0 &
\displaystyle
\ve_{sb}
&
\displaystyle
\frac{\,|V_{tb}|\,|V_{ts}|}{1 - |V_{td}|^{2}}
\\[1.4em]
0 &
\displaystyle
-\,\ve_{sb}\frac{\,|V_{ts}|^{3}}
{|V_{tb}|\,\bigl(1- |V_{td}|^{2}\bigr)}
&
\displaystyle
\frac{\,|V_{tb}|^2}{1 - |V_{td}|^{2}}
\end{pmatrix}.
\]

%============================================================
\section{$\Delta F=2$ Wilson Coefficients}
\label{app:PVfuncs}
%============================================================

In this Appendix, we provide the expressions for the relevant one-loop neutral and charged scalar contributions to the Wilson coefficients for the $\Delta F=2$ operators. As discussed in Sec.~\ref{sec:deltaf}, we neglect one-loop contributions proportional to $\xi_{ij}^d$. The general expressions for a type-III 2HDM have been calculated (in general $R_{\xi}$ gauge) in Ref.~\cite{Crivellin:2013wna} and are expressed in terms of the standard loop functions:
\begin{align}
B_{0} \left( m^{2}_{1},m^{2}_{2} \right) &= 1 +
  \dfrac{1}{\varepsilon} - \gamma_{E}+\ln \left(4\pi\right) 
  + \dfrac{m^{2}_{1} \ln\left( \dfrac{\mu^{2}}{m^{2}_{1}}\right) - m^{2}_{2}
  \ln\left( \dfrac{\mu^{2}}{m^{2}_{2}}\right)  }{ m^{2}_{1} - m^{2}_{2} } + O\left( \varepsilon \right)  \,, \\
C_{0} \left(  m^{2}_{1},m^{2}_{2} ,m^{2}_{3} \right) &= 
  \dfrac{B_{0} \left(  m^{2}_{1},m^{2}_{2} \right) - B_{0} \left(
  m^{2}_{1},m^{2}_{3} \right)  }{ m^{2}_{2} - m^{2}_{3}  } \notag \\
&= \dfrac{  m^{2}_{1} m^{2}_{2} \ln\left(
  \dfrac{m^{2}_{1}}{m^{2}_{2}}\right)   + m^{2}_{3} m^{2}_{2}
  \ln\left( \dfrac{m^{2}_{2}}{m^{2}_{3}}\right)  +  m^{2}_{3}
  m^{2}_{1} \ln\left( \dfrac{m^{2}_{3}}{m^{2}_{1}}\right)   }{  \left(
  m^{2}_{1} - m^{2}_{2} \right) \,  \left( m^{2}_{3} - m^{2}_{1}
  \right) \,  \left( m^{2}_{2} - m^{2}_{3} \right)      }   \, , \\ \notag \\
D_{0} \left(  m^{2}_{1},m^{2}_{2} ,m^{2}_{3},m^{2}_{4} \right) &= 
  \dfrac{C_{0} \left(  m^{2}_{1},m^{2}_{2},m^{2}_{3} \right) - C_{0}
  \left(  m^{2}_{1},m^{2}_{2},m^{2}_{4} \right)  }{ m^{2}_{3} -
  m^{2}_{4}  } \,.
\end{align}
It is also useful to define the functions $C_2$ and $D_2$, which appear in certain Wilson coefficients:
\begin{eqnarray}
&& C_{2} \left(  m^{2}_{1},m^{2}_{2} ,m^{2}_{3} \right) =
B_{0} \left(  m^{2}_{1},m^{2}_{2} \right) + m^{2}_{3} \, C_{0}
\left(  m^{2}_{1},m^{2}_{2} ,m^{2}_{3}\right)  \, , \nonumber \\ [0.25cm]
&& D_{2} \left(  m^{2}_{1},m^{2}_{2} ,m^{2}_{3},m^{2}_{4} \right) =
C_{0} \left(  m^{2}_{1},m^{2}_{2},m^{2}_{3} \right) + m^{2}_{4} \, D_{0}
\left(  m^{2}_{1},m^{2}_{2} ,m^{2}_{3},m^{2}_{4} \right)  \, .
\end{eqnarray}
Note that while $C_2$ inherits a divergence from $B_0$, it only enters the Wilson coefficients through the difference of two $C_2$ functions in Eq.~\eqref{eq:chargedHiggs_mixed} such that the pole cancels.

\paragraph{Neutral Higgs Box:}
First, the neutral Higgs box diagrams (e.g. Fig.~\ref{fig:neutral-scalar-box}), which we only include for $D$-mixing, generate the following contributions to the Wilson coefficients
\begin{align}
\mC_1(\mu_W) &=
-\frac{1}{128\pi^2}
\sum_{i,j=1}^{3}\sum_{X^0_a,X^0_b}
\bigl[
C_{ci}^{X^0_a*}\,C_{ui}^{X^0_b}\,
C_{cj}^{X^0_b*}\,C_{uj}^{X^0_a}
\, D_2(m_{u_i}^2, m_{u_j}^2, m_{X^0_a}^2, m_{X^0_b}^2)
\bigr],
\nonumber\\[0.5em]
\mC_1'(\mu_W) &=
-\frac{1}{128\pi^2}
\sum_{i,j=1}^{3}\sum_{X^0_a,X^0_b}
\bigl[
C_{ic}^{X^0_a}\,C_{iu}^{X^0_b*}\,
C_{jc}^{X^0_b}\,C_{ju}^{X^0_a*}
\, D_2(m_{u_i}^2, m_{u_j}^2, m_{X^0_a}^2, m_{X^0_b}^2)
\bigr],
\nonumber\\[0.5em]
\mC_2(\mu_W) &=
-\frac{1}{32\pi^2}
\sum_{i,j=1}^{3}\sum_{X^0_a,X^0_b}
m_{u_i} m_{u_j}\,
\bigl[
(C^{X^0_a*}_{iu} C^{X^0_b*}_{ci})
(C^{X^0_a*}_{ju} C^{X^0_b*}_{cj})
\, D_0(m_{u_i}^2, m_{u_j}^2, m_{X^0_a}^2, m_{X^0_b}^2)
\bigr],
\nonumber\\[0.5em]
\mC_4(\mu_W) &=
-\frac{1}{16\pi^2}
\sum_{i,j=1}^{3}\sum_{X^0_a,X^0_b}
m_{u_i} m_{u_j}\,
\bigl[
(C^{X^0_a*}_{iu} C^{X^0_b*}_{ci})
(C^{X^0_b}_{jc} C^{X^0_a}_{uj})
\, D_0(m_{u_i}^2, m_{u_j}^2, m_{X^0_a}^2, m_{X^0_b}^2)
\bigr],
\nonumber\\[0.5em]
\mC_5(\mu_W) &=
+\frac{1}{32\pi^2}
\sum_{i,j=1}^{3}\sum_{X^0_a,X^0_b}
\bigl[
C^{X^0_a*}_{iu} C^{X^0_b}_{ic}
C^{X^0_a}_{uj} C^{X^0_b*}_{cj}
\, D_2(m_{u_i}^2, m_{u_j}^2, m_{X^0_a}^2, m_{X^0_b}^2)
\bigr],
\label{eq:neutralHiggs_boxes}
\end{align}
where $i,j$ run over the internal up-type quarks and $X^0_a,X^0_b=\{h_u,H_u,A_u\}$ run over the neutral scalars in the propagators.

\paragraph{Charged Higgs Contributions:}
These include both the pure $H^\pm$ box diagrams (e.g. Fig.~\ref{fig:hh-box}) and the mixed $H^+ - W^\pm$ diagrams (e.g. Fig.~\ref{fig:wh-box}), with the charged-Higgs vertices following from Eq.~\eqref{eq:Lyuk}. For a down-type meson with valence quarks $d_i,d_j$ and $i \neq j$, the charged-Higgs contributions to the Wilson coefficients are given below, where $k,l$ run over the internal up-type quarks.  Note that we set $C^{H_d^+}_{ij} = (m_{d_i}t_{\beta}/v)\delta_{ij}$ in the following expressions, since we neglect contributions proportional to $\xi_{ij}^d$. 

\paragraph{Pure $H^+$ diagrams:}
\begin{align}
\mC_1(\mu_W) &=
-\frac{1}{32\pi^2}
\sum_{k,l=1}^{3}
\bigl(C^{H^-_u\dagger}V\bigr)^{*}_{ki}
\bigl(C^{H^-_u\dagger}V\bigr)_{kj}
\bigl(C^{H^-_u\dagger}V\bigr)^{*}_{li}
\bigl(C^{H^-_u\dagger}V\bigr)_{lj}\,
D_2(m_{u_k}^2, m_{u_l}^2, m_{H^\pm}^2, m_{H^\pm}^2),
\nonumber\\[0.5em]
\mC_1'(\mu_W) &=
-\frac{1}{32\pi^2}
\sum_{k,l=1}^{3}
\bigl(V C^{H^+_d}\bigr)^{*}_{ki}
\bigl(V C^{H^+_d}\bigr)_{kj}
\bigl(V C^{H^+_d}\bigr)^{*}_{li}
\bigl(V C^{H^+_d}\bigr)_{lj}\,
D_2(m_{u_k}^2, m_{u_l}^2, m_{H^\pm}^2, m_{H^\pm}^2),
\nonumber\\[0.5em]
\mC_2(\mu_W) &=
-\frac{1}{8\pi^2}
\sum_{k,l=1}^{3}
m_{u_k} m_{u_l}\,
\bigl(V C^{H^+_d}\bigr)^{*}_{ki}
\bigl(C^{H^-_u\dagger}V\bigr)_{kj}
\bigl(V C^{H^+_d}\bigr)^{*}_{li}
\bigl(C^{H^-_u\dagger}V\bigr)_{lj}\,
D_0(m_{u_k}^2, m_{u_l}^2, m_{H^\pm}^2, m_{H^\pm}^2),
\nonumber\\[0.5em]
\mC_2'(\mu_W) &=
-\frac{1}{8\pi^2}
\sum_{k,l=1}^{3}
m_{u_k} m_{u_l}\,
\bigl(C^{H^-_u\dagger}V\bigr)^{*}_{ki}
\bigl(V C^{H^+_d}\bigr)_{kj}
\bigl(C^{H^-_u\dagger}V\bigr)^{*}_{li}
\bigl(V C^{H^+_d}\bigr)_{lj}\,
D_0(m_{u_k}^2, m_{u_l}^2, m_{H^\pm}^2, m_{H^\pm}^2),
\nonumber\\[0.5em]
\mC_4(\mu_W) &=
-\frac{1}{4\pi^2}
\sum_{k,l=1}^{3}
m_{u_k} m_{u_l}\,
\bigl(V C^{H^+_d}\bigr)^{*}_{ki}
\bigl(C^{H^-_u\dagger}V\bigr)_{kj}
\bigl(C^{H^-_u\dagger}V\bigr)^{*}_{li}
\bigl(V C^{H^+_d}\bigr)_{lj}\,
D_0(m_{u_k}^2, m_{u_l}^2, m_{H^\pm}^2, m_{H^\pm}^2),
\nonumber\\[0.5em]
\mC_5(\mu_W) &=
+\frac{1}{8\pi^2}
\sum_{k,l=1}^{3}
\bigl(V C^{H^+_d}\bigr)^{*}_{ki}
\bigl(V C^{H^+_d}\bigr)_{kj}
\bigl(C^{H^-_u\dagger}V\bigr)^{*}_{li}
\bigl(C^{H^-_u\dagger}V\bigr)_{lj}\,
D_2(m_{u_k}^2, m_{u_l}^2, m_{H^\pm}^2, m_{H^\pm}^2).
\label{eq:chargedHiggs_boxes}
\end{align}

\paragraph{Sum of mixed $W^+$--$H^+$ and $G^+$--$H^+$ diagrams:}
The sum of the $W^+$--$H^+$ and $G^+$--$H^+$ diagrams contributes only to the operators $O_1$ and $O_4$. The contributions to the Wilson coefficients are
\begin{align}
\mC_1(\mu_W) =
\frac{g_2^2}{16\pi^2}
&\sum_{k,l=1}^{3}
m_{u_k} m_{u_l}\,
V_{ki}^{*}\, V_{lj}\,
\bigl(C^{H^-_u\dagger}V\bigr)_{kj}
\bigl(C^{H^-_u\dagger}V\bigr)_{li}^{*}
\nonumber \\&\times\frac{
4M_W^2\, D_0(M_W^2, m_{H^\pm}^2, m_{u_k}^2, m_{u_l}^2)
- D_2(M_W^2, m_{H^\pm}^2, m_{u_k}^2, m_{u_l}^2)
}{4M_W^2},
\nonumber
\end{align}
\begin{align}
\mC_4(\mu_W) &=
\frac{g_2^2}{16\pi^2 M_W^2}
\sum_{k,l=1}^{3}
V_{kj}\, V_{li}^{*}\,
\bigl(V C^{H^+_d}\bigr)^{*}_{ki}
\bigl(V C^{H^+_d}\bigr)_{lj}
\bigl[
C_2(\xi_W M_W^2, m_{H^\pm}^2, m_{u_k}^2)
\nonumber \\
 &- C_2(m_{H^\pm}^2, m_{u_k}^2, m_{u_l}^2)
+ m_{u_l}^2\, C_0( \xi_W M_W^2, m_{H^\pm}^2, m_{u_l}^2)
\bigr], \nonumber \\
 &= -\frac{g_2^2\, t_\beta^2\, m_{d_i} m_{d_j}}{16\pi^2 M_W^2 v^2}
\sum_{k,l=1}^3  V_{kj} V_{ki}^{*} V_{lj} V_{li}^{*} \, m_{u_l}^2\, C_0\bigl(m_{H^\pm}^2, m_{u_k}^2, m_{u_l}^2\bigr),
\label{eq:chargedHiggs_mixed}
\end{align}
where $g_2$ the weak coupling constant and $\xi_W$ is the $R_\xi$ gauge parameter. In the final line we have substituted $C^{H_d^+}_{ij} = (m_{d_i}t_{\beta}/v)\delta_{ij}$ and used CKM unitarity to make manifest the gauge-independence of the $\mC_4$ Wilson coefficient once terms proportional to $\xi^d$ are neglected. To calculate the one-loop contributions proportional to $\xi^d$, one would also need to include the one-loop vertex and self-energy corrections to neutral scalar exchange (which cancel the gauge-dependence).

%============================================================
\section{$\bar{B} \to X_s \gamma$ Wilson Coefficients}
\label{app:bsgamma}
%============================================================

Here, we collect the neutral and charged scalar contributions to the Wilson coefficients of the dipole operators in Eq.~\eqref{eq:bsg_decomposition}, obtained via one-loop matching. 

First, it is convenient to define $x_k\equiv m_{u_k}^2/m_{H^\pm}^2$, $y_k^{\phi}\equiv m_{d_k}^2/m_{\phi}^2$ and $Q_d=-1/3$. The relevant loop functions are~\cite{Botella:2014ska}
\begin{align}
A_{H}^{(0)}(x) &= \frac{2+3x-6x^2+x^3+6x\ln x}{24(1-x)^4}\,,
&
A_{H}^{(1)}(x) &= \frac{-3+4x-x^2-2\ln x}{4(1-x)^3}\,, \notag \\
A_{H}^{(2)}(x) &= \frac{-7+5x+8x^2}{36(1-x)^3}+\frac{x(3x-2)\ln x}{6(1-x)^4}\,,
&
A_{H}^{(3)}(x) &= \frac{-3+8x-5x^2+(6x-4)\ln x}{6(1-x)^3}\,.
\end{align}
The charged-scalar contributions then read
\begin{align}
\Delta \mC_7^{H^\pm}(\Lambda)
&=\frac{v^2}{2\,V_{ts}^\ast V_{tb}}\sum_{k=1}^3
\frac{1}{m_{H^\pm}^2}\Bigg[
\bigl(V^\dagger C^{H^-_u}\bigr)_{sk}
\bigl(V^\dagger C^{H^-_u}\bigr)_{bk}^{\!\ast}
A_H^{(2)}(x_k)
\nonumber\\
&\hspace{10em}
-\bigl(V^\dagger C^{H^-_u}\bigr)_{sk}
\bigl(C^{H^+_d\dagger} V^\dagger\bigr)_{bk}^{\!\ast}
\frac{m_{u_k}}{m_b}\,A_H^{(3)}(x_k)
\Bigg],
\label{eq:DeltaC7Hpm}
\\[4pt]
\Delta \mC_7^{\prime\,H^\pm}(\Lambda)
&=\frac{v^2}{2\,V_{ts}^\ast V_{tb}\,m_{H^\pm}^2}\sum_{k=1}^3
\bigl(C^{H^+_d\dagger} V^\dagger\bigr)_{sk}
\bigl(C^{H^+_d\dagger} V^\dagger\bigr)_{bk}^{\!\ast}A_H^{(2)}(x_k),
\label{eq:DeltaC7pHpm}
\end{align}
and the neutral-scalar FCNC contributions, controlled by the flavour off-diagonal couplings $C^{\phi_d}_{ij}$, are
\begin{align}
\Delta \mC_7^{\phi}(\mu_{\phi})
&= -\frac{v^2\,Q_d}{2\,V_{ts}^\ast V_{tb}\,m_\phi^2}\sum_{k=1}^3
\Bigl[
C^{\phi_d}_{sk}\bigl(C^{\phi_d}_{bk}\bigr)^\ast
A_H^{(0)}(y_k^\phi)
+\sigma_\phi\,C^{\phi_d}_{sk}\,C^{\phi_d}_{kb}\,
\frac{m_{d_k}}{m_b}\,
A_H^{(1)}(y_k^\phi)
\Bigr],
\label{eq:DeltaC7neutral}
\\[4pt]
\Delta \mC_7^{\prime\,\phi}(\mu_{\phi})
&= -\frac{v^2\,Q_d}{2\,V_{ts}^\ast V_{tb}\,m_\phi^2}\sum_{k=1}^3
\bigl(C^{\phi_d}_{ks}\bigr)^\ast C^{\phi_d}_{kb}\,
A_H^{(0)}(y_k^\phi),
\label{eq:DeltaC7pneutral}
\end{align}
with $\sigma_\phi=+1$ for the CP-even scalars $\phi=h,H$ and $\sigma_A=-1$ for the CP-odd scalar. The chromomagnetic coefficients follow from Eqs.~\eqref{eq:DeltaC7Hpm}--\eqref{eq:DeltaC7pneutral}
through the replacements
\begin{equation}
\Delta \mC_8^{(\prime)\,H^\pm}:\;
A_H^{(2)}\to -2A_H^{(0)},\quad A_H^{(3)}\to 2A_H^{(1)};
\qquad
\Delta \mC_8^{(\prime)\,\phi}:\;
Q_d\to 1 .
\end{equation}

\bibliographystyle{JHEP}
\bibliography{biblio.bib}

\providecommand{\href}[2]{#2}\begingroup\raggedright\begin{thebibliography}{100}

\bibitem{Abel:2020pzs}
C.~Abel et~al., \emph{{Measurement of the Permanent Electric Dipole Moment of the Neutron}}, \href{https://doi.org/10.1103/PhysRevLett.124.081803}{\emph{Phys. Rev. Lett.} {\bfseries 124} (2020) 081803} [\href{https://arxiv.org/abs/2001.11966}{{\ttfamily 2001.11966}}].

\bibitem{Peccei:1977ur}
R.D.~Peccei and H.R.~Quinn, \emph{{Constraints Imposed by CP Conservation in the Presence of Instantons}}, \href{https://doi.org/10.1103/PhysRevD.16.1791}{\emph{Phys. Rev. D} {\bfseries 16} (1977) 1791}.

\bibitem{Peccei:1977hh}
R.D.~Peccei and H.R.~Quinn, \emph{{CP Conservation in the Presence of Instantons}}, \href{https://doi.org/10.1103/PhysRevLett.38.1440}{\emph{Phys. Rev. Lett.} {\bfseries 38} (1977) 1440}.

\bibitem{Kim:1979if}
J.E.~Kim, \emph{{Weak Interaction Singlet and Strong CP Invariance}}, \href{https://doi.org/10.1103/PhysRevLett.43.103}{\emph{Phys. Rev. Lett.} {\bfseries 43} (1979) 103}.

\bibitem{Shifman:1979if}
M.A.~Shifman, A.I.~Vainshtein and V.I.~Zakharov, \emph{{Can Confinement Ensure Natural CP Invariance of Strong Interactions?}}, \href{https://doi.org/10.1016/0550-3213(80)90209-6}{\emph{Nucl. Phys. B} {\bfseries 166} (1980) 493}.

\bibitem{Dine:1981rt}
M.~Dine, W.~Fischler and M.~Srednicki, \emph{{A Simple Solution to the Strong CP Problem with a Harmless Axion}}, \href{https://doi.org/10.1016/0370-2693(81)90590-6}{\emph{Phys. Lett. B} {\bfseries 104} (1981) 199}.

\bibitem{Srednicki:1985xd}
M.~Srednicki, \emph{{Axion Couplings to Matter. 1. CP Conserving Parts}}, \href{https://doi.org/10.1016/0550-3213(85)90054-9}{\emph{Nucl. Phys. B} {\bfseries 260} (1985) 689}.

\bibitem{Zhitnitsky:1980tq}
A.R.~Zhitnitsky, \emph{{On Possible Suppression of the Axion Hadron Interactions. (In Russian)}}, {\emph{Sov. J. Nucl. Phys.} {\bfseries 31} (1980) 260}.

\bibitem{Davidson:1981zd}
A.~Davidson and K.C.~Wali, \emph{{Minimal Flavor Unification via Multi-Generational Peccei-Quinn Symmetry}}, \href{https://doi.org/10.1103/PhysRevLett.48.11}{\emph{Phys. Rev. Lett.} {\bfseries 48} (1982) 11}.

\bibitem{Davidson:1983fy}
A.~Davidson, V.P.~Nair and K.C.~Wali, \emph{{{Peccei-Quinn} Symmetry as Flavor Symmetry and Grand Unification}}, \href{https://doi.org/10.1103/PhysRevD.29.1504}{\emph{Phys. Rev. D} {\bfseries 29} (1984) 1504}.

\bibitem{Davidson:1984ik}
A.~Davidson and M.A.H.~Vozmediano, \emph{{The Horizontal Axion Alternative: The Interplay of Vacuum Structure and Flavor Interactions}}, \href{https://doi.org/10.1016/0550-3213(84)90616-3}{\emph{Nucl. Phys. B} {\bfseries 248} (1984) 647}.

\bibitem{Geng:1988nc}
C.Q.~Geng and J.N.~Ng, \emph{{Flavor Connections and Neutrino Mass Hierarchy Invariant Invisible Axion Models Without Domain Wall Problem}}, \href{https://doi.org/10.1103/PhysRevD.39.1449}{\emph{Phys. Rev. D} {\bfseries 39} (1989) 1449}.

\bibitem{Sikivie:1982qv}
P.~Sikivie, \emph{{Of Axions, Domain Walls and the Early Universe}}, \href{https://doi.org/10.1103/PhysRevLett.48.1156}{\emph{Phys. Rev. Lett.} {\bfseries 48} (1982) 1156}.

\bibitem{Vilenkin:1982ks}
A.~Vilenkin and A.E.~Everett, \emph{{Cosmic Strings and Domain Walls in Models with Goldstone and Pseudo-Goldstone Bosons}}, \href{https://doi.org/10.1103/PhysRevLett.48.1867}{\emph{Phys. Rev. Lett.} {\bfseries 48} (1982) 1867}.

\bibitem{Cox:2023squ}
P.~Cox, M.J.~Dolan, M.~Hayat, A.~Thamm and R.R.~Volkas, \emph{{Classification of three-family flavoured DFSZ axion models that have no domain wall problem}}, \href{https://doi.org/10.1007/JHEP02(2024)011}{\emph{JHEP} {\bfseries 02} (2024) 011} [\href{https://arxiv.org/abs/2310.16348}{{\ttfamily 2310.16348}}].

\bibitem{Bjorkeroth:2018ipq}
F.~Bj\"orkeroth, L.~Di~Luzio, F.~Mescia and E.~Nardi, \emph{{$U(1)$ flavour symmetries as Peccei-Quinn symmetries}}, \href{https://doi.org/10.1007/JHEP02(2019)133}{\emph{JHEP} {\bfseries 02} (2019) 133} [\href{https://arxiv.org/abs/1811.09637}{{\ttfamily 1811.09637}}].

\bibitem{Rocha:2025ade}
J.R.~Rocha, H.B.~C\^amara and F.R.~Joaquim, \emph{{Flavored Peccei-Quinn symmetries in the minimal $\nu$DFSZ model}},  \href{https://arxiv.org/abs/2504.00088}{{\ttfamily 2504.00088}}.

\bibitem{Rocha:2024twm}
J.R.~Rocha, H.B.~C\^amara, R.G.~Felipe and F.R.~Joaquim, \emph{{Minimal U(1) two-Higgs-doublet models for quark and lepton flavor}}, \href{https://doi.org/10.1103/PhysRevD.110.035027}{\emph{Phys. Rev. D} {\bfseries 110} (2024) 035027} [\href{https://arxiv.org/abs/2406.03331}{{\ttfamily 2406.03331}}].

\bibitem{Ludl:2015lta}
P.O.~Ludl and W.~Grimus, \emph{{A complete survey of texture zeros in general and symmetric quark mass matrices}}, \href{https://doi.org/10.1016/j.physletb.2015.03.033}{\emph{Phys. Lett. B} {\bfseries 744} (2015) 38} [\href{https://arxiv.org/abs/1501.04942}{{\ttfamily 1501.04942}}].

\bibitem{Ramond:1993kv}
P.~Ramond, R.G.~Roberts and G.G.~Ross, \emph{{Stitching the Yukawa quilt}}, \href{https://doi.org/10.1016/0550-3213(93)90159-M}{\emph{Nucl. Phys. B} {\bfseries 406} (1993) 19} [\href{https://arxiv.org/abs/hep-ph/9303320}{{\ttfamily hep-ph/9303320}}].

\bibitem{DAmbrosio:2002vsn}
G.~D'Ambrosio, G.F.~Giudice, G.~Isidori and A.~Strumia, \emph{{Minimal flavor violation: An Effective field theory approach}}, \href{https://doi.org/10.1016/S0550-3213(02)00836-2}{\emph{Nucl. Phys. B} {\bfseries 645} (2002) 155} [\href{https://arxiv.org/abs/hep-ph/0207036}{{\ttfamily hep-ph/0207036}}].

\bibitem{Botella:2009pq}
F.J.~Botella, G.C.~Branco and M.N.~Rebelo, \emph{{Minimal Flavour Violation and Multi-Higgs Models}}, \href{https://doi.org/10.1016/j.physletb.2010.03.014}{\emph{Phys. Lett. B} {\bfseries 687} (2010) 194} [\href{https://arxiv.org/abs/0911.1753}{{\ttfamily 0911.1753}}].

\bibitem{Branco:1996bq}
G.C.~Branco, W.~Grimus and L.~Lavoura, \emph{{Relating the scalar flavor changing neutral couplings to the CKM matrix}}, \href{https://doi.org/10.1016/0370-2693(96)00494-7}{\emph{Phys. Lett. B} {\bfseries 380} (1996) 119} [\href{https://arxiv.org/abs/hep-ph/9601383}{{\ttfamily hep-ph/9601383}}].

\bibitem{Branco:1988iq}
G.C.~Branco, L.~Lavoura and F.~Mota, \emph{{Nearest Neighbor Interactions and the Physical Content of Fritzsch Mass Matrices}}, \href{https://doi.org/10.1103/PhysRevD.39.3443}{\emph{Phys. Rev. D} {\bfseries 39} (1989) 3443}.

\bibitem{Alves:2017xmk}
J.M.~Alves, F.J.~Botella, G.C.~Branco, F.~Cornet-Gomez and M.~Nebot, \emph{{Controlled Flavour Changing Neutral Couplings in Two Higgs Doublet Models}}, \href{https://doi.org/10.1140/epjc/s10052-017-5156-3}{\emph{Eur. Phys. J. C} {\bfseries 77} (2017) 585} [\href{https://arxiv.org/abs/1703.03796}{{\ttfamily 1703.03796}}].

\bibitem{Wolfenstein:1983yz}
L.~Wolfenstein, \emph{{Parametrization of the Kobayashi-Maskawa Matrix}}, \href{https://doi.org/10.1103/PhysRevLett.51.1945}{\emph{Phys. Rev. Lett.} {\bfseries 51} (1983) 1945}.

\bibitem{Ahn:2011fg}
Y.H.~Ahn, H.-Y.~Cheng and S.~Oh, \emph{{Wolfenstein Parametrization at Higher Order: Seeming Discrepancies and Their Resolution}}, \href{https://doi.org/10.1016/j.physletb.2011.08.047}{\emph{Phys. Lett. B} {\bfseries 703} (2011) 571} [\href{https://arxiv.org/abs/1106.0935}{{\ttfamily 1106.0935}}].

\bibitem{DiLuzio:2023ndz}
L.~Di~Luzio, A.W.M.~Guerrera, X.P.~D\'\i{}az and S.~Rigolin, \emph{{On the IR/UV flavour connection in non-universal axion models}}, \href{https://doi.org/10.1007/JHEP06(2023)046}{\emph{JHEP} {\bfseries 06} (2023) 046} [\href{https://arxiv.org/abs/2304.04643}{{\ttfamily 2304.04643}}].

\bibitem{Crivellin:2013wna}
A.~Crivellin, A.~Kokulu and C.~Greub, \emph{{Flavor-phenomenology of two-Higgs-doublet models with generic Yukawa structure}}, \href{https://doi.org/10.1103/PhysRevD.87.094031}{\emph{Phys. Rev. D} {\bfseries 87} (2013) 094031} [\href{https://arxiv.org/abs/1303.5877}{{\ttfamily 1303.5877}}].

\bibitem{Botella:2015hoa}
F.J.~Botella, G.C.~Branco, M.~Nebot and M.N.~Rebelo, \emph{{Flavour Changing Higgs Couplings in a Class of Two Higgs Doublet Models}}, \href{https://doi.org/10.1140/epjc/s10052-016-3993-0}{\emph{Eur. Phys. J. C} {\bfseries 76} (2016) 161} [\href{https://arxiv.org/abs/1508.05101}{{\ttfamily 1508.05101}}].

\bibitem{Celis:2014jua}
A.~Celis, J.~Fuentes-Mart{\'\i}n and H.~Ser{\^o}dio, \emph{{A class of invisible axion models with FCNCs at tree level}}, \href{https://doi.org/10.1007/JHEP12(2014)167}{\emph{JHEP} {\bfseries 12} (2014) 167} [\href{https://arxiv.org/abs/1410.6218}{{\ttfamily 1410.6218}}].

\bibitem{Weinberg:1977ma}
S.~Weinberg, \emph{{A New Light Boson?}}, \href{https://doi.org/10.1103/PhysRevLett.40.223}{\emph{Phys. Rev. Lett.} {\bfseries 40} (1978) 223}.

\bibitem{BaBar:2004xlo}
{\scshape BaBar} collaboration, \emph{{A search for the decay $B^+ \to K^+ \nu \bar{\nu}$}}, \href{https://doi.org/10.1103/PhysRevLett.94.101801}{\emph{Phys. Rev. Lett.} {\bfseries 94} (2005) 101801} [\href{https://arxiv.org/abs/hep-ex/0411061}{{\ttfamily hep-ex/0411061}}].

\bibitem{BaBar:2013npw}
{\scshape BaBar} collaboration, \emph{{Search for $B \to K^{(*)} \nu \overline \nu$ and invisible quarkonium decays}}, \href{https://doi.org/10.1103/PhysRevD.87.112005}{\emph{Phys. Rev. D} {\bfseries 87} (2013) 112005} [\href{https://arxiv.org/abs/1303.7465}{{\ttfamily 1303.7465}}].

\bibitem{CLEO:2008ffk}
{\scshape CLEO} collaboration, \emph{{Precision Measurement of $B(D^+ \to \mu^+ \nu)$ and the Pseudoscalar Decay Constant $f(D^+)$}}, \href{https://doi.org/10.1103/PhysRevD.78.052003}{\emph{Phys. Rev. D} {\bfseries 78} (2008) 052003} [\href{https://arxiv.org/abs/0806.2112}{{\ttfamily 0806.2112}}].

\bibitem{E949:2007xyy}
{\scshape E949, E787} collaboration, \emph{{Measurement of the $K^+ \to \pi^+ \nu \nu$ branching ratio}}, \href{https://doi.org/10.1103/PhysRevD.77.052003}{\emph{Phys. Rev. D} {\bfseries 77} (2008) 052003} [\href{https://arxiv.org/abs/0709.1000}{{\ttfamily 0709.1000}}].

\bibitem{NA62:2025upx}
{\scshape NA62} collaboration, \emph{{Searches for hidden sectors using $K^+\to \pi^+X$ decays}}, \href{https://doi.org/10.1007/JHEP11(2025)143}{\emph{JHEP} {\bfseries 11} (2025) 143} [\href{https://arxiv.org/abs/2507.17286}{{\ttfamily 2507.17286}}].

\bibitem{MartinCamalich:2020dfe}
J.~Martin~Camalich, M.~Pospelov, P.N.H.~Vuong, R.~Ziegler and J.~Zupan, \emph{{Quark Flavor Phenomenology of the QCD Axion}}, \href{https://doi.org/10.1103/PhysRevD.102.015023}{\emph{Phys. Rev. D} {\bfseries 102} (2020) 015023} [\href{https://arxiv.org/abs/2002.04623}{{\ttfamily 2002.04623}}].

\bibitem{Carenza:2019pxu}
P.~Carenza, T.~Fischer, M.~Giannotti, G.~Guo, G.~Mart\'\i{}nez-Pinedo and A.~Mirizzi, \emph{{Improved axion emissivity from a supernova via nucleon-nucleon bremsstrahlung}}, \href{https://doi.org/10.1088/1475-7516/2019/10/016}{\emph{JCAP} {\bfseries 10} (2019) 016} [\href{https://arxiv.org/abs/1906.11844}{{\ttfamily 1906.11844}}].

\bibitem{Beznogov:2018fda}
M.V.~Beznogov, E.~Rrapaj, D.~Page and S.~Reddy, \emph{{Constraints on Axion-like Particles and Nucleon Pairing in Dense Matter from the Hot Neutron Star in HESS J1731-347}}, \href{https://doi.org/10.1103/PhysRevC.98.035802}{\emph{Phys. Rev. C} {\bfseries 98} (2018) 035802} [\href{https://arxiv.org/abs/1806.07991}{{\ttfamily 1806.07991}}].

\bibitem{GrillidiCortona:2015jxo}
G.~Grilli~di Cortona, E.~Hardy, J.~Pardo~Vega and G.~Villadoro, \emph{{The QCD axion, precisely}}, \href{https://doi.org/10.1007/JHEP01(2016)034}{\emph{JHEP} {\bfseries 01} (2016) 034} [\href{https://arxiv.org/abs/1511.02867}{{\ttfamily 1511.02867}}].

\bibitem{Gorghetto:2020qws}
M.~Gorghetto, E.~Hardy and G.~Villadoro, \emph{{More axions from strings}}, \href{https://doi.org/10.21468/SciPostPhys.10.2.050}{\emph{SciPost Phys.} {\bfseries 10} (2021) 050} [\href{https://arxiv.org/abs/2007.04990}{{\ttfamily 2007.04990}}].

\bibitem{Hoof:2021jft}
S.~Hoof, J.~Riess and D.J.E.~Marsh, \emph{{Statistical Uncertainties of the $N_\text{DW} = 1$ QCD Axion Mass Window from Topological Defects}},  \href{https://arxiv.org/abs/2108.09563}{{\ttfamily 2108.09563}}.

\bibitem{Buschmann:2021sdq}
M.~Buschmann, J.W.~Foster, A.~Hook, A.~Peterson, D.E.~Willcox, W.~Zhang et~al., \emph{{Dark matter from axion strings with adaptive mesh refinement}}, \href{https://doi.org/10.1038/s41467-022-28669-y}{\emph{Nature Commun.} {\bfseries 13} (2022) 1049} [\href{https://arxiv.org/abs/2108.05368}{{\ttfamily 2108.05368}}].

\bibitem{Sopov:2022bog}
A.H.~Sopov and R.R.~Volkas, \emph{{VISH$\nu$: a unified solution to five SM shortcomings with a protected electroweak scale}},  \href{https://arxiv.org/abs/2206.11598}{{\ttfamily 2206.11598}}.

\bibitem{Hook:2026grn}
A.~Hook, R.~Mondal and S.~Mukherjee, \emph{{Putting the Brakes on Axion Strings: Friction and Its Impact on the QCD Axion Abundance}},  \href{https://arxiv.org/abs/2603.00237}{{\ttfamily 2603.00237}}.

\bibitem{DiLuzio:2017ogq}
L.~Di~Luzio, F.~Mescia, E.~Nardi, P.~Panci and R.~Ziegler, \emph{{Astrophobic Axions}}, \href{https://doi.org/10.1103/PhysRevLett.120.261803}{\emph{Phys. Rev. Lett.} {\bfseries 120} (2018) 261803} [\href{https://arxiv.org/abs/1712.04940}{{\ttfamily 1712.04940}}].

\bibitem{DiLuzio:2017pfr}
L.~Di~Luzio, F.~Mescia and E.~Nardi, \emph{{Window for preferred axion models}}, \href{https://doi.org/10.1103/PhysRevD.96.075003}{\emph{Phys. Rev. D} {\bfseries 96} (2017) 075003} [\href{https://arxiv.org/abs/1705.05370}{{\ttfamily 1705.05370}}].

\bibitem{ADMX:2025vom}
{\scshape ADMX} collaboration, \emph{{Search for Axion Dark Matter from 1.1 to 1.3~GHz with ADMX}}, \href{https://doi.org/10.1103/d7mg-6sqq}{\emph{Phys. Rev. Lett.} {\bfseries 135} (2025) 191001} [\href{https://arxiv.org/abs/2504.07279}{{\ttfamily 2504.07279}}].

\bibitem{HAYSTAC:2024jch}
{\scshape HAYSTAC} collaboration, \emph{{Dark Matter Axion Search with HAYSTAC Phase II}}, \href{https://doi.org/10.1103/PhysRevLett.134.151006}{\emph{Phys. Rev. Lett.} {\bfseries 134} (2025) 151006} [\href{https://arxiv.org/abs/2409.08998}{{\ttfamily 2409.08998}}].

\bibitem{QUAX:2024fut}
{\scshape QUAX} collaboration, \emph{{Search for axion dark matter with the QUAX{\textendash}LNF tunable haloscope}}, \href{https://doi.org/10.1103/PhysRevD.110.022008}{\emph{Phys. Rev. D} {\bfseries 110} (2024) 022008} [\href{https://arxiv.org/abs/2402.19063}{{\ttfamily 2402.19063}}].

\bibitem{TASEH:2022vvu}
{\scshape TASEH} collaboration, \emph{{First Results from the Taiwan Axion Search Experiment with a Haloscope at 19.6{\,}{\,}{\ensuremath{\mu}}eV}}, \href{https://doi.org/10.1103/PhysRevLett.129.111802}{\emph{Phys. Rev. Lett.} {\bfseries 129} (2022) 111802} [\href{https://arxiv.org/abs/2205.05574}{{\ttfamily 2205.05574}}].

\bibitem{Yi:2022fmn}
A.K.~Yi et~al., \emph{{Axion Dark Matter Search around 4.55{\,}{\,}{\ensuremath{\mu}}eV with Dine-Fischler-Srednicki-Zhitnitskii Sensitivity}}, \href{https://doi.org/10.1103/PhysRevLett.130.071002}{\emph{Phys. Rev. Lett.} {\bfseries 130} (2023) 071002} [\href{https://arxiv.org/abs/2210.10961}{{\ttfamily 2210.10961}}].

\bibitem{OHare:2020AxionLimits}
C.A.J.~O'Hare, \emph{{AxionLimits}: Data and plots of constraints on axions and axion-like particles},  July, 2020.
\newblock 10.5281/zenodo.3932430.

\bibitem{Bernon:2015qea}
J.~Bernon, J.F.~Gunion, H.E.~Haber, Y.~Jiang and S.~Kraml, \emph{{Scrutinizing the alignment limit in two-Higgs-doublet models: m$_h$=125 GeV}}, \href{https://doi.org/10.1103/PhysRevD.92.075004}{\emph{Phys. Rev. D} {\bfseries 92} (2015) 075004} [\href{https://arxiv.org/abs/1507.00933}{{\ttfamily 1507.00933}}].

\bibitem{ATLAS:2021vrm}
{\scshape ATLAS} collaboration, \emph{{Combined measurements of Higgs boson production and decay using up to $139$ fb$^{-1}$ of proton-proton collision data at $\sqrt{s}= 13$ TeV collected with the ATLAS experiment}},  Tech. Rep. \href{https://cds.cern.ch/record/2789544}{ATLAS-CONF-2021-053}, CERN, Geneva (2021).

\bibitem{Gunion:2002zf}
J.F.~Gunion and H.E.~Haber, \emph{{The CP conserving two Higgs doublet model: The Approach to the decoupling limit}}, \href{https://doi.org/10.1103/PhysRevD.67.075019}{\emph{Phys. Rev. D} {\bfseries 67} (2003) 075019} [\href{https://arxiv.org/abs/hep-ph/0207010}{{\ttfamily hep-ph/0207010}}].

\bibitem{Carena:2013ooa}
M.~Carena, I.~Low, N.R.~Shah and C.E.M.~Wagner, \emph{{Impersonating the Standard Model Higgs Boson: Alignment without Decoupling}}, \href{https://doi.org/10.1007/JHEP04(2014)015}{\emph{JHEP} {\bfseries 04} (2014) 015} [\href{https://arxiv.org/abs/1310.2248}{{\ttfamily 1310.2248}}].

\bibitem{Bernon:2015wef}
J.~Bernon, J.F.~Gunion, H.E.~Haber, Y.~Jiang and S.~Kraml, \emph{{Scrutinizing the alignment limit in two-Higgs-doublet models. II. m$_H$=125 GeV}}, \href{https://doi.org/10.1103/PhysRevD.93.035027}{\emph{Phys. Rev. D} {\bfseries 93} (2016) 035027} [\href{https://arxiv.org/abs/1511.03682}{{\ttfamily 1511.03682}}].

\bibitem{Herrero-Garcia:2019mcy}
J.~Herrero-Garcia, M.~Nebot, F.~Rajec, M.~White and A.G.~Williams, \emph{{Higgs Quark Flavor Violation: Simplified Models and Status of General Two-Higgs-Doublet Model}}, \href{https://doi.org/10.1007/JHEP02(2020)147}{\emph{JHEP} {\bfseries 02} (2020) 147} [\href{https://arxiv.org/abs/1907.05900}{{\ttfamily 1907.05900}}].

\bibitem{Chiang:2015cba}
C.-W.~Chiang, H.~Fukuda, M.~Takeuchi and T.T.~Yanagida, \emph{{Flavor-Changing Neutral-Current Decays in Top-Specific Variant Axion Model}}, \href{https://doi.org/10.1007/JHEP11(2015)057}{\emph{JHEP} {\bfseries 11} (2015) 057} [\href{https://arxiv.org/abs/1507.04354}{{\ttfamily 1507.04354}}].

\bibitem{Abbas:2015cua}
G.~Abbas, A.~Celis, X.-Q.~Li, J.~Lu and A.~Pich, \emph{{Flavour-changing top decays in the aligned two-Higgs-doublet model}}, \href{https://doi.org/10.1007/JHEP06(2015)005}{\emph{JHEP} {\bfseries 06} (2015) 005} [\href{https://arxiv.org/abs/1503.06423}{{\ttfamily 1503.06423}}].

\bibitem{Posch:2010hx}
P.~Posch, \emph{{Enhancement of h ---{\ensuremath{>}} gamma gamma in the Two Higgs Doublet Model Type I}}, \href{https://doi.org/10.1016/j.physletb.2011.01.003}{\emph{Phys. Lett. B} {\bfseries 696} (2011) 447} [\href{https://arxiv.org/abs/1001.1759}{{\ttfamily 1001.1759}}].

\bibitem{Altmannshofer:2019ogm}
W.~Altmannshofer, B.~Maddock and D.~Tuckler, \emph{{Rare Top Decays as Probes of Flavorful Higgs Bosons}}, \href{https://doi.org/10.1103/PhysRevD.100.015003}{\emph{Phys. Rev. D} {\bfseries 100} (2019) 015003} [\href{https://arxiv.org/abs/1904.10956}{{\ttfamily 1904.10956}}].

\bibitem{Baum:2008qm}
I.~Baum, G.~Eilam and S.~Bar-Shalom, \emph{{Scalar flavor changing neutral currents and rare top quark decays in a two H iggs doublet model 'for the top quark'}}, \href{https://doi.org/10.1103/PhysRevD.77.113008}{\emph{Phys. Rev. D} {\bfseries 77} (2008) 113008} [\href{https://arxiv.org/abs/0802.2622}{{\ttfamily 0802.2622}}].

\bibitem{Chen:2013qta}
K.-F.~Chen, W.-S.~Hou, C.~Kao and M.~Kohda, \emph{{When the Higgs meets the Top: Search for $t \to ch^0$ at the LHC}}, \href{https://doi.org/10.1016/j.physletb.2013.07.060}{\emph{Phys. Lett. B} {\bfseries 725} (2013) 378} [\href{https://arxiv.org/abs/1304.8037}{{\ttfamily 1304.8037}}].

\bibitem{Altunkaynak:2015twa}
B.~Altunkaynak, W.-S.~Hou, C.~Kao, M.~Kohda and B.~McCoy, \emph{{Flavor Changing Heavy Higgs Interactions at the LHC}}, \href{https://doi.org/10.1016/j.physletb.2015.10.024}{\emph{Phys. Lett. B} {\bfseries 751} (2015) 135} [\href{https://arxiv.org/abs/1506.00651}{{\ttfamily 1506.00651}}].

\bibitem{Chiang:2017fjr}
C.-W.~Chiang, H.~Fukuda, M.~Takeuchi and T.T.~Yanagida, \emph{{Current Status of Top-Specific Variant Axion Model}}, \href{https://doi.org/10.1103/PhysRevD.97.035015}{\emph{Phys. Rev. D} {\bfseries 97} (2018) 035015} [\href{https://arxiv.org/abs/1711.02993}{{\ttfamily 1711.02993}}].

\bibitem{ATLAS:2024mih}
{\scshape ATLAS} collaboration, \emph{{Search for flavour-changing neutral-current couplings between the top quark and the Higgs boson in multi-lepton final states in 13~TeV pp collisions with the ATLAS detector}}, \href{https://doi.org/10.1140/epjc/s10052-024-12994-1}{\emph{Eur. Phys. J. C} {\bfseries 84} (2024) 757} [\href{https://arxiv.org/abs/2404.02123}{{\ttfamily 2404.02123}}].

\bibitem{Kamenik:2023hvi}
J.F.~Kamenik, A.~Korajac, M.~Szewc, M.~Tammaro and J.~Zupan, \emph{{Flavor-violating Higgs and Z boson decays at a future circular lepton collider}}, \href{https://doi.org/10.1103/PhysRevD.109.L011301}{\emph{Phys. Rev. D} {\bfseries 109} (2024) L011301} [\href{https://arxiv.org/abs/2306.17520}{{\ttfamily 2306.17520}}].

\bibitem{Arroyo-Urena:2025mju}
M.A.~Arroyo-Ure{\~n}a, D.~Carre{\~n}o, A.I.~Garc{\'\i}a-Gutierrez, M.G.~Villanueva-Utrilla and T.A.~Valencia-P{\'e}rez, \emph{{Beyond the Standard Model Higgs physics: Hunting $h \rightarrow bs$ with Higgs-strahlung at CEPC and FCC-ee}},  \href{https://arxiv.org/abs/2507.01141}{{\ttfamily 2507.01141}}.

\bibitem{HFLAV:2022esi}
{\scshape HFLAV} collaboration, \emph{{Averages of b-hadron, c-hadron, and {\ensuremath{\tau}}-lepton properties as of 2021}}, \href{https://doi.org/10.1103/PhysRevD.107.052008}{\emph{Phys. Rev. D} {\bfseries 107} (2023) 052008} [\href{https://arxiv.org/abs/2206.07501}{{\ttfamily 2206.07501}}].

\bibitem{ParticleDataGroup:2024cfk}
{\scshape Particle Data Group} collaboration, \emph{{Review of particle physics}}, \href{https://doi.org/10.1103/PhysRevD.110.030001}{\emph{Phys. Rev. D} {\bfseries 110} (2024) 030001}.

\bibitem{HeavyFlavorAveragingGroupHFLAV:2024ctg}
{\scshape Heavy Flavor Averaging Group (HFLAV)} collaboration, \emph{{Averages of $b$-hadron, $c$-hadron, and $\tau$-lepton properties as of 2023}},  \href{https://arxiv.org/abs/2411.18639}{{\ttfamily 2411.18639}}.

\bibitem{Gabbiani:1996hi}
F.~Gabbiani, E.~Gabrielli, A.~Masiero and L.~Silvestrini, \emph{{A Complete analysis of FCNC and CP constraints in general SUSY extensions of the standard model}}, \href{https://doi.org/10.1016/0550-3213(96)00390-2}{\emph{Nucl. Phys. B} {\bfseries 477} (1996) 321} [\href{https://arxiv.org/abs/hep-ph/9604387}{{\ttfamily hep-ph/9604387}}].

\bibitem{FermilabLattice:2016ipl}
{\scshape Fermilab Lattice, MILC} collaboration, \emph{{$B^0_{(s)}$-mixing matrix elements from lattice QCD for the Standard Model and beyond}}, \href{https://doi.org/10.1103/PhysRevD.93.113016}{\emph{Phys. Rev. D} {\bfseries 93} (2016) 113016} [\href{https://arxiv.org/abs/1602.03560}{{\ttfamily 1602.03560}}].

\bibitem{Greljo:2022jac}
A.~Greljo, J.~Salko, A.~Smolkovi{\v{c}} and P.~Stangl, \emph{{Rare b decays meet high-mass Drell-Yan}}, \href{https://doi.org/10.1007/JHEP05(2023)087}{\emph{JHEP} {\bfseries 05} (2023) 087} [\href{https://arxiv.org/abs/2212.10497}{{\ttfamily 2212.10497}}].

\bibitem{Garron:2016mva}
{\scshape RBC/UKQCD} collaboration, \emph{{Neutral Kaon Mixing Beyond the Standard Model with $n_f=2+1$ Chiral Fermions Part 1: Bare Matrix Elements and Physical Results}}, \href{https://doi.org/10.1007/JHEP11(2016)001}{\emph{JHEP} {\bfseries 11} (2016) 001} [\href{https://arxiv.org/abs/1609.03334}{{\ttfamily 1609.03334}}].

\bibitem{FlavourLatticeAveragingGroupFLAG:2024oxs}
{\scshape Flavour Lattice Averaging Group (FLAG)} collaboration, \emph{{FLAG Review 2024}},  \href{https://arxiv.org/abs/2411.04268}{{\ttfamily 2411.04268}}.

\bibitem{Inami:1980fz}
T.~Inami and C.S.~Lim, \emph{{Effects of Superheavy Quarks and Leptons in Low-Energy Weak Processes k(L) ---{\ensuremath{>}} mu anti-mu, K+ ---{\ensuremath{>}} pi+ Neutrino anti-neutrino and K0 {\ensuremath{<}}---{\ensuremath{>}} anti-K0}}, \href{https://doi.org/10.1143/PTP.65.297}{\emph{Prog. Theor. Phys.} {\bfseries 65} (1981) 297}.

\bibitem{Albrecht:2024oyn}
J.~Albrecht, F.~Bernlochner, A.~Lenz and A.~Rusov, \emph{{Lifetimes of b-hadrons and mixing of neutral B-mesons: theoretical and experimental status}}, \href{https://doi.org/10.1140/epjs/s11734-024-01124-3}{\emph{Eur. Phys. J. ST} {\bfseries 233} (2024) 359} [\href{https://arxiv.org/abs/2402.04224}{{\ttfamily 2402.04224}}].

\bibitem{Dowdall:2019bea}
R.J.~Dowdall, C.T.H.~Davies, R.R.~Horgan, G.P.~Lepage, C.J.~Monahan, J.~Shigemitsu et~al., \emph{{Neutral B-meson mixing from full lattice QCD at the physical point}}, \href{https://doi.org/10.1103/PhysRevD.100.094508}{\emph{Phys. Rev. D} {\bfseries 100} (2019) 094508} [\href{https://arxiv.org/abs/1907.01025}{{\ttfamily 1907.01025}}].

\bibitem{King:2019lal}
D.~King, A.~Lenz and T.~Rauh, \emph{{B$_{s}$ mixing observables and |V$_{td}$/V$_{ts}$| from sum rules}}, \href{https://doi.org/10.1007/JHEP05(2019)034}{\emph{JHEP} {\bfseries 05} (2019) 034} [\href{https://arxiv.org/abs/1904.00940}{{\ttfamily 1904.00940}}].

\bibitem{Brod:2011ty}
J.~Brod and M.~Gorbahn, \emph{{Next-to-Next-to-Leading-Order Charm-Quark Contribution to the $CP$ Violation Parameter $\epsilon_K$ and $\Delta M_K$}}, \href{https://doi.org/10.1103/PhysRevLett.108.121801}{\emph{Phys. Rev. Lett.} {\bfseries 108} (2012) 121801} [\href{https://arxiv.org/abs/1108.2036}{{\ttfamily 1108.2036}}].

\bibitem{Botella:2014ska}
F.J.~Botella, G.C.~Branco, A.~Carmona, M.~Nebot, L.~Pedro and M.N.~Rebelo, \emph{{Physical Constraints on a Class of Two-Higgs Doublet Models with FCNC at tree level}}, \href{https://doi.org/10.1007/JHEP07(2014)078}{\emph{JHEP} {\bfseries 07} (2014) 078} [\href{https://arxiv.org/abs/1401.6147}{{\ttfamily 1401.6147}}].

\bibitem{Ferreira:2019aps}
P.M.~Ferreira and L.~Lavoura, \emph{{No strong $CP$ violation up to the one-loop level in a two-Higgs-doublet model}}, \href{https://doi.org/10.1140/epjc/s10052-019-7053-4}{\emph{Eur. Phys. J. C} {\bfseries 79} (2019) 552} [\href{https://arxiv.org/abs/1904.08438}{{\ttfamily 1904.08438}}].

\bibitem{Carrasco:2015pra}
{\scshape ETM} collaboration, \emph{{{\ensuremath{\Delta}}S=2 and {\ensuremath{\Delta}}C=2 bag parameters in the standard model and beyond from N$_f$=2+1+1 twisted-mass lattice QCD}}, \href{https://doi.org/10.1103/PhysRevD.92.034516}{\emph{Phys. Rev. D} {\bfseries 92} (2015) 034516} [\href{https://arxiv.org/abs/1505.06639}{{\ttfamily 1505.06639}}].

\bibitem{Golowich:2005pt}
E.~Golowich and A.A.~Petrov, \emph{{Short distance analysis of D0 - anti-D0 mixing}}, \href{https://doi.org/10.1016/j.physletb.2005.08.023}{\emph{Phys. Lett. B} {\bfseries 625} (2005) 53} [\href{https://arxiv.org/abs/hep-ph/0506185}{{\ttfamily hep-ph/0506185}}].

\bibitem{Dulibic:2025emg}
L.~Dulibi{\'c}, B.~Meli{\'c} and A.A.~Petrov, \emph{{Nonperturbative dynamics in D-meson mixing}}, \href{https://doi.org/10.1007/JHEP12(2025)112}{\emph{JHEP} {\bfseries 12} (2025) 112} [\href{https://arxiv.org/abs/2508.16337}{{\ttfamily 2508.16337}}].

\bibitem{DiCarlo:2025mnm}
M.~Di~Carlo, F.~Erben and M.T.~Hansen, \emph{{Long distance contributions to neutral D-meson mixing from lattice QCD}}, \href{https://doi.org/10.1007/JHEP07(2025)229}{\emph{JHEP} {\bfseries 07} (2025) 229} [\href{https://arxiv.org/abs/2504.16189}{{\ttfamily 2504.16189}}].

\bibitem{Bazavov:2017weg}
A.~Bazavov et~al., \emph{{Short-distance matrix elements for $D^0$-meson mixing for $N_f=2+1$ lattice QCD}}, \href{https://doi.org/10.1103/PhysRevD.97.034513}{\emph{Phys. Rev. D} {\bfseries 97} (2018) 034513} [\href{https://arxiv.org/abs/1706.04622}{{\ttfamily 1706.04622}}].

\bibitem{Buras:2008nn}
A.J.~Buras and D.~Guadagnoli, \emph{{Correlations among new CP violating effects in $\Delta$ F = 2 observables}}, \href{https://doi.org/10.1103/PhysRevD.78.033005}{\emph{Phys. Rev. D} {\bfseries 78} (2008) 033005} [\href{https://arxiv.org/abs/0805.3887}{{\ttfamily 0805.3887}}].

\bibitem{Brod:2019rzc}
J.~Brod, M.~Gorbahn and E.~Stamou, \emph{{Standard-Model Prediction of $\epsilon_K$ with Manifest Quark-Mixing Unitarity}}, \href{https://doi.org/10.1103/PhysRevLett.125.171803}{\emph{Phys. Rev. Lett.} {\bfseries 125} (2020) 171803} [\href{https://arxiv.org/abs/1911.06822}{{\ttfamily 1911.06822}}].

\bibitem{Haber:1999zh}
H.E.~Haber and H.E.~Logan, \emph{{Radiative corrections to the Z b anti-b vertex and constraints on extended Higgs sectors}}, \href{https://doi.org/10.1103/PhysRevD.62.015011}{\emph{Phys. Rev. D} {\bfseries 62} (2000) 015011} [\href{https://arxiv.org/abs/hep-ph/9909335}{{\ttfamily hep-ph/9909335}}].

\bibitem{Hernandez-Sanchez:2012vxa}
J.~Hernandez-Sanchez, S.~Moretti, R.~Noriega-Papaqui and A.~Rosado, \emph{{Off-diagonal terms in Yukawa textures of the Type-III 2-Higgs doublet model and light charged Higgs boson phenomenology}}, \href{https://doi.org/10.1007/JHEP07(2013)044}{\emph{JHEP} {\bfseries 07} (2013) 044} [\href{https://arxiv.org/abs/1212.6818}{{\ttfamily 1212.6818}}].

\bibitem{Misiak:2006ab}
M.~Misiak and M.~Steinhauser, \emph{{NNLO QCD corrections to the anti-B ---{\ensuremath{>}} X(s) gamma matrix elements using interpolation in m(c)}}, \href{https://doi.org/10.1016/j.nuclphysb.2006.11.027}{\emph{Nucl. Phys. B} {\bfseries 764} (2007) 62} [\href{https://arxiv.org/abs/hep-ph/0609241}{{\ttfamily hep-ph/0609241}}].

\bibitem{Buras:2011zb}
A.J.~Buras, L.~Merlo and E.~Stamou, \emph{{The Impact of Flavour Changing Neutral Gauge Bosons on $\bar{B} -> X_s \gamma$}}, \href{https://doi.org/10.1007/JHEP08(2011)124}{\emph{JHEP} {\bfseries 08} (2011) 124} [\href{https://arxiv.org/abs/1105.5146}{{\ttfamily 1105.5146}}].

\bibitem{Blanke:2011ry}
M.~Blanke, A.J.~Buras, K.~Gemmler and T.~Heidsieck, \emph{{Delta F = 2 observables and $B \to X_q \gamma$ decays in the Left-Right Model: Higgs particles striking back}}, \href{https://doi.org/10.1007/JHEP03(2012)024}{\emph{JHEP} {\bfseries 03} (2012) 024} [\href{https://arxiv.org/abs/1111.5014}{{\ttfamily 1111.5014}}].

\bibitem{Blanke:2012tv}
M.~Blanke, B.~Shakya, P.~Tanedo and Y.~Tsai, \emph{{The Birds and the Bs in RS: The $b to s \gamma$ penguin in a warped extra dimension}}, \href{https://doi.org/10.1007/JHEP08(2012)038}{\emph{JHEP} {\bfseries 08} (2012) 038} [\href{https://arxiv.org/abs/1203.6650}{{\ttfamily 1203.6650}}].

\bibitem{Misiak:2020vlo}
M.~Misiak, A.~Rehman and M.~Steinhauser, \emph{{Towards $ \overline{B}\to {X}_s\gamma $ at the NNLO in QCD without interpolation in m$_{c}$}}, \href{https://doi.org/10.1007/JHEP06(2020)175}{\emph{JHEP} {\bfseries 06} (2020) 175} [\href{https://arxiv.org/abs/2002.01548}{{\ttfamily 2002.01548}}].

\bibitem{Papaefstathiou:2017xuv}
A.~Papaefstathiou and G.~Tetlalmatzi-Xolocotzi, \emph{{Rare top quark decays at a 100 TeV proton{\textendash}proton collider: $t \rightarrow bWZ$ and $t\rightarrow hc$}}, \href{https://doi.org/10.1140/epjc/s10052-018-5701-8}{\emph{Eur. Phys. J. C} {\bfseries 78} (2018) 214} [\href{https://arxiv.org/abs/1712.06332}{{\ttfamily 1712.06332}}].

\bibitem{ATLAS:2021upq}
{\scshape ATLAS} collaboration, \emph{{Search for charged Higgs bosons decaying into a top quark and a bottom quark at $ \sqrt{\mathrm{s}} $ = 13 TeV with the ATLAS detector}}, \href{https://doi.org/10.1007/JHEP06(2021)145}{\emph{JHEP} {\bfseries 06} (2021) 145} [\href{https://arxiv.org/abs/2102.10076}{{\ttfamily 2102.10076}}].

\bibitem{ATLAS:2024dkv}
{\scshape ATLAS} collaboration, \emph{{Search for heavy neutral Higgs bosons decaying to a top quark pair in 140 fb$^{-1}$ of proton-proton collision data at $\sqrt{s}=13$ TeV}},  Tech. Rep. \href{https://cds.cern.ch/record/2891813}{ATLAS-CONF-2024-001}, CERN, Geneva (2024).

\bibitem{CMS:2025dzq}
{\scshape CMS} collaboration, \emph{{Search for heavy pseudoscalar and scalar bosons decaying to a top quark pair in proton{\textendash}proton collisions at $\sqrt{s} = 13\,\textrm{TeV}$}}, \href{https://doi.org/10.1088/1361-6633/ae2207}{\emph{Rept. Prog. Phys.} {\bfseries 88} (2025) 127801} [\href{https://arxiv.org/abs/2507.05119}{{\ttfamily 2507.05119}}].

\bibitem{Cepeda:2019klc}
M.~Cepeda et~al., \emph{{Report from Working Group 2}: {Higgs Physics at the HL-LHC and HE-LHC}}, \href{https://doi.org/10.23731/CYRM-2019-007.221}{\emph{CERN Yellow Rep. Monogr.} {\bfseries 7} (2019) 221} [\href{https://arxiv.org/abs/1902.00134}{{\ttfamily 1902.00134}}].

\bibitem{Bahl:2020kwe}
H.~Bahl, P.~Bechtle, S.~Heinemeyer, S.~Liebler, T.~Stefaniak and G.~Weiglein, \emph{{HL-LHC and ILC sensitivities in the hunt for heavy Higgs bosons}}, \href{https://doi.org/10.1140/epjc/s10052-020-08472-z}{\emph{Eur. Phys. J. C} {\bfseries 80} (2020) 916} [\href{https://arxiv.org/abs/2005.14536}{{\ttfamily 2005.14536}}].

\bibitem{Aebischer:2018bkb}
J.~Aebischer, J.~Kumar and D.M.~Straub, \emph{{Wilson: a Python package for the running and matching of Wilson coefficients above and below the electroweak scale}}, \href{https://doi.org/10.1140/epjc/s10052-018-6492-7}{\emph{Eur. Phys. J. C} {\bfseries 78} (2018) 1026} [\href{https://arxiv.org/abs/1804.05033}{{\ttfamily 1804.05033}}].

\bibitem{Branco:2011iw}
G.C.~Branco, P.M.~Ferreira, L.~Lavoura, M.N.~Rebelo, M.~Sher and J.P.~Silva, \emph{{Theory and phenomenology of two-Higgs-doublet models}}, \href{https://doi.org/10.1016/j.physrep.2012.02.002}{\emph{Phys. Rept.} {\bfseries 516} (2012) 1} [\href{https://arxiv.org/abs/1106.0034}{{\ttfamily 1106.0034}}].

\bibitem{Deshpande:1977rw}
N.G.~Deshpande and E.~Ma, \emph{{Pattern of Symmetry Breaking with Two Higgs Doublets}}, \href{https://doi.org/10.1103/PhysRevD.18.2574}{\emph{Phys. Rev. D} {\bfseries 18} (1978) 2574}.

\bibitem{Barroso:2012mj}
A.~Barroso, P.M.~Ferreira, I.P.~Ivanov, R.~Santos and J.P.~Silva, \emph{{Evading death by vacuum}}, \href{https://doi.org/10.1140/epjc/s10052-013-2537-0}{\emph{Eur. Phys. J. C} {\bfseries 73} (2013) 2537} [\href{https://arxiv.org/abs/1211.6119}{{\ttfamily 1211.6119}}].

\bibitem{Ivanov:2015nea}
I.P.~Ivanov and J.P.~Silva, \emph{{Tree-level metastability bounds for the most general two Higgs doublet model}}, \href{https://doi.org/10.1103/PhysRevD.92.055017}{\emph{Phys. Rev. D} {\bfseries 92} (2015) 055017} [\href{https://arxiv.org/abs/1507.05100}{{\ttfamily 1507.05100}}].

\bibitem{Lee:1977eg}
B.W.~Lee, C.~Quigg and H.B.~Thacker, \emph{{Weak Interactions at Very High-Energies: The Role of the Higgs Boson Mass}}, \href{https://doi.org/10.1103/PhysRevD.16.1519}{\emph{Phys. Rev. D} {\bfseries 16} (1977) 1519}.

\bibitem{Lee:1977yc}
B.W.~Lee, C.~Quigg and H.B.~Thacker, \emph{{The Strength of Weak Interactions at Very High-Energies and the Higgs Boson Mass}}, \href{https://doi.org/10.1103/PhysRevLett.38.883}{\emph{Phys. Rev. Lett.} {\bfseries 38} (1977) 883}.

\bibitem{Grimus:2007if}
W.~Grimus, L.~Lavoura, O.M.~Ogreid and P.~Osland, \emph{{A Precision constraint on multi-Higgs-doublet models}}, \href{https://doi.org/10.1088/0954-3899/35/7/075001}{\emph{J. Phys. G} {\bfseries 35} (2008) 075001} [\href{https://arxiv.org/abs/0711.4022}{{\ttfamily 0711.4022}}].

\bibitem{Grimus:2008nb}
W.~Grimus, L.~Lavoura, O.M.~Ogreid and P.~Osland, \emph{{The Oblique parameters in multi-Higgs-doublet models}}, \href{https://doi.org/10.1016/j.nuclphysb.2008.04.019}{\emph{Nucl. Phys. B} {\bfseries 801} (2008) 81} [\href{https://arxiv.org/abs/0802.4353}{{\ttfamily 0802.4353}}].

\bibitem{Haber:2010bw}
H.E.~Haber and D.~O'Neil, \emph{{Basis-independent methods for the two-Higgs-doublet model III: The CP-conserving limit, custodial symmetry, and the oblique parameters S, T, U}}, \href{https://doi.org/10.1103/PhysRevD.83.055017}{\emph{Phys. Rev. D} {\bfseries 83} (2011) 055017} [\href{https://arxiv.org/abs/1011.6188}{{\ttfamily 1011.6188}}].

\bibitem{Huang:2020hdv}
G.-y.~Huang and S.~Zhou, \emph{{Precise Values of Running Quark and Lepton Masses in the Standard Model}}, \href{https://doi.org/10.1103/PhysRevD.103.016010}{\emph{Phys. Rev. D} {\bfseries 103} (2021) 016010} [\href{https://arxiv.org/abs/2009.04851}{{\ttfamily 2009.04851}}].

\bibitem{Arkani-Hamed:2026uos}
N.~Arkani-Hamed, C.~Figueiredo, L.J.~Hall and C.A.~Manzari, \emph{{The Very Nearly Right Theory of Flavor}},  \href{https://arxiv.org/abs/2607.27315}{{\ttfamily 2607.27315}}.

\end{thebibliography}\endgroup

\end{document}